\documentclass{tcibook}
\pdfoutput=1
\usepackage{fancyhea}
\usepackage{work}
\usepackage{bm}       
\usepackage{graphicx}
\usepackage{hyperref} 
\usepackage{xspace}
\usepackage{placeins}
\usepackage{multirow}

\newcommand{\nc}{\newcommand}  

\def\ibid{{\it ibid}.}

\def\beq{\begin{equation}}
\def\eeq#1{\label{#1}\end{equation}}
\def\eeqn{\end{equation}}

\newenvironment{Eqnarray}%
   {\arraycolsep 0.14em\begin{eqnarray}}{\end{eqnarray}}
\def\beqa{\begin{Eqnarray}}
\def\eeqa#1{\label{#1}\end{Eqnarray}}
\def\eeqan{\end{Eqnarray}}

\nc{\ra}{\rightarrow}  
\nc{\slsh}{\slash\hspace*{-0.22cm}}
\def\Re{{\cal R \mskip-4mu \lower.1ex \hbox{\it e}\,}}
\def\Im{{\cal I \mskip-5mu \lower.1ex \hbox{\it m}\,}}

\nc{\vev}[1]{ \left\langle {#1} \right\rangle }
\nc{\bra}[1]{ \langle {#1} | }
\nc{\ket}[1]{ | {#1} \rangle }
\nc{\fb}{\,{\rm fb}^{-1}}
\nc{\ev}{{\rm eV}}
\nc{\kev}{{\rm keV}}
\nc{\Mev}{{\rm MeV}}
\nc{\gev}{{\rm GeV}}
\nc{\tev}{{\rm TeV}}
\nc{\mev}{{\rm MeV}}

\def\del{\partial}
\def\Dslash{\not{\hbox{\kern-4pt $D$}}}
\def\dslash{\not{\hbox{\kern-2pt $\del$}}}
\def\pslash{\not{\hbox{\kern-2pt $p$}}}
\def\ETmiss{ \not{\hbox{\kern-4pt $E$}}_T }

\def\BR{\mbox{\rm BR}}

\def\msb{{\bar{\ssstyle M \kern -1pt S}}}

\def\babar{\mbox{\sl B\hspace{-0.4em} {\small\sl A}\hspace{-0.37em} \sl B\hspace{-0.4em} {\small\sl A\hspace{-0.02em}R}}}

\newcommand{\sn}[2]{\ensuremath{#1\times10^{#2}}\xspace}

\begin{document}

\def\bibname{References}
\bibliographystyle{plain}

\raggedbottom

\pagenumbering{roman}

\parindent=0pt
\parskip=8pt
\setlength{\evensidemargin}{0pt}
\setlength{\oddsidemargin}{0pt}
\setlength{\marginparsep}{0.0in}
\setlength{\marginparwidth}{0.0in}
\marginparpush=0pt


\pagenumbering{arabic}

\renewcommand{\chapname}{chap:intro_}
\renewcommand{\chapterdir}{.}
\renewcommand{\arraystretch}{1.25}
\addtolength{\arraycolsep}{-3pt}




%
\begin{center}
{\Huge\bf Charged Leptons}
\end{center}


\begin{center}\begin{boldmath}



\begin{center}

\begin{large} {\bf Conveners: B.~C.~K.~Casey$^{9}$, Y.~Grossman$^{7}$, D.~G.~Hitlin$^{5}$} \end{large}

J.~Albrecht$^{17}$, 
M.~Artuso$^{16}$, 
K.~Babu$^{14}$,
R.H.~Bernstein$^{9}$,
T.~Blum$^{6}$,
D.N.~Brown$^{11}$, 
B.C.K.~Casey$^{9}$,
C.-h.~Cheng$^{5}$,
V.~Cirigliano$^{12}$,
A.~Cohen$^{2}$,
A.~Deshpande$^{15}$,
E.C.~Dukes$^{19}$, 
B.~Echenard$^{5}$,
A.~Gaponenko$^{9}$,
D.~Glenzinski$^{9}$,
M.~Gonzalez-Alonso$^{20}$, 
F.~Grancagnolo$^{10}$,
Y.~Grossman$^{7}$,
R.C.~Group$^{9,19}$,
R.~Harnik$^{9}$,
D.G.~Hitlin$^{5}$, 
B.~Kiburg$^{9}$,
K.~Knoepfel$^{9}$,
K.~Kumar$^{13}$,
G.~Lim$^{4}$,
Z.-T.~Lu$^{1}$,
D.~McKeen$^{18}$,
J.P.~Miller$^{2}$,
M.~Ramsey-Musolf$^{13,20}$, 
R.~Ray$^{9}$,
B.L.~Roberts$^{2}$,
M.~Rominsky$^{9}$,
Y.~Semertzidis$^{3}$,
D.~Stoeckinger$^{8}$,
R.~Talman$^{7}$,
R.~Van De Water$^{9}$,
P.~Winter$^{1}$

\end{center}

\begin{center}

$^{1}$Argonne National Laboratory, Argonne, IL 60439, USA\\
$^{2}$Boston University, Boston, MA 02215, USA\\
$^{3}$Brookhaven National Lab, Upton, NY, 11973-5000, USA\\
$^{4}$University of California, Irvine, Irvine, CA 92698, USA\\
$^{5}$California Institute of Technology, Pasadena, CA 91125, USA\\
$^{6}$University of Connecticut, Storrs, CT 06269-3046, USA\\
$^{7}$Cornell University, Laboratory for Elementary Particle Physics, Ithaca, NY 14853, USA\\
$^{8}$Dresden University of Technology, Dresden, Germany\\
$^{9}$Fermi National Accelerator Laboratory, Batavia, IL 60510, USA\\
$^{10}$INFN Sezione di Lecce and Universit\'a del Salento, Lecce, Italy\\
$^{11}$Lawrence Berkeley National Laboratory, Berkeley, CA 94720, USA\\
$^{12}$Los Alamos National Laboratory, Los Alamos, NM 87545, USA\\
$^{13}$University of Massachusetts, Amherst, MA 01003, USA\\
$^{14}$Oklahoma State University, Stillwater, OK 74078, USA\\
$^{15}$Stony Brook University, Stony Brook, NY 11794, USA\\
$^{16}$Syracuse University, Syracuse, NY, 13244-5040, USA\\
$^{17}$Technical University of Dortmund, Dortmund, Germany\\
$^{18}$University of Victoria, Victoria, BC V8N 1M5, Canada\\
$^{19}$University of Virginia, Charlottesville, VA 22904-4714, USA\\
$^{20}$University of Wisconsin, Madison, WI 53706, USA\\

\end{center}

\begin{center}
{\large\bf Abstract}

\parbox{\linewidth}{
This is the report of the Intensity Frontier Charged Lepton Working Group of the 2013 Community Summer Study ``Snowmass on the Mississippi'', summarizing the current status and future experimental opportunities in muon and tau lepton studies and their sensitivity to new physics. These include searches for charged lepton flavor violation, measurements of magnetic and electric dipole moments, and precision measurements of the decay spectrum and parity-violating asymmetries.}

\end{center}



\end{boldmath}\end{center}

\makeatletter
\renewcommand{\paragraph}{\@startsection{paragraph}{4}{0ex}%
   {-3.25ex plus -1ex minus -0.2ex}%
   {1.5ex plus 0.2ex}%
   {\normalfont\normalsize\bfseries}}
\makeatother

\stepcounter{secnumdepth}
\stepcounter{tocdepth}

\tableofcontents

\chapter{Executive Summary}\label{sec:cl:execsum}
\medskip

The enormous physics potential of the charged
lepton experimental program was very much in evidence at the Workshop. There are discovery opportunities in experiments that will be conducted over the coming decade using existing facilities and in more sensitive experiments possible with future facilities such as Project X.
Exquisitly sensitive searches for rare decays of muons and tau leptons, together with precision measurements of their properties will either elucidate the scale and dynamics of flavor generation, or
limit the scale of flavor generation to well above $10^4$ TeV.  

The crown jewel of the program is the discovery potential of muon and tau decay experiments searching for charged lepton flavor violation (CLFV) with several orders-of-magnitude improvement in sensitivity in
multiple processes.  There is an
international program of CLFV searches, with experiments recently completed, currently running, and
soon to be constructed in the United States, Japan, and Europe.  These include the completion of the MEG experiment at PSI, an upgrade of MEG,  the proposed mu3e search at PSI, new searches from muon to electron conversion (Mu2e at Fermilab, COMET at J-PARC), SuperKEKB, and over the longer term, experiments exploiting megawatt proton sources such as Project X.

\begin{figure}[ht]
\begin{minipage}[b]{0.48\linewidth}
\centering
\includegraphics[trim = 45mm 130mm 50mm 10mm, clip, width=\linewidth]{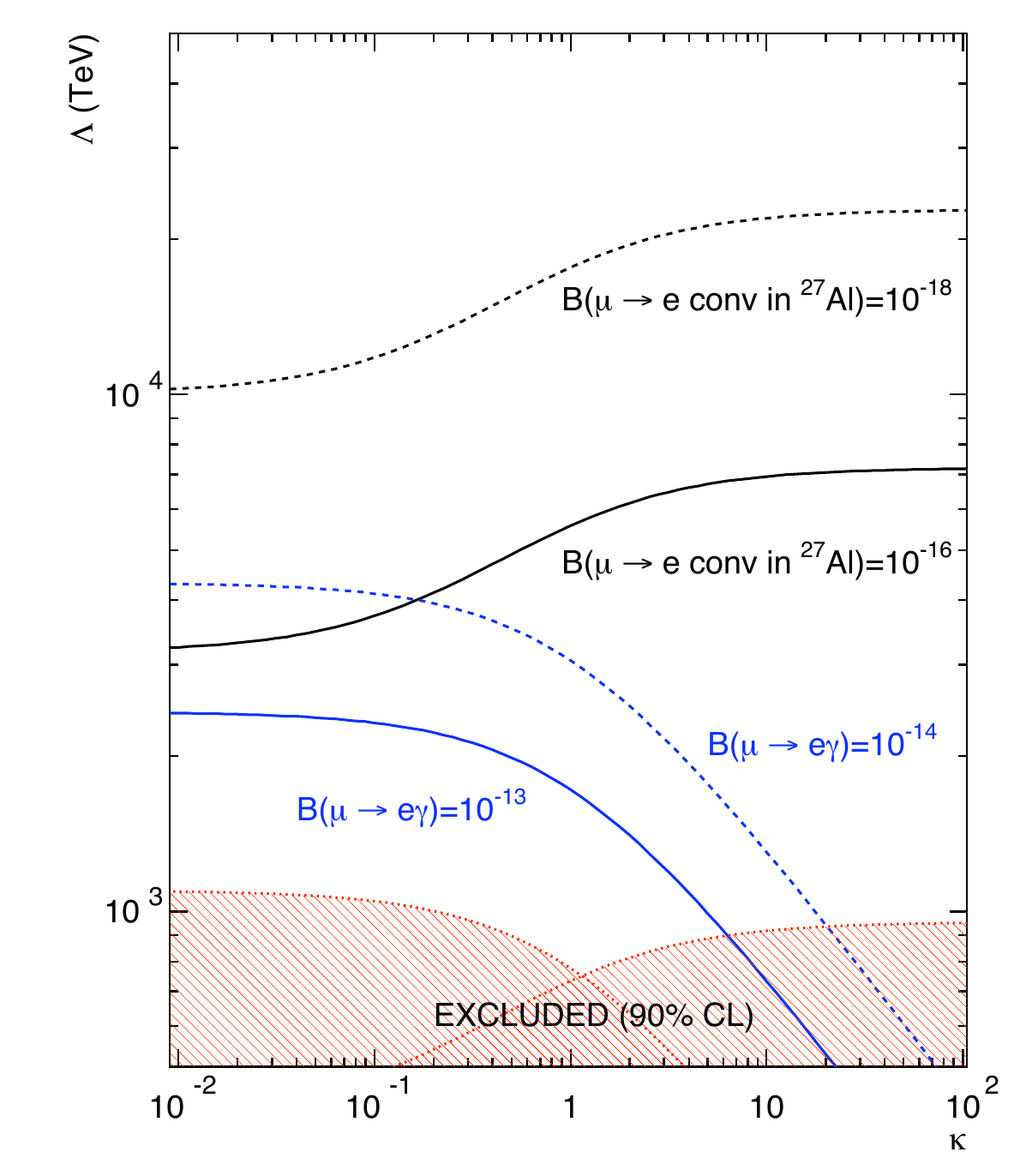}
  \caption{{Sensitivity of a $\mu\to e$ conversion in $^{27}$Al experiment that can probe a normalized capture 
rate of $10^{−16}$ and $10^{−18}$, and of a $\mu \to e \gamma$ search that is sensitive to a branching ratio of $10^{−13}$ and 
$10^{−14}$, to the new physics scale $\Lambda$ as a function of $\kappa$, as defined in the text. Also depicted is the 
currently excluded region of this parameter space.
}}
\label{fig:cl:p7}
\end{minipage}
\hspace{0.3cm}
\begin{minipage}[b]{0.48\linewidth}
\centering
    \includegraphics[trim = 45mm 130mm 50mm 10mm, clip, width=\linewidth]{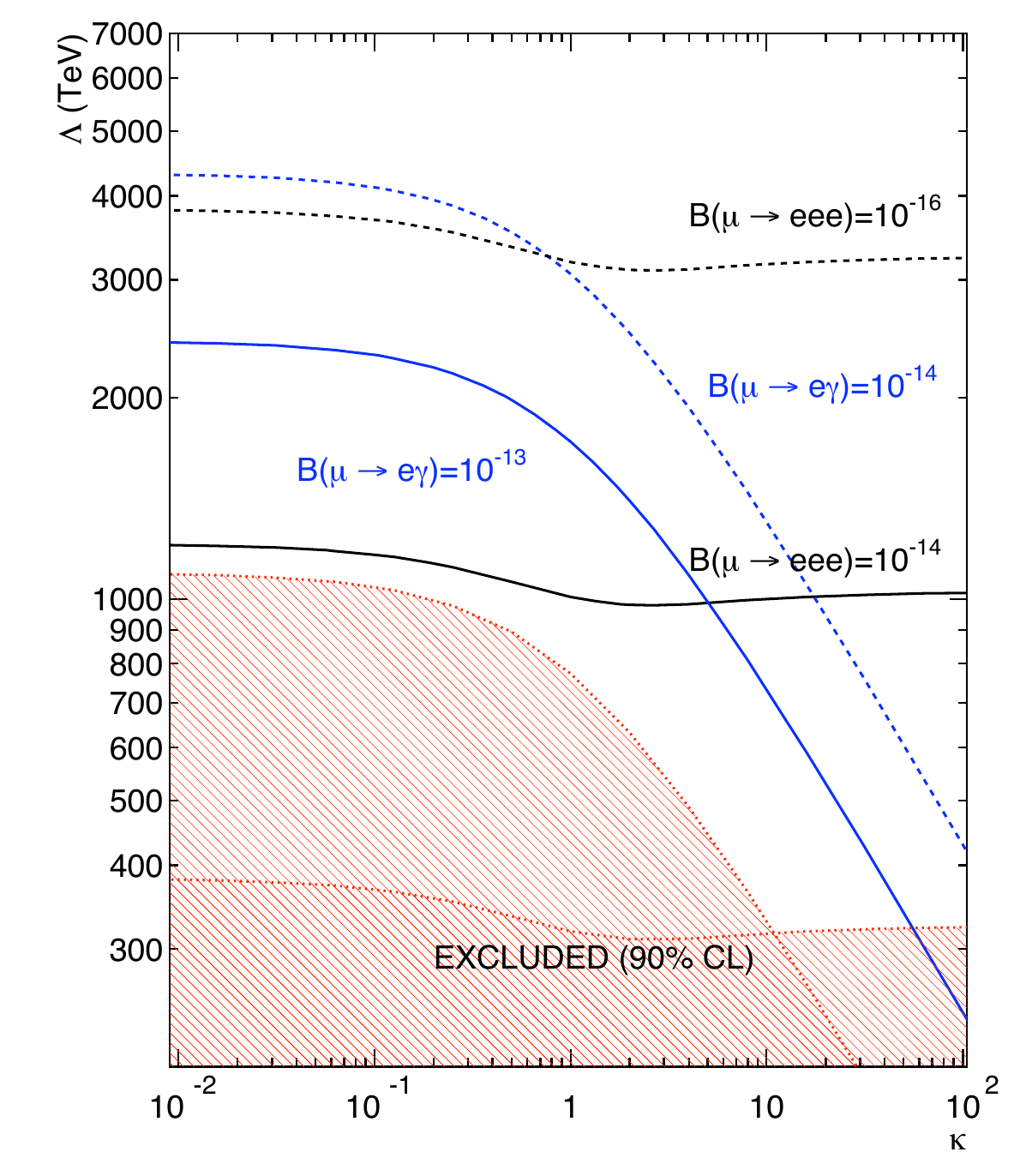}
  \caption{{Sensitivity of a $\mu \to eee$ experiment that is sensitive to branching ratios $10^{−14}$ and 
$10^{−16}$, and of a $\mu \to e \gamma$ search that is sensitive to a branching ratio of $10^{−13}$ and $10^{−14}$, to the new 
physics scale $\Lambda$ as a function of $\kappa$, as defined in the text.  Also depicted is the
currently excluded region of this parameter space.
}}
  \label{fig:cl:p8}
\end{minipage}
\end{figure}

Over the next decade
gains of up to five orders-of-magnitude are feasible in
muon-to-electron conversion and in the $\mu \to 3 e$
searches, while gains of at least two orders-of-magnitude
are possible in $\mu \to e\gamma$ and $\tau \to 3\ell$ decay and more than one order of magnitude in $\tau \to \ell\gamma$ CLFV
searches.  The question of which of these processes is the more sensitive was addressed in some detail at the Workshop; the answer is that the relative sensitivity depends on the type of new physics amplitude responsible for lepton flavor violation. 
The four-fermion operators that mediate these decays or conversions can be characterized by two parameters, $\Lambda$ which determines the mass scale of the four fermion amplitude $\kappa$, which governs the ratio of the four fermion amplitude and the dipole amplitude. 
For $\kappa << 1$ the dipole-type operator dominates CLFV phenomena, while for $\kappa >> 1$ the four-fermion
operators are dominant. Figures~\ref{fig:cl:p7} and \ref{fig:cl:p8}, from A.~de~Gouvea and P.~Vogel, Progress in Particle and Nuclear Physics
{\bf 71},  75 (2013), show these relationships and the capability of new experimental searches, which can extend our knowledge quite dramatically in the next decade.

Thus the pattern of violation that emerges thus yields quite specific information about new physics in the lepton sector. Existing searches already place strong constraints on
many models of physics beyond the standard model; the contemplated improvements increase these constraints significantly, covering substantial regions of the parameter space of many new physics models.
These improvements are important regardless of the outcome of new particle searches of the
LHC; the next generation of CLFV searches are an essential
component of the particle physics road map going forward.  If the LHC finds new
physics, then CLFV searches will confront the lepton sector in ways
that are not possible at the LHC, while if the LHC uncovers no sign of
new physics, CLFV may provide the path to discovery.

In general, muon measurements have the best
sensitivity over the largest range of the parameter space of many new
physics models. There are, however, models
in which  rare tau decays could provide the discovery
channel. $\tau$ flavor violation searches will have their sensitivity extended by around an order of magnitude at new $e^+e^-$ flavor factories. Polarized electron beams can provide an additional gain in sensitivity.  It was clear from the discussion that as many different
CLFV searches as feasible should be conducted, since the best discovery
channel is model-dependent and the model is not yet known.  Should a
signal be observed in any channel, searches and measurements in as
many CLFV channels as possible will be crucial to determining the nature
of the underlying physics, since correlations between the rates
expected in different channels provide a powerful discriminator between
physics model.

The new muon $g\!\!-\!\!2$ experiment will measure the anomaly to close to 100 parts per billion precision
with different experimental techniques. This will be an important measurement whether or not the LHC sees new physics. If the LHC sees SUSY-like new physics, $g\!\!-\!\!2$ will be used as a constraint in determining which model we see. The LHC will be particularly sensitive to color super-partners, while $g\!\!-\!\!2$ can pin down the flavor sector. The sensitivity of $g\!\!-\!\!2$ to $\tan\beta$ will provide a test the universality of that parameter. If the LHC does not see new physics, then $g\!\!-\!\!2$ can be used to constrain other models, such as theories involving dark photons and extra dimensions. Any new physics model will have to explain the discrepancy between the theoretical and experimental values of $g\!\!-\!\!2$.
The reduction of theory errors in the calculation of $g\!\!-\!\!2$ is thus also of great importance, particularly the contribution of light-by-light scattering.  New data from the KLOE and BES-III experiments  will put the candidate models on firmer ground, as will lattice calculations to be undertaken by, among others, the USQCD collaboration. A Super $B$ Factory with a polarized electron beam can measure, for the first time, the anomalous moment of the $\tau$, using new variables involving the polarization.

The search for EDMs will also play an important role in new physics
searches. The achievable limit on the electron EDM is the most stringent, but searches for muon and tau EDMs are nonetheless of interest, since new physics contributions scale as the lepton mass. These can be
important: if an electron EDM were to be found, the value of second and third generation EDMs would be of great interest.  Parasitic measurements with the new Fermilab $g\!\!-\!\!2$ experiment will improve the $\mu$ EDM limit by two
orders of magnitude. Improvement of this limit would also help to rule out
the possibility that the muon EDM is the cause of the current discrepancy in the
$g\!\!-\!\!2$ measurement. New dedicated experiments now being discussed
could bring the limit down to the $10^{-24}$ $e$cm level, making it
competitive with the electron EDM constraints. In the same vein, a Super $B$ Factory with a polarized electron beam can reach a sensitivity below $10^{-21}$ $e$cm.
Additional symmetry tests will also be possible, including sensitive searches for $C\!P$ violation in $\tau$ decay and 
tests of electroweak parity violation using electron scattering and $e^+e^-$ collisions. 

An exciting program of sensitive searches for new physics using the large samples of $\mu$ and $\tau$ decays in experiments at the intensity frontier awaits us. These experiments will likely be central to our understanding of physics beyond the Standard Model.

\chapter{Overview}\label{sec:cl:over}
\medskip
The theme of the ``Snowmass on the Mississippi'' exercise can be simply summed up as ``How do we find physics beyond the Standard Model?''. The Intensity Frontier answer evokes the power and reach of virtual processes in both finding evidence for New Physics and constraining its properties. Experiments in the lepton sector of the Intensity Frontier, by searching for rare decay processes involving lepton flavor violation and $C\!P$-violation, and by making precision measurements of quantities whose value is extremely well-predicted in the Standard Model, can advance our understanding of the most basic features of the Standard Model for which we currently have no rationale. Why are there three lepton families? Since lepton flavor conservation is violated in the neutrino sector, is it violated in the charged lepton sector as well? Why are the patterns of lepton and quark flavor mixing so different?

 Charged leptons are unique in several ways:
\begin{itemize}
\item
They directly probe the couplings of new particles to leptons.  This is unique in that the current energy frontier machine, the 
CERN LHC, is a hadron
collider. It is very effective at probing the quark sector, but
is significantly more limited in the lepton sector.
\item
Very
precise measurements  and sensitive searches can be made at a level that is difficult to achieve in
other sectors.
\item
They can be studied using a diverse set of independent
processes. The combination of these studies can provide additional
insights into the structure of the lepton sector.
\item
Hadronic uncertainties in the Standard Model predictions are either insignificant, or in the case of muon $g-2$, are obtained using independent data sets and estimates from theory
\item
There are  many cases, in particular charged lepton flavor violation (CLFV), where any signal would be an indisputable discovery of physics beyond the Standard Model (henceforth BSM physics).
\end{itemize}

There are important charged lepton observables that are best studied using electrons, most notably the electron electric dipole moment (EDM).  In most cases, these experiments are performed using outer-shell or shared electrons in either atoms or molecules.  These topics are covered in detail by the Nucleons/Nuclei/Atoms working group; we refer the reader to that chapter of these proceedings.

The program of studies of charged leptons is diverse, encompassing highly optimized, single-purpose experiments that focus on near-forbidden interactions of muons and multi-purpose experiments that take advantage of the large $\tau$-pair production cross section at $B$ or $\tau$/charm factories .
Very large
improvements in sensitivity are possible in the near future; even larger sensitivity gains can be made at Project X.  New
experiments such as Mu2e can probe rare processes at rates four orders
of magnitude more sensitive than current bounds. At this level of sensitivity many models predict that there will be observations of SM-forbidden processes, not just limits. These 
improvements will be a significant part of the program to understand new short-distance dynamics or new ultra-weak interactions.

Aside from being an intergral part of the broader Intensity Frontier program, studies of the  charged lepton sector provide  a vital link to the Energy Frontier. In the same way that, taken together, the results of individual charged
lepton experiment are more sensitive to BSM physics, charged lepton sector results as a whole are more
powerful when considered in concert with other Intensity and Energy Frontier experiments.  In
particular, there are three domains in which such combined results are a
crucial probe of BSM physics.   First, since neutrinos and charged leptons form a 
natural doublet, one would expect any BSM effects
 in neutrinos to also be seen in sufficiently sensitive charged lepton experiments.  
 Second, any complete theory of 
 flavor generation and the observed matter-antimatter asymmetry of the universe must 
 relate flavor and $C\!P$ (or $T$) violation in the heavy quark, neutrino, and charged lepton sectors.  
 Third,  any theory that predicts new particles or interactions at the LHC must also account for the virtual effects of those particles on decays and interactions of charged leptons and heavy quarks.
Thus, the major expansion in the study of charged leptons now underway is a natural extension of the successful heavy quark, neutrino and energy frontier programs of the previous decades.

A fourth domain is the probe of new ultra-weak, low energy interactions, referred to collectively  as hidden or dark sectors.  Here, charged lepton experiments overlap  with a wide variety of
experiments at the Intensity, Cosmic, and Energy Frontiers.  A large
experimental program is now under way to directly probe for new
hidden sectors, particularly in regions of parameter space
consistent with the muon $g-2$ anomaly.  This program is covered
in detail in the ``New Light Weakly-Coupled Particles'' chapter of this report.

\begin{figure}[t!]
\begin{center}
\includegraphics[width=10cm]{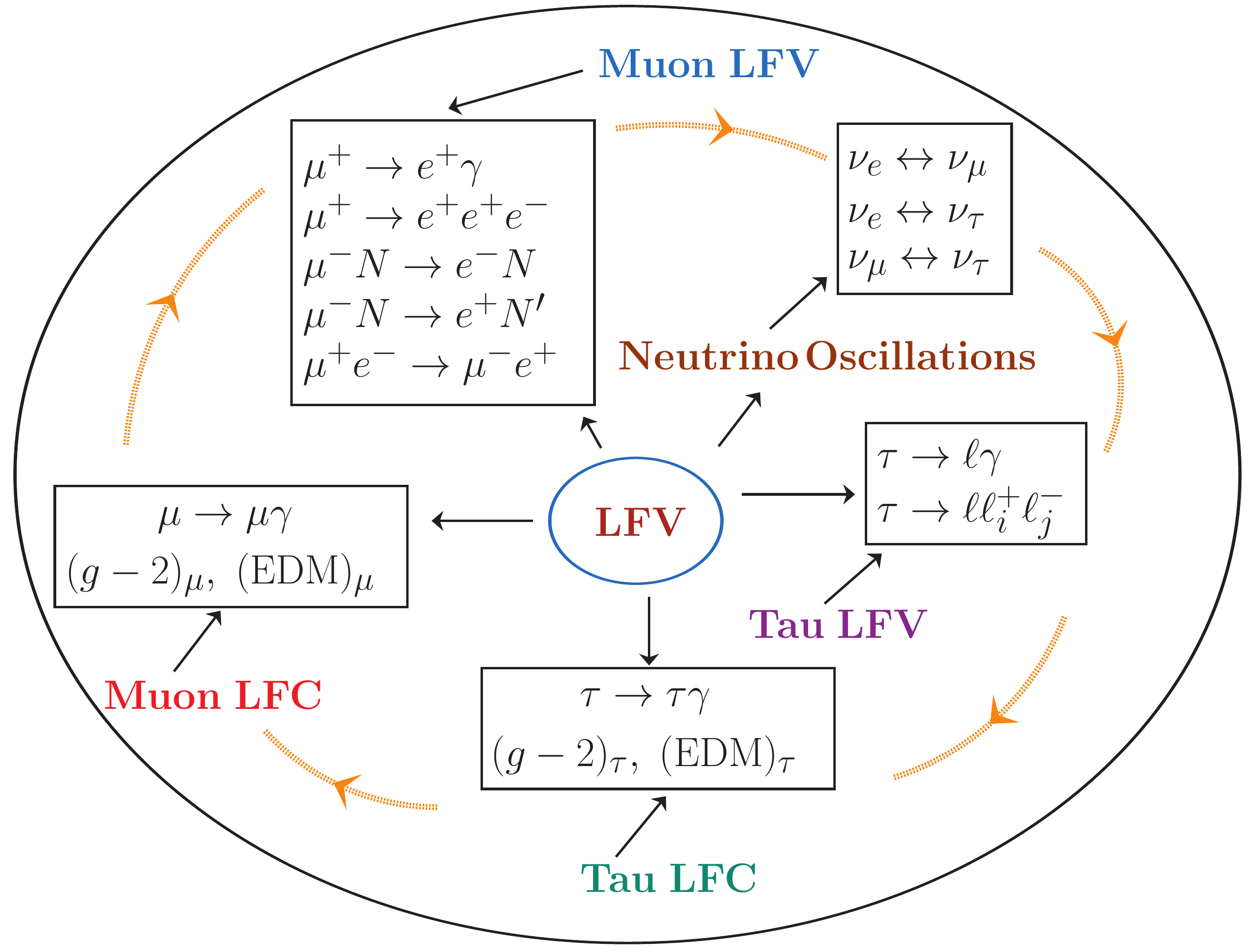}
\caption{\label{CL:chart}Interconnection between various lepton flavor violating and  lepton flavor conserving processes.}
\end{center}
\end{figure}

Fig. \ref{CL:chart} schematically depicts the interconnection between
various flavor-conserving and -violating processes in the lepton
sector~\cite{IF_review}.  In an underlying theory, neutrino flavor oscillations,
charged lepton flavor violation,  the anomalous magnetic moments, and permanent electric
dipole moments are all related.  Each experimental avenue we pursue allows us to uncover further attributes of the underlying theory. 

There are many important physical observables potentially sensitive to
BSM effects in charged lepton processes. Below, they are  split into  flavor violating observables and flavor conserving observables  such as $g-2$, EDMs, and parity violation measurements. Tau decays offer a unique opportunity to simultaneously study flavor-conserving, flavor-violating, $C\!P$-violating,  and $T$-violating effects and are discussed in
 their own section below.

\chapter{Flavor-Violating Processes}

\section{Theory Overview}\label{sec:cl:fvt}
\vskip -12pt
Neutrino flavor oscillations are well established. This requires charged lepton flavor violation at some level as well.  However flavor violation in charged lepton interactions has never been observed.
If neutrino mass is the only source of new physics, and if
the mass generation occurs at a very high energy scale, CLFV processes are highly suppressed.  For example, if neutrinos are 
Dirac particles, the branching ratio for $\mu\rightarrow e\gamma$ is
\begin{equation}
BR(\mu \rightarrow e \gamma)=\frac{3\alpha}{32\pi}\left|\sum_i U_{\mu
i}^* U_{e i}\frac{m_{\nu_{i}}^2}{m_{W}^2}\right|^2 \sim 10^{-52}
\end{equation}
where $U_{ei}$ are the leptonic mixing matrix elements. This value,
which suffers from extreme suppression from the small neutrino masses,
is experimentally inaccessible. In many extensions of the Standard Model
however, there are much larger contributions to CLFV and current experimental bounds set strict limits on the parameter space available for new physics models.

The effective Lagrangian relevant for the $\mu \rightarrow e\gamma$ and $\mu^+ \rightarrow e^+e^-e^+$ decays can be parametrized,
regardless of the origin of CLFV, as
\begin{eqnarray}
{\cal L}_{\mu\rightarrow e\gamma, eee} &=& -{4G_{F}\over\sqrt{2}}
\left[  {m_{\mu }}{A_R}\overline{\mu_{R}}
        {{\sigma }^{\mu \nu}{e_L}{F_{\mu \nu}}}
       + {m_{\mu }}{A_L}\overline{\mu_{L}}
        {{\sigma }^{\mu \nu}{e_R}{F_{\mu \nu}}} \right. \nonumber \\
    && + {g_1}(\overline{{{\mu }_R}}{e_L})
              (\overline{{e_R}}{e_L})
       + {g_2}(\overline{{{\mu }_L}}{e_R})
              (\overline{{e_L}}{e_R}) \nonumber \\
    &&   +{g_3}(\overline{{{\mu }_R}}{{\gamma }^{\mu }}{e_R})
              (\overline{{e_R}}{{\gamma }_{\mu }}{e_R})
       + {g_4}(\overline{{{\mu }_L}}{{\gamma }^{\mu }}{e_L})
              (\overline{{e_L}}{{\gamma }_{\mu }}{e_L})  \nonumber \\
    && \left.  +{g_5}(\overline{{{\mu }_R}}{{\gamma }^{\mu }}{e_R})
              (\overline{{e_L}}{{\gamma }_{\mu }}{e_L})
       + {g_6}(\overline{{{\mu }_L}}{{\gamma }^{\mu }}{e_L})
              (\overline{{e_R}}{{\gamma }_{\mu }}{e_R})
       +  h.c. \right].
       \label{CL:int}
\end{eqnarray}
The decay $\mu \rightarrow e\gamma$ is mediated by the first two terms 
of Eq. (\ref{CL:int}), the dipole terms.  These
terms, as well as the remaining contact terms, all contribute to the 
decay $\mu^+ \rightarrow e^+e^-e^+$.  The relative strength of
these decay rates depends on the relative strength of the dipole and contact terms. 
 Turning this around, searches for
these two decays reveal much about the underlying flavor structure.  
In some models the dipole contribution 
dominates both decays.  In this case, a simple relation
exists for the relative branching ratio:
\begin{equation}
\frac{B(\mu^{+}\rightarrow e^{+}e^{-}e^{+})}{B(\mu^{+} \rightarrow
e^{+} \gamma)} \simeq
\frac{\alpha}{3\pi}\left(\ln(\frac{m_{\mu}^{2}}{m_{e}^{2}})-\frac{11}{4}\right)
= 0.006.
\label{CL:branching}
\end{equation}
However, contact terms arise
frequently in popular models where the relation
Eq. (\ref{CL:branching}) does not hold.  A good example is the type II seesaw mechanism for small neutrino masses.  Here, one
does not add right-handed neutrinos to the spectrum, rather one
includes an iso--triplet scalar $\Delta = (\Delta^{++},\,\Delta^+,\,\Delta^0)$ with quantum numbers
$(1,3,+2)$ under $SU(3)_c \times SU(2)_L \times U(1)_Y$.  Neutrino
masses are generated via the Yukawa coupling $\frac{f_{ij}}{2}
\ell_i^T C \ell_j \Delta$, once a nonzero $\langle
\Delta^0\rangle$ develops.  The doubly charged scalar
$\Delta^{++}$ could mediate the decay $\mu^+
\rightarrow e^+e^-e^+$ at tree level.  In this case, the branching ratios for $\mu
\rightarrow e\gamma$ and $\mu^+ \rightarrow e^+e^-e^+$ become comparable.


\begin{figure}[t]
\begin{center}
\includegraphics[width=5cm]{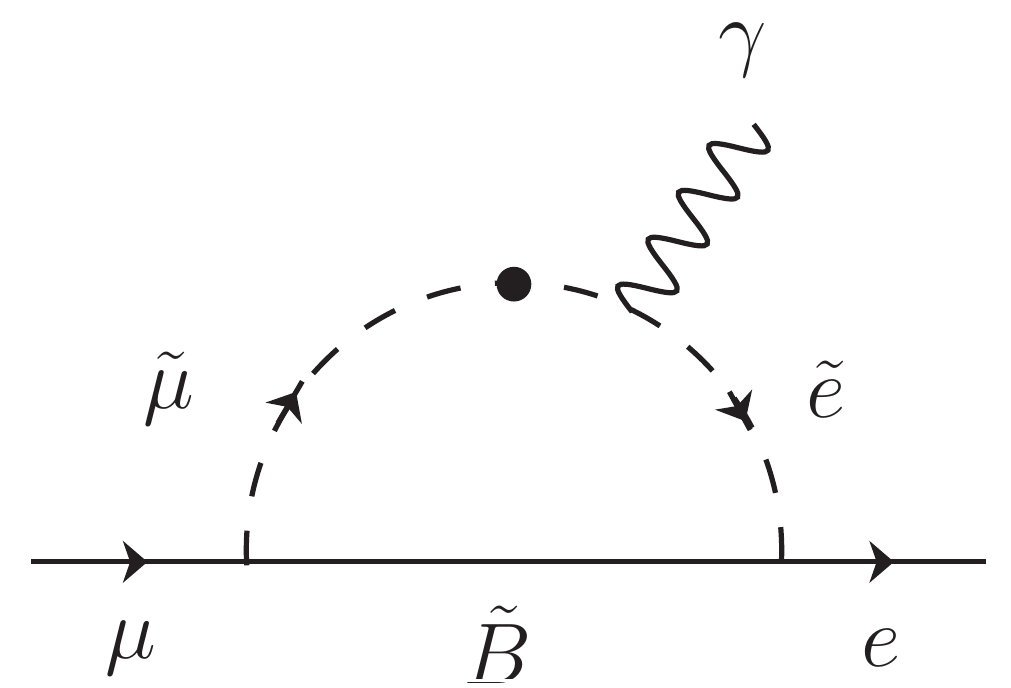} \hspace*{2cm}
\includegraphics[width=5cm]{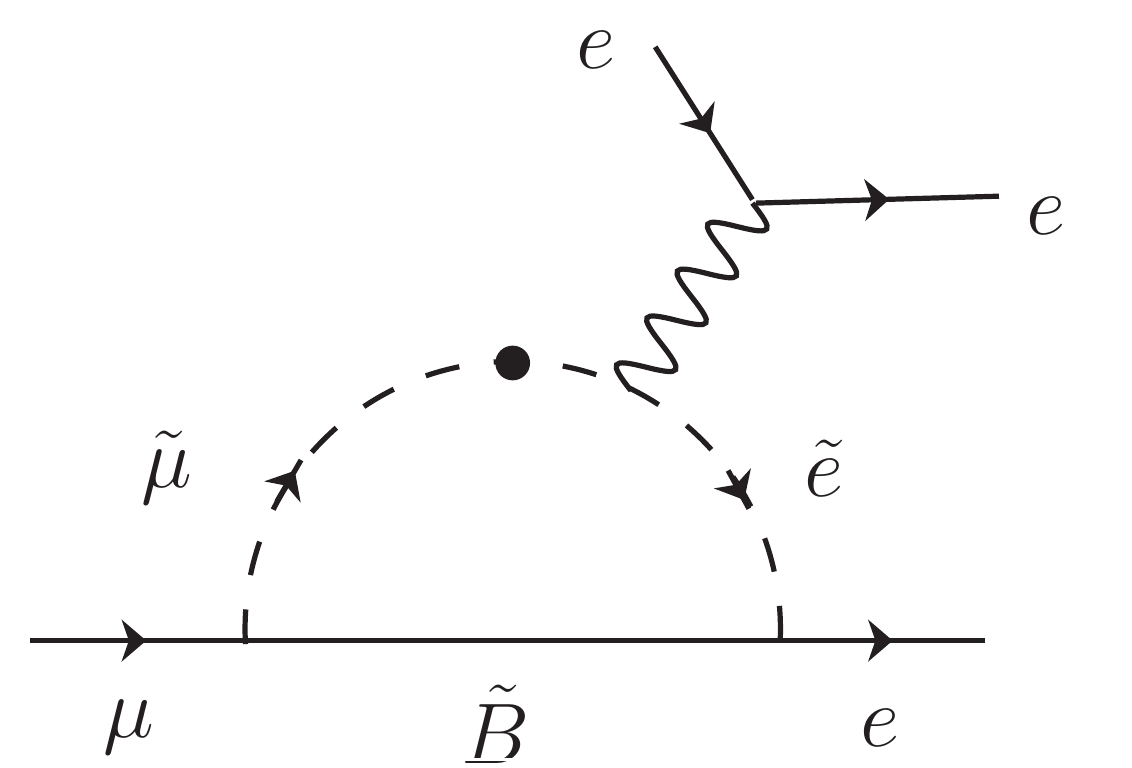}
\caption{\label{CL:mutoegamma}$\mu \rightarrow e\gamma$ decay mediated by SUSY particles (left panel), and $\mu \rightarrow 3e$ decay (right panel).}
\end{center}
\end{figure}

Equally important as the decays $\mu \rightarrow e\gamma$ and $\mu^+
\rightarrow e^+e^-e^+$ is the coherent $\mu^-N \rightarrow e^-N$ conversion
process in nuclei.  Muonic atoms are formed
when negative muons are stopped in matter.  In the ground state of these
atoms, the muon can decay in orbit or be captured with the emission of
a neutrino via the process $\mu^{-} + (A,Z) \rightarrow
\nu_{\mu} + (A,Z-1)$.  If there are new sources of CLFV, muon capture
without the emission of a neutrino can occur: $\mu^{-} + (A,Z)
\rightarrow e^{-} + (A,Z)$.   This would occur in supersymmetry (SUSY)  via
the diagram of Fig.~\ref{CL:mutoegamma}, when the photon is attached to
a quark line.  Like $\mu^+\rightarrow e^+e^-e^+$, this process can occur through dipole
interactions or through contact interactions.  Such contact
interactions arise naturally in leptoquark models at the tree level,
while in  SUSY the dipole interactions dominate.  If only the
dipole couplings are important, one can obtain a relation for the ratio of rates
\begin{equation}
\frac{B(\mu^{+}\rightarrow e^{+}\gamma)}
{B(\mu^{-}N\rightarrow e^{-}N)}
=\frac{96\pi^{3} \alpha}{G_F^2 m_{\mu}^4}\cdot
{1\over{3\cdot 10^{12}B(A,Z)}}\simeq \frac{428}{B(A,Z)},
\end{equation}
where $B(A,Z)$ is a function of the atomic number and atomic weight,
with its value ranging from 1.1 to 2.2 for Al, Ti and Pb atoms.  
The best limits on these processes are $B(\mu^- + {\rm Ti}
\rightarrow e^- + {\rm Ti}) < 4.3 \times 10^{-12}$~\cite{Dohmen:1993mp} and $B(\mu^- + {\rm Au}
\rightarrow e^- + {\rm Au}) < 7.0 \times 10^{-13}$~\cite{Bertl:2006up} from experiments conducted at PSI.
For these searches the limits quoted are with respect to the muon capture process $\mu^{-} + (A,Z) \rightarrow \nu_{\mu} + (A,Z-1)$.
Future
experiments can improve tremendously on these limits down to  $10^{-18}$.

A related process is the incoherent, lepton number violating
process $\mu^{-} + (A,Z) \rightarrow e^{+} + (A,Z-2)^{*}$, which
occurs in left--right symmetric models via the exchange of
right--handed neutrinos and $W_R^\pm$ gauge bosons.  The best limit
presently on this process is $B(\mu^- + {\rm Ti} \rightarrow
e^+ + {\rm Ca}) < 1.7 \times 10^{-12}$, also from PSI.  TeV scale
left--right symmetry predicts observable rates for this
transition.

CLFV could also be seen in other muonic systems.  Muonium is a  ($\mu^+ e^-$) bound state   analogous to the hydrogen atom, which
in the presence of a CLFV interaction can oscillate into antimuonium
($\mu^-e^+$).  The doubly charged scalar of the seesaw model, the left--right
symmetric model, or the radiative neutrino mass model would all lead to
this process.  If
the Lagrangian for this process is parametrized as
\begin{equation}
H_{\rm Mu\overline{Mu}} = \left({G_{\rm Mu\overline{Mu}} \over
\sqrt{2}} \right)
\overline{\mu}\gamma_{\lambda}(1-\gamma_5){e}
\overline{\mu}\gamma^{\lambda}(1-\gamma_5){e} + h.c.,
\end{equation}
the current limit from PSI experiments is $G_{\rm Mu\overline{Mu}} < 0.003
\,G_F$, with room for improvement by several orders of magnitude in
the near future.

\subsection{CLFV Decays in Specific New Physics Models}
\vskip -12pt

If we assume that neutrino mass is generated by a seesaw mechanism \cite{Minkowski:1977sc}, we can see effects in CLFV if the seesaw scale is low~\cite{deGouvea:2007uz}. It is perhaps more natural that the seesaw mechanism
is realized at a very high energy scale, $M_R \sim 10^{10} - 10^{14}$
GeV. In this case there can be significant CLFV provided that there is
some new physics at the TeV scale. Below we briefly mention two such
scenarios, SUSY and Randall-Sundrum warped extra dimensions (RS).

Implementing the seesaw mechanism within the context of SUSY leads to
a new source of CLFV. In a momentum range  between $M_R$ and  $M_{\rm
Pl}$, where $M_{\rm Pl}$ is the fundamental Planck scale, the
right-handed neutrinos are active and their Dirac
Yukawa couplings with the lepton doublets induce flavor
violation among the sleptons.  The sleptons must have masses of order
TeV or less, if SUSY is to solve the hierarchy problem, and they carry
information on flavor violation originating from the seesaw.
Specifically, the squared masses of the sleptons would receive flavor
violating contributions given by
\begin{equation}
(m_{\tilde{l}_{L}}^{2})_{ij} \simeq -\frac{1}{8\pi^{2}}
(Y_{\nu}^\dagger Y_{\nu})_{ij} (3 m_{0}^{2} + |A_{0}|^{2})\ln\left({M_{\rm
Pl}\over M_{R}}\right)
\label{CL:SUSY-LFV}
\end{equation}
where $Y_\nu$ is the Dirac Yukawa coupling of the neutrinos, and $m_0$
and $A_0$ are SUSY breaking mass parameters of order $100$ GeV.  In
SUSY GUTs, even without neutrino masses, there is an independent
contribution to CLFV, originating from the grouping of quarks and
leptons in the same GUT multiplet.   The squared masses of
the right-handed sleptons would receive contributions to CLFV in this momentum range from the GUT
scale particles that are active, given by
\begin{equation}
(m_{\tilde{e}_{R}}^{2})_{ij} \simeq -\frac{3}{8\pi^{2}}
V_{3i}V_{3j}^{*}|Y_t|^{2}(3 m_{0}^{2}  +
|A_{0}|^{2})\ln\left(\frac{M_{\rm Pl}}{M_{{\rm GUT}}}\right).
\label{CL:GUT_LFV}
\end{equation}
Here $V_{ij}$ denote the known CKM quark mixing matrix elements, and
$Y_t$ is the top quark Yukawa coupling.  Unlike
Eq.~(\ref{CL:SUSY-LFV}), which has some ambiguity since $Y_\nu$ is not
fully known, the CLFV contribution from Eq.~(\ref{CL:GUT_LFV}) is
experimentally determined, apart from the SUSY parameters. 

We next consider RS models with
bulk gauge fields and fermions.   In these models, our universe
 is localized on one (ultraviolet) membrane of a multidimensional 
 space while the Higgs field is localized on a different (infrared) membrane. 
  Each particle has a wave function that is localized near the Higgs 
  membrane for heavy particles or near our membrane 
  for light particles. Thus localization of different wave functions between
   the membranes  generates flavor.
For a given fermion mass spectrum, there are only two free parameters,
 an energy scale to set the Yukawa couplings and a length scale of 
 compactification that sets the level of Kaluza-Klein  (KK) excitations.  The two scales can be accessed using a combination of tree induced CLFV processes that occur in $\mu N\rightarrow e N$ or $\mu\rightarrow 3e$ and loop induced interactions such as $\mu\rightarrow e\gamma$ \cite{Agashe:2006iy},\cite{Csaki:2010a}. The amplitude of loop-induced
flavor-changing decays, such as $\mu\to e \gamma$, is given by a
positive power of the Yukawa and a negative power of the KK
scale.  Tree-level flavor-changing diagrams, on the other hand, come
from four-fermion interactions whose flavor-changing vertices come
from the non-universal profile of an intermediate KK gauge boson. This
non-universality is an effect of electroweak symmetry breaking so that
the flavor-changing part of the KK gauge boson profile is localized
near the IR brane and the size of flavor-changing effects depend on
the size of the zero mode fermion profile towards the IR
brane. However, in order to maintain the Standard Model fermion
spectrum the zero-mode fermion profiles must be pushed away from the
Higgs vacuum expectation value on the IR brane as the anarchic Yukawa scale is
increased. Thus the tree-level flavor changing amplitudes go like a
negative power of the anarchic Yukawa scale. For a given KK scale,
experimental constraints on lepton flavor-changing processes at tree
and loop level thus set lower and upper bounds on the Yukawa scale,
respectively.

A version of minimal flavor violation exists in RS models where the new
 scales have a very small effect on low energy flavor changing processes.  
 It was noted in \cite{Agashe:2006iy} and \cite{Agashe:2004cp} that
certain flavor changing diagrams are suppressed in the RS scenario
because the particular structure of zero mode wave functions and
Yukawa matrices is the same as the zero mode mass terms induced by
electroweak symmetry breaking. When passing to the physical basis
of light fermions these processes are also nearly diagonalized, or
\textit{aligned}, and off-diagonal elements of these transitions are
suppressed. These flavor-changing processes are not completely zero 
since the fermion bulk masses are an additional flavor spurion
in these theories. In other words, the $U(3)^3$ lepton flavor 
symmetry is not restored in the limit where the Yukawa terms vanish. 
The full one-loop calculation of $\mu\to e \gamma$ in Randall-Sundrum
models including these misalignment effects and a proof of finiteness 
was performed in \cite{Csaki:2010a}.



\section{Muon Experimental Overview}\label{sec:cl:muexp}

\subsection{Muon Flavor Violation Experiments in this Decade}
\vskip -12pt
\subsubsection{The Mu2e Experiment}
\vskip -12pt


\label{cl:sec:mu2e}

The Mu2e experiment~\cite{Abrams:2012er}, to be hosted at Fermilab, a flagship component
of the U.S.\ Intensity Frontier program~\cite{IF_review}, will
search for the charged-lepton-flavor-violating process of coherent
muon-to-electron conversion in the presence of a nucleus ($\mu^-N
\rightarrow e^-N$).  Mu2e will improve sensitivity
compared to current experimental by four orders of
magnitude and will set a limit on $R_{\mu e}$, defined
as,

\begin{eqnarray}
  R_{\mu e} &=& \frac
  {\Gamma(\mu^{-}\;  N(A,Z) \to e^{-}\; N(A,Z)}  {\Gamma(\mu^{-}\; N(A,Z)\to \nu_{\mu}\; N(A,Z-1))},\end{eqnarray}

where $N(A,Z)$ denotes a nucleus with mass number $A$ and atomic
number $Z$.  The numerator corresponds to the rate for the CLFV
conversion process and the denominator corresponds to the rate for
ordinary muon capture on the same nucleus.  The current best limit is
$R_{\mu{}e}<7\times10^{-13}$\cite{Bertl:2006up}.

There is no observable Standard Model contribution to the $\mu^-N
\rightarrow e^-N$ signal at Mu2e.  Neutrino oscillations imply that
muon-to-electron conversion can proceed via a penguin diagram that
contains a $W$ and an oscillating neutrino. However, the rate for this
conversion process is more than 30 orders of magnitude below the
projected sensitivity of the Mu2e experiment.  Any signal observed at
Mu2e would be an unambiguous indication of BSM
physics~\cite{Marciano:2008zz,deGouvea:2013zba}.

\begin{figure}
\centering
\includegraphics[width=0.8\textwidth]{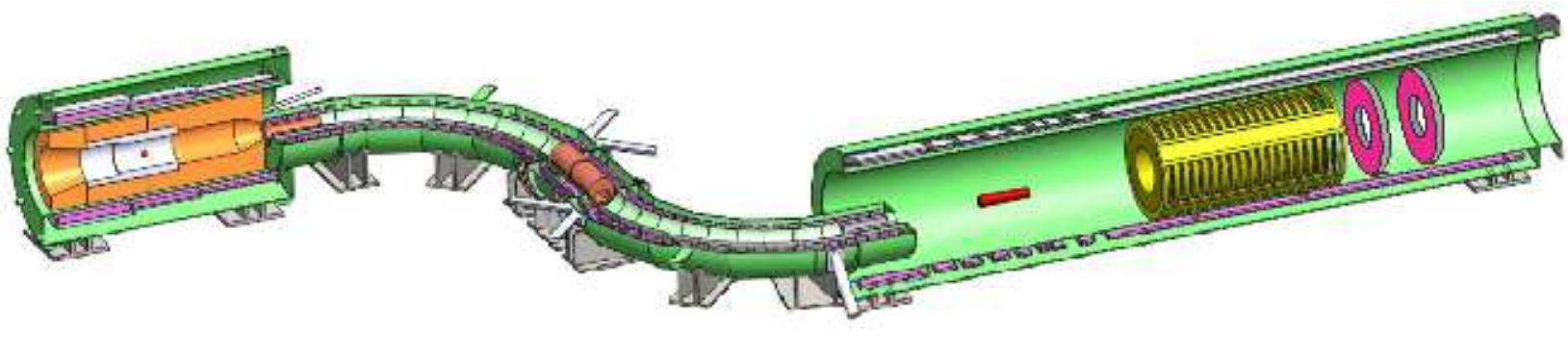}
\caption{The Mu2e experimental setup. The pulsed proton beam enters
the production solenoid (far left) from the top right.  Muons
produced are captured by the production solenoid and transported
through the S-shaped transport solenoid to the aluminum stopping
target (small red cylinder).  Electrons produced in the stopping
target are captured by the magnetic field of the detector solenoid
(right) and transported through the tracker (yellow) where the
momentum is measured.  The electrons then strike the electromagnetic
calorimeter (pink annuli), which provides particle identification and an independent measurement of momentum.
A cosmic ray-veto system and some other parts of the apparatus are not
shown.
}
\label{cl:fig:mu2e}
\end{figure}

The approach of the Mu2e experiment is to stop low-momentum muons from
a pulsed beam on an aluminum target to form muonic atoms and then to
measure the resulting electron spectrum.  The signal would produce a
mono-energetic electron with an energy of about 105~MeV.  In order to
reach the design sensitivity (single-event sensitivity of $2\times
10^{-17}$), about $10^{18}$ muons must be stopped.  Keeping the
background expectation to less than one event in this high-intensity
experiment is obviously quite a challenge and results in the unique
experimental setup summarized below and depicted in
Fig.~\ref{cl:fig:mu2e}.

The first step in the experiment is to produce the low-momentum pulsed
muon beam.  Recycled Tevatron infrastructure will deliver 8~GeV
protons with 1695~ns bunch spacing to the experiment, the
revolution period of the Debuncher Ring. This spacing is well-suited to Mu2e, given that the lifetime of a muonic aluminum atom is about
864~ns.  Pions and muons produced inside the production solenoid are
collected and passed to the S-shaped muon beamline where absorbers and
collimators are optimized to eliminate positively-charged particles
and anti-protons while efficiently transmitting low-energy
negatively-charge pions and muons.  Most of the pions will decay
inside the 13~m long beamline, while about 40\% of surviving muons
will be stopped in an aluminum stopping target.  Simulations estimate
that Mu2e will produce $0.0016$ stopped muons per proton on target.

Muons stopped in the target form a muonic atom.  As they settle into
the $K$-shell a cascade of X-rays will be emitted.  By detecting these
X-rays the rate of stopped muons can be measured, thereby establishing
the denominator of $R_{\mu e}$. About 60\% of stopped muons will
undergo muon capture on the nucleus while the other 40\% will decay in
orbit (DIO).  The DIO process produces an electron with a continuous
Michel distribution including a long tail due to photon exchange with
the nucleus.  In the limit where the neutrinos carry no energy from
the Michel decay, the electron carries the maximum energy of 105~MeV~\cite{czarnecki}.
In this limit, the DIO electron is indistinguishable from the $\mu^-N
\rightarrow e^-N$ conversion signal.  In addition, mis-measurements of
the DIO electron momentum contribute to an irreducible background.
In order to combat the DIO background, the Mu2e experiment requires a
tracking detector with momentum resolution of about $0.1$\%.

To achieve this momentum resolution, the Mu2e tracker will
use straw tubes in vacuum.  The inner radius of the tracker is empty
so that only tracks with transverse momenta above $53$~MeV/c will pass
through the straw tubes. The Mu2e scintillating-crystal (LYSO)
calorimeter will provide cross checks of signal candidates and
particle identification.  The calorimeter also has an empty inner-radius 
region.  The empty inner regions effectively make the tracker
and calorimeter blind to the bulk of the DIO electrons and  to muons that
don't stop in the target, and allow these detectors to cope with the
high-intensity environment of the experiment.

The primary backgrounds for the Mu2e experiment can be classified into
several categories: intrinsic muon-induced backgrounds, late-arriving
backgrounds, and miscellaneous backgrounds, primarily anti-proton-induced 
and cosmic ray-induced backgrounds. The muon-induced
backgrounds arise from muon DIO and radiative muon capture.  The
kinematic endpoint of the electron energy distribution is slightly
below 105 MeV, so this background, like the DIO, can be mitigated by
minimizing non-Gaussian contributions to the tails of the momentum
resolution.  The dominant late-arriving background arises from
radiative pion capture and subsequent conversion of the $\gamma$ in
the stopping target material to produce a 105~MeV electron.
Late-arriving backgrounds like this can be controlled by taking
advantage of the long muon lifetime and by optimizing the properties
of the pulsed beam.  After a beam pulse there is therefore a delay of about
700~ns before the signal timing window begins.  In order to avoid
late-arriving backgrounds in the signal time window, Mu2e requires the
fraction of protons outside the beam pulse to be less than $10^{-10}$,
which will be achieved and monitored with dedicated systems.  Beam
electrons, muon decay in flight, and pion decay in flight are other
late-arriving backgrounds that are suppressed through the combination of
the pulsed beam and delayed signal window.

Cosmic ray muons interacting within the detector solenoid region can
produce background electrons.  Passive shielding and an active
cosmic ray-veto system are employed to ensure that cosmic rays are a
sub-dominant background.

The total background expectation in the Mu2e experiment for a
three-year run at 8~kW beam power is less than 0.5 events and
summarized in Table~\ref{cl:tab:PXBgd}.


\subsection{Muon Flavor Violation: The Next Generation}

\subsubsection{Mu2e at Project X}
\vskip -12pt


We summarize here a feasibility study~\cite{Mu2eII} of a
next-generation Mu2e experiment (Mu2e-II) that uses much of the
currently-planned facility and Project~X~\cite{ProjectX} beams to
achieve a sensitivity that is about a factor of ten beyond that of the
Mu2e experiment described in Section~\ref{cl:sec:mu2e}. A factor of
ten improvement will be interesting regardless of the outcome of Mu2e.
If the Mu2e experiment observes events completely consistent with
background expectations, then another factor of ten improvement in
sensitivity extends the reach to additional beyond-the-standard-model
parameter space.  If Mu2e observes a $3\sigma$ excess, then a Mu2e-II
upgrade would be able to definitively resolve the situation.  And if
Mu2e discovers charged-lepton-flavor-violating physics, then a Mu2e-II
upgrade could explore different stopping targets in an effort to
untangle the underlying physics.  By measuring the signal rate using
nuclear targets at various $Z$, Mu2e-II would have the unique ability
to resolve information about the underlying effective operators that
are responsible for the lepton-flavor-violating
signal~\cite{Kitano:2002mt,Cirigliano:2009bz}.

To estimate the signal acceptance and background prediction for
Mu2e-II scenarios we use \texttt{G4Beamline v2\_12}~\cite{g4bl}, which
is a simplified version of Geant4~\cite{GEANT4}.  Three sets of
simulated experiments are studied: the 8~\gev\ case which corresponds
to the Mu2e configuration, and potential Project~X upgrades
corresponding to protons with 1 or 3~\gev\ of kinetic energy.  In all
instances the full Mu2e solenoid system is simulated including all
collimators, the Production Solenoid heat and radiation shield, the
antiproton window, and the magnetic field.  The stopping target
geometry is described in~\cite{Abrams:2012er} and is left unchanged for the
different scenarios.

The timing distribution of the proton pulse in \texttt{G4Beamline} is
modeled as a delta function located at $t = 0$~ns.  In order to get a
more accurate estimate of the experimental sensitivity we convolute
the relevant timing distributions with the expected shape of the
proton pulse as estimated using dedicated simulations of the Mu2e
proton beam.  For Project~X, the width of the proton pulses are
expected to be $\pm 50$~ns for 100 kW of beam power~\cite{bobT}, which is
about a factor of two narrower than what will be used for Mu2e.  So,
for the Project~X studies we assume the same proton pulse shape as
supplied for Mu2e, but reduce the width of the pulse by a factor of
two.  Most pions decay before reaching the stopping target, and due to
this short lifetime it isn't practical to study the pion backgrounds
if the decay is simulated.  Instead, in these studies all charged
pions pions are propagated through to the stopping target and events
are weighted by the survival probability.

The stopping-time distributions for muons and pions are studied in
Ref.~\cite{Mu2eII} and are used for the estimations of background and
signal acceptance below.  We find that the stopping-time distributions
are largely independent of the stopping target material.

\begin{table}[tb]
  \centering
  \begin{tabular}{rccc} \hline\hline
           & lifetime (ns) & capture fraction & decay fraction \\ \hline
  Aluminum & 864           & 0.61            & 0.39 \\
  Titanium & 297           & 0.85            & 0.15 \\ \hline\hline
  Gold     &  72           & 0.97            & 0.03     \\ \hline\hline
  \end{tabular}
  \caption{The lifetime of the bound muon and the muon capture and decay     
    fractions for various stopping target nuclei that affect the 
    sensitivity estimates for Mu2e-II. 
  }
  \label{cl:tab:nucleistuff}
\end{table}
  
\begin{figure}[htb]
  \centering
  \includegraphics[width=0.75\textwidth]{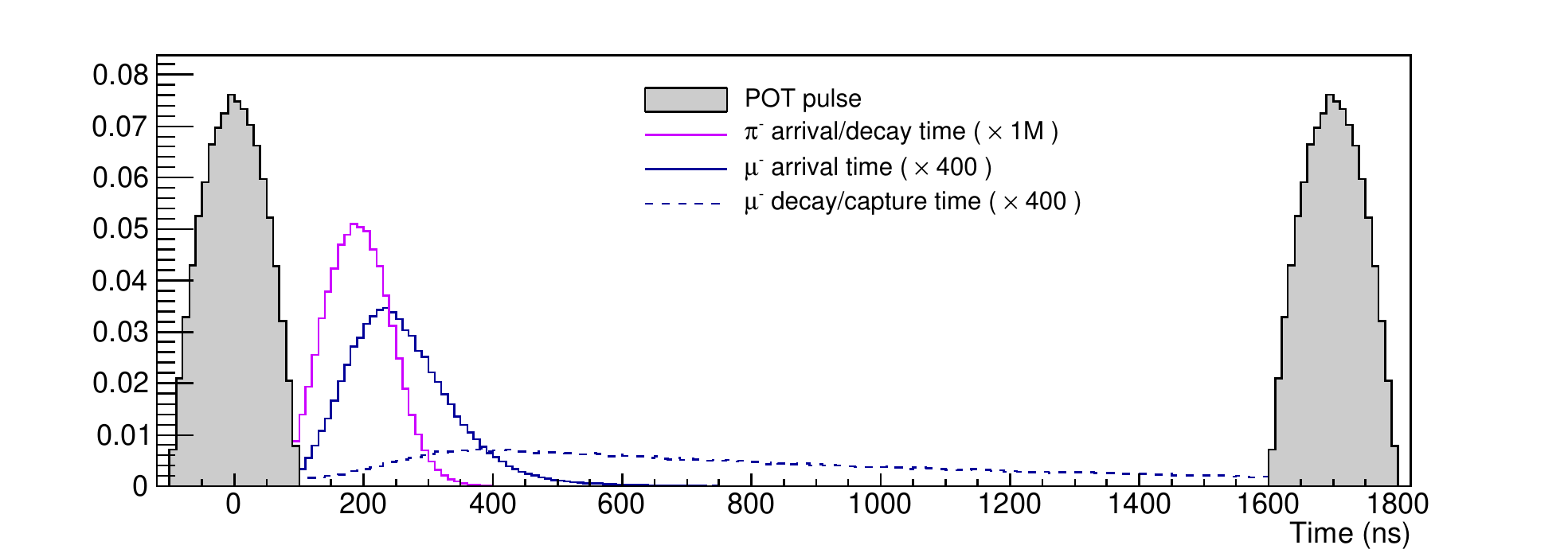}
  \includegraphics[width=0.75\textwidth]{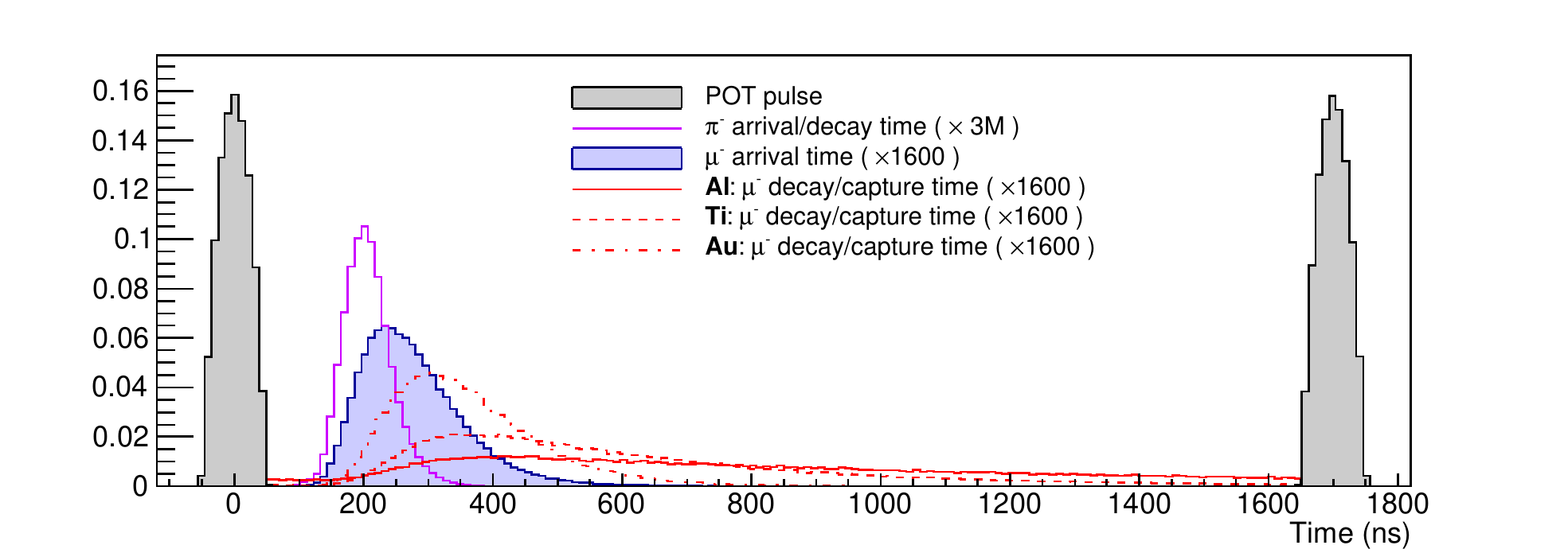}
  \caption{Timing distribution for the 8-GeV Mu2e case (top) and the 3-GeV Project~X case (bottom).  Shown
  here is a figurative POT pulse width, the arrival time of the
  charged pions and muons, and the decay-time distribution of the
  muons on aluminum, titanium, and gold.}
  \label{cl:fig:timing}
\end{figure}

Relevant timing distributions for the Mu2e experiment are shown in the
top panel of Fig.~\ref{cl:fig:timing} for an aluminum stopping
target. Since the $\pi^-$-N interaction is strong, the pion
capture-time distribution is assumed to be the same as the pion
arrival-time distribution.  For muons, the capture/decay rate is
characterized by a falling exponential.  The fractions of bound muons
that are captured or that decay-in-orbit (DIO) are also nuclei
dependent.  These nuclei-dependent characteristics affect the
sensitivity of a given experiment and are listed in
Table~\ref{cl:tab:nucleistuff} for the stopping-target nuclei we
considered~\cite{Suzuki:1987jf}.  The muon decay-time distribution is
shown in the top panel of Fig.~\ref{cl:fig:timing} as the blue dashed
line.  In this figure the timing distributions are folded over modulo
1695~ns in order to account for contributions from previous proton
pulses.

The bottom panel of Fig.~\ref{cl:fig:timing} shows the Project~X
3~\gev\ scenario, where the proton pulse width is half that of the
8~\gev\ configuration, and both aluminum, titanium, and gold are
considered as a stopping target~\footnote{Note that the arrival times
for stopped particles are slightly earlier for titanium than for
aluminum although it isn't depicted in the figure.}. Because the proton
pulse width is narrower, the live gate can be increased in the
Project~X scenario~\cite{MoveLiveGate}.  The same exercise was
performed for the Project~X 1~\gev\ scenario and with the exception of
the stopping rates, the parameters of the timing distributions for the
1~\gev\ case are very similar to those of the 3~\gev\ case.  The
relevant quantities associated with Fig.~\ref{cl:fig:timing} are
included in Ref.~\cite{Mu2eII} and used below to predict the
background and signal rates for each scenario.

We also produced the muon decay-time distribution for a gold stopping
target as shown in Fig.~\ref{cl:fig:timing}.  Since the lifetime of
the bound muon is so small (72~ns) the fraction of muon
decays/captures that occur within the live gate is quite small -- only
1.22\% for a live gate of $670 < t < 1645$~ns.  Since the muon
decay-time distribution has such a large overlap with the pion arrival
time distribution, it's clear that it will be difficult to achieve a
reasonable signal acceptance while sufficiently suppressing the RPC
background.  To achieve the necessary pion/muon separation requires a
dedicated study of alternative transport systems and is beyond the
scope of this study. We do not further consider the gold stopping
target.

We estimate the backgrounds for several Mu2e-II scenarios as described in Ref.~\cite{Mu2eII}.  These estimates assume that the detector performance for Mu2e-II is unchanged relative to the currently planned Mu2e detector.  Simulation studies of detector performance at the higher Mu2e-II intensities indicate that this can be achieved with modest upgrades as discussed below.


For each beam scenario target nuclei we estimate the number of protons
on target ($N_{\mathrm{POT}}$) required to improve the signal
expectation by a factor of ten. This is done by scaling from the Mu2e
experiment by the number of stopped muons per POT, the muon capture
fraction, and the fraction of muon capture/decay that occur in the
signal timing window.  We then take the resulting number of POT and
calculate the required beam power, and the number of muon and pion
stops per kW, assuming the same run time for Mu2e-II as Mu2e, namely a
three-year run with $\sn{2}{7}$ seconds of run time per year with
proton pulse spacing of 1695~ns.  These studies predict that between
\sn{8.7}{21} and \sn{5.3}{22} $N_{\mathrm{POT}}$ are required
depending on the beam and target scenario which corresponds to 70--150
kW of beam power.  The detailed results of these calculations are
given in Ref.~\cite{Mu2eII}.


Several background sources will be reduced or remain minimal at
Mu2e-II due to the improved beam characteristics compared to Mu2e.
For example, there are no antiproton-induced backgrounds for the
Project~X scenarios since both 1~\gev\ and 3~\gev\ kinetic energy
protons are below the proton-antiproton production threshold.  The
radiative pion capture background is kept under control for Mu2e-II
owing to the narrower proton pulse widths at Project~X and owing to
the significantly improved extinction provided by Project~X beams.
Late-arriving backgrounds, such as $\pi$ decay-in-flight, are also
kept under control by the decrease in the extinction ratio relative to
those same quantities for Mu2e.
   
The cosmic ray background is independent of $N_{\mathrm{POT}}$ and
beam power and depends only on the live time of the experiment. Since
we assume that Mu2e-II will have the same run time (three-year run
with $\sn{2}{7}$ s/year) and proton pulse spacing (1695~ns) as Mu2e,
the only difference is in the duty factor (which increases from 30\%
to 90\%) and the live-gate fraction. We scale the current Mu2e
estimate for the cosmic-ray-induced backgrounds to account for these
differences.  We assume that the veto efficiency for Mu2e-II is
unchanged relative to Mu2e (99.99\% ). For a Project~X (PX) driven
Mu2e-II experiment, the resulting cosmic-ray-induced background is
$0.16$ events.

Some backgrounds scale linearly with the number of stopped muons, such
as radiative muon captures and muon decay in orbit, and these can't be
mitigated through improved beam characteristics.  Radiative muon
capture is a negligible background for Mu2e and remains so for
Mu2e-II.

The $\mu$ decay-in-orbit background is estimated using dedicated simulations that include the effects of increased occupancy expected for the Mu2e-II scenarios considered.  It was found that in order to keep these backgrounds under control it will be necessary to improve the spectrometer resolution by reducing the material budget.  In the simulations we assumed the Mu2e tracker could be replaced with a similar geometry but using straws with half the wall thickness (ie. rebuild the tracker reducing the straw walls from 15~$\mu m$ to 8~$\mu m$ thick).  For the increased sensitivity of a Mu2e-II experiment an improved momentum-scale calibration may also be required to keep systematic uncertainties affecting the DIO background estimate from degrading the sensitivity.


The summary of the background estimates for Mu2e-II are given in
Table~\ref{cl:tab:PXBgd} along with the estimate for Mu2e.

\begin{table}[t]
  \centering
  \begin{tabular}{llcrcrcr} \hline\hline
    & & \hspace*{0.15in} & \multicolumn{1}{c}{Mu2e} &\hspace{0.15in} &\multicolumn{3}{c}{Mu2e-II} \\
    & & & \multicolumn{1}{c}{8~\gev } & &\multicolumn{3}{c}{1 or 3~\gev } \\
    & & & \multicolumn{1}{c}{Al} & & \multicolumn{1}{c}{Al}  &\hspace*{0.1in} & \multicolumn{1}{c}{Ti} \\ \hline
    Category      & Source                     
    & &\multicolumn{5}{c}{Events} \\ \hline
    \multirow{2}{*}{Intrinsic} 
                  & $\mu$ decay in orbit       
    & & 0.22 & & 0.26 & & $0.92$  \\ 
                  & radiative $\mu$ capture    
    & & $<0.01$ & & $<0.01$ & & $<0.01$  \\ \hline
    \multirow{4}{*}{Late Arriving}
                  & radiative $\pi$ capture    
    & & 0.03 & & 0.04 & & 0.05  \\
                  & beam electrons             
    & & $<0.01$ & & $<0.01$ & & $<0.01$  \\
                  & $\mu$ decay in flight      
    & & 0.01 & & $<0.01$ & & $<0.01$   \\
                  & $\pi$ decay in flight      
    & & $<0.01$ & & $<0.01$ & & $<0.01$  \\ \hline
    \multirow{3}{*}{Miscellaneous}
                  & anti-proton induced        
    & & 0.10 & & -- & & --  \\
                  & cosmic-ray induced         
    & & 0.05 & & 0.16 & & 0.16   \\
                  & pat. recognition errors    
    & & $<0.01$ & & $<0.01$ & & $<0.01$  \\ \hline
    Total Background &                         
    & & 0.41 & & 0.46 & & 1.13  \\ \hline\hline
  \end{tabular}
\medskip
  \caption{A summary of the current Mu2e background estimate and 
    estimates of how the backgrounds would scale for a next generation
    Mu2e experiment, Mu2e-II, that employs Project~X beams at 1 or 
    3~\gev\ and an aluminum or titanium stopping target. For a given 
    stopping target, the difference in background yields between a 
    1~\gev\ or a 3~\gev\ proton beam is about 10\%.  The total 
    uncertainty on the total Mu2e background is estimated to be 
    about 20\%. Note that the DIO estimates for the Mu2e-II case
    assume the tracker has been upgraded to use thinner walled straws.
    If the tracker is not upgraded, the DIO estimates would increase as
    discussed in reference~\cite{Mu2eII}
  }
  \label{cl:tab:PXBgd}
\end{table}

Several components of the Mu2e experiment may need to be upgraded to
handle the higher rates and physics requirements of Mu2e-II.  To
accommodate beam power in the 80kW-110kW range estimated for Mu2e-II
several aspects of the Target and Target Hall would have to be
upgraded.  The proton beam dump would need improved cooling and the
production target would need to be redesigned.  A new production
target design would likely require modifications to the remote target
handling system.  Depending on the proton beam energy and the PS
configuration, it may be necessary to redesign some portions of the
Mu2e beam line just upstream of the PS. Since the extinction in the
Project~X scenarios is a factor 100 smaller than for Mu2e, the
Extinction Monitor system would have to be upgraded to increased
acceptance/sensitivity.  It is expected that even at the increased
rates and beam power discussed earlier, the transport solenoid and
detector solenoid should be able to operate reliably for a Mu2e-II
run. For the production solenoid, the limiting factors are the peak
power and radiation damage it would incur at the increased beam power.
For example, the Production Solenoid heat and radiation shield , which
is currently planned to be fabricated from bronze, could sustain
higher peak power if replaced with tungsten~\cite{Mu2eII}.

Although the beam power increases by a factor of 10 or more, the
instantaneous rates only increase by a factor of 3-5 owing to the
increased duty factor expected from Project~X.  This has important
consequences for the viability of reusing much of the currently
planned Mu2e detector apparatus for a next generation Mu2e-II.  For
example, the current Mu2e tracker is being designed to handle
instantaneous rates higher than what is currently estimated.
Simulation studies in which the instantaneous rates are increased by
factors of two or four have been performed.  These studies indicate
that at four times the nominal rates the tracker reconstruction
efficiency only falls by about 5\% while maintaining the same momentum
resolution.  Thus, these studies indicate that the Mu2e tracker would
be able to handle the Project~X rates.  Simulation studies are
necessary to quantify the degree to which the increased rates affect
the calorimeter's performance.  If it turns out that the expected
performance is not sufficient to meet the Mu2e-II physics
requirements, then it may be necessary to upgrade to faster readout
electronics or to a faster crystal ({\it e.g.} BaF$_2$).  Regardless of the
outcome of the simulation studies, the photo sensors will likely
require replacement owing to radiation damage incurred during Mu2e
running.  The performance of the LYSO crystals is not expected to be
significantly affected by the radiation dose incurred during Mu2e
running.

The current design of the cosmic ray veto system should be adaptable
to the case with a factor of 3--5 increase in rate with only minor
upgrades required and these may only be needed in the most intense
radiation regions of the veto system.  Experience from the first run
of the Mu2e experiment will be crucial in quantifying limitations of
the system for future intensity upgrades.  Improved shielding will
likely be required in the highest radiation regions to reduce
incidental rates in the veto counters.

We investigated the feasibility of a next generation Mu2e experiment
(Mu2e-II) that uses as much of the currently planned facility as
possible and Project~X beams to achieve a sensitivity that's about a
factor of ten better than Mu2e.  Based on these studies we conclude
that a Mu2e-II experiment that reuses a large fraction of the
currently planned Mu2e apparatus and provides a $\times 10$ improved
sensitivity is feasible at Project~X with 100-150 kW of 1 or 3~\gev\
proton beams.  Aside from the DIO background, which requires improved
momentum resolution to mitigate, the remaining backgrounds are kept
under control due to important features of Project~X.  The narrower
proton pulse width and expected excellent intrinsic extinction are
both important to mitigating the RPC background.  The excellent
intrinsic extinction is also important in mitigating backgrounds from
late arriving protons such as $\mu$ and $\pi$ decay-in-flight and beam
election backgrounds.  A beam energy below the proton-antiproton production
threshold eliminates the antiproton induced backgrounds.  The high
duty factor expected for Project~X is important since it enables a
ten-fold improvement in sensitivity over a reasonable timescale while
necessitating only a modest increase (a factor of 3-5, depending on
scenario) in instantaneous rates at the detectors.  Because the
instantaneous rates increase only modestly, we believe that Mu2e-II
could reuse the currently planned Mu2e apparatus with only modest
upgrades necessary.

\subsubsection{A surface $\mu^+$ Beam}
\vskip -12pt
The production of muons typically proceeds through the interaction of a proton beam in a target where positive and negative pions are created. Their decays generate positive and negative muons that can be collected in a subsequent secondary beamline in order to provide the muon beam to the experimental setup. In most cases, the muons are stopped in a target material at the center of the detection system. Experiments searching for charged lepton flavor violating (CLFV) decay channels of the muon such as $\mu\to e \gamma$ or $\mu \to e e e$ require the precise reconstructionof the decay prodiucts' four momentum. For that reason, the muon stopping target should be as thin as possible to suppress multiple scattering of the decay products. On the other hand, the target's thickness affects the stopping efficiency of muons. The range $R_\mu$ of a muon beam with momentum $p$ is proportional to $p^{3.5}$. The total range spread $\Delta R_\mu$ of the stopped muons in a target is given by the momentum spread $\Delta p$ of the beam and a constant term due to the range straggling \cite{Pifer:1976ia}:
\[
\frac{\Delta R_\mu}{R_\mu} = \sqrt{\left(3.5\frac{\Delta p}{p}\right)^2 + \left(0.1\right)^2}.
\]
A lower beam momentum hence increases the muon stopping efficiency and allows the use of thin targets.

\begin{figure}[ht]
\includegraphics[width=0.51\textwidth]{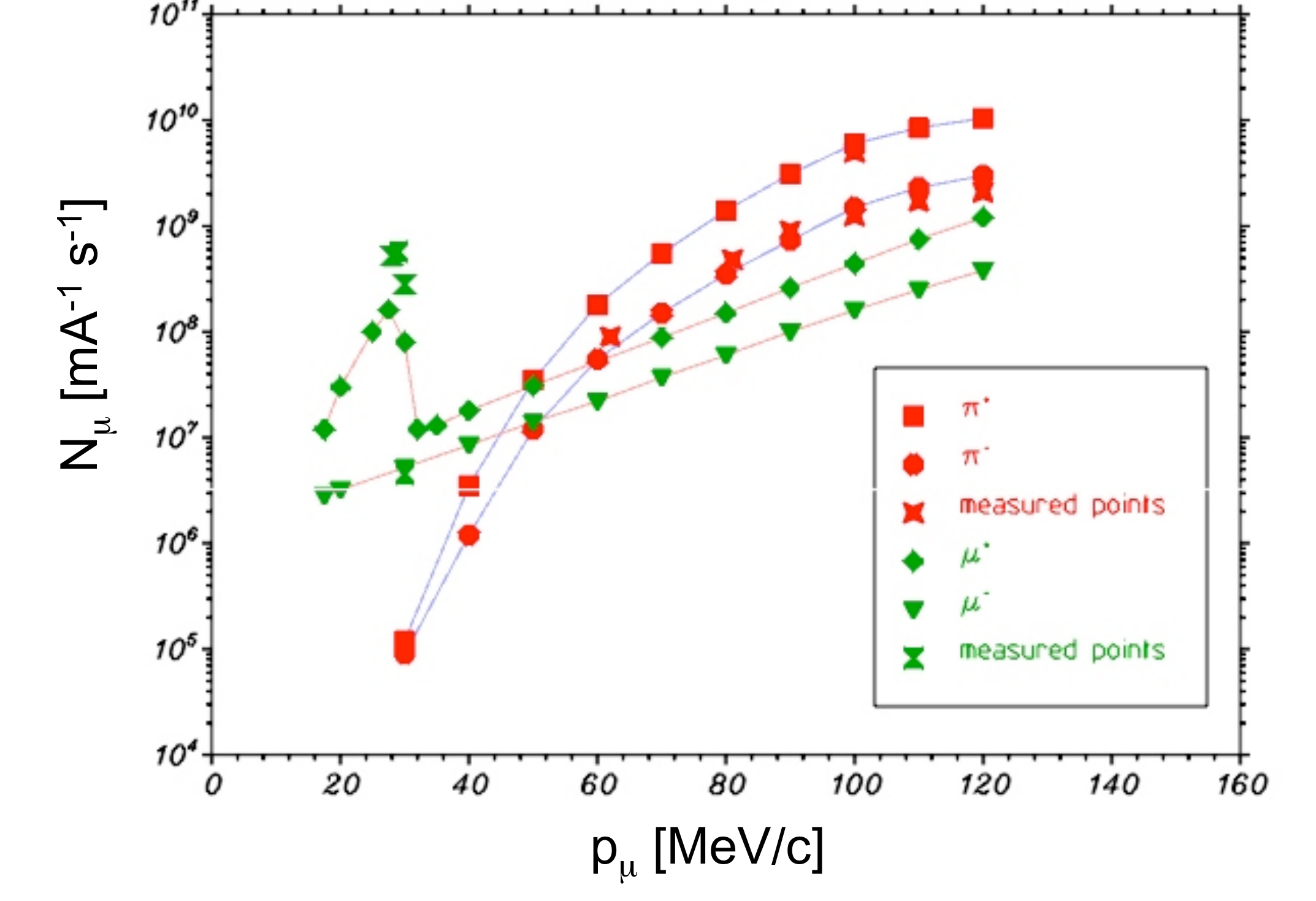}\hfill
\includegraphics[width=0.46\textwidth]{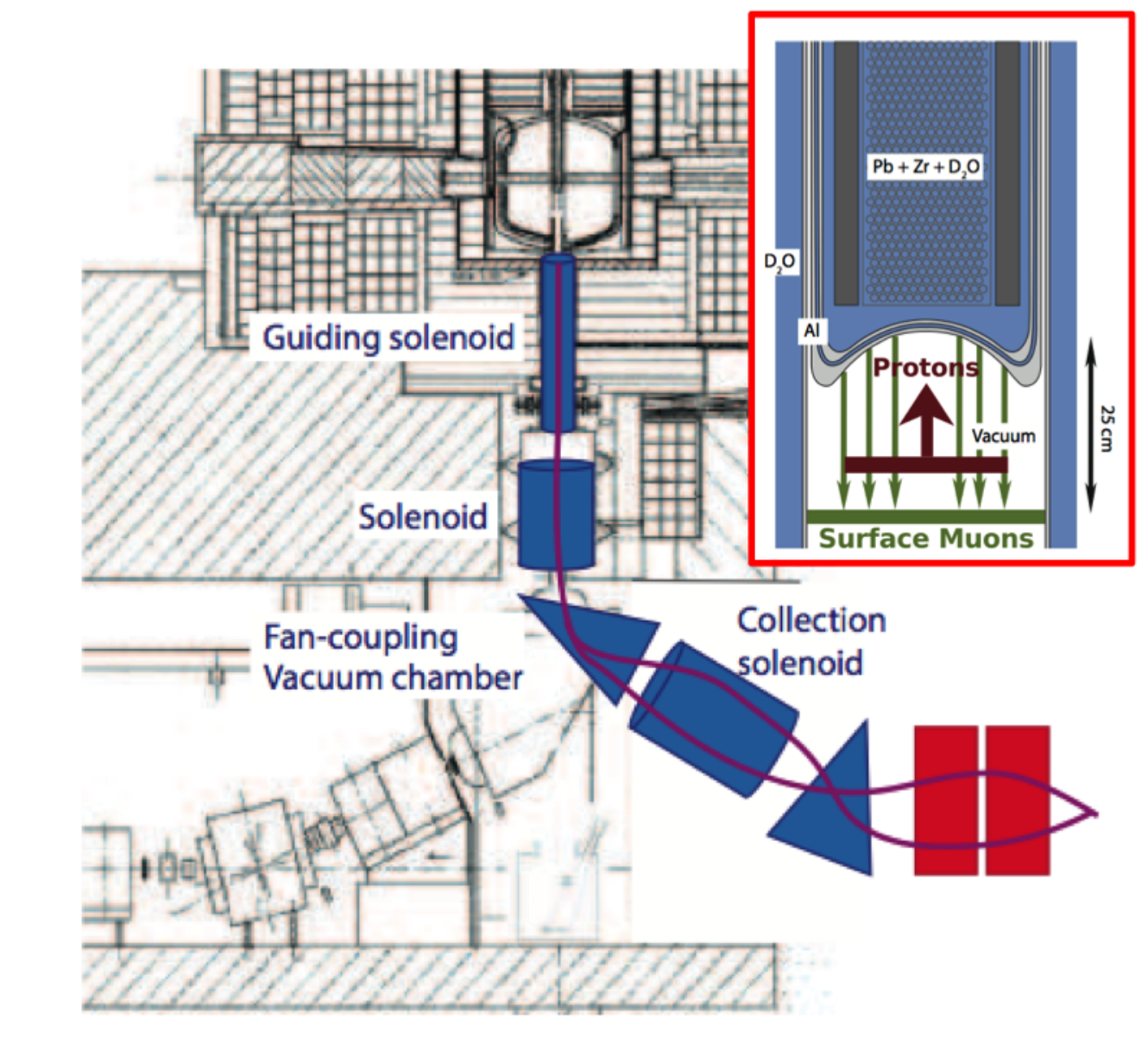}
\caption{Left panel: Pion and muon rates versus particle momentum for both charges predicted for the $\pi$E5 beamline at PSI together with a few measured data points. Right panel: Conceptual design of a possible high intensity muon beamline at the spallation source target at PSI. Red inlet is a zoom of the neutron target region.\label{fig:cl:muonrates}}
\end{figure}

The pions produced in the proton target provide two distinct sources for muons: i) so-called surface muons and ii) cloud muons. The surface muons originate from pions that stop near the surface of the proton target. Due to the fixed kinematics of the two body decay $\pi^\pm \to \mu^\pm \nu_\mu$ at rest, the outgoing muon has a fixed momentum of 29.8\,MeV/c. Surface muons hence have momenta below this specific momentum depending on the pion's decay location inside the target. If the pion has enough momentum to escape the proton target, its decay in flight produces muons with momenta exceeding 29.8\,MeV/c. These muons are called cloud muons, as they have their origin in the moving pions near the target surface. The left panel of Figure \ref{fig:cl:muonrates} shows the predicted production rates of pions (red) and muons (green) for the $\pi E5$ beamline at the Paul Scherrer Institute (PSI) in Switzerland. A clear surface muon peak in the $\mu^+$ yields just below the momentum of 29.8\,MeV/c is visible. Since negative pions undergo nuclear capture if stopped in the target, negative surface muons are not produced. Cloud muon rates scale typically with $p^3$ due to a constant momentum bite of the muon beamline channel and the phase space scaling with $p^2$ \cite{VanDyck:1979xr}. The Figure also shows that positive particle rates exceed the negative particle rates due to the positive charge of the proton beam.  

The high rates of surface positive muons and their low momentum leading to a high stopping density as explained above make these an ideal choice for the searches of CLFV via $\mu^+\to e^+ \gamma$ or $\mu^+ \to e^+ e^+ e^-$. Other physics experiments such as the MuLan experiment \cite{Tishchenko:2012ie} also relied on surface muon beams due to the high rates achievable with a thin stopping target. Table \ref{table:cl:muonexperiments} shows some examples of finalized and future particle physics experiments that all use surface muons (except for the future Mu2e experiment at FNAL which requires negative muons). From the beam rates column one can see that future experiments require the generation surface muon beams with rates an order of magnitude larger than the current available maximum rates at PSI. 

\begin{table}[ht]
\begin{center}
\caption{Overview of some muon experiments and their beam parameters. Experiment marked with $^*$ are future experiments.\label{table:cl:muonexperiments}}
{ \begin{tabular}{lllll}\\
\hline
{\bf Experiment} & {\bf Beam} & {\bf Momentum} & {\bf Rates} & B{\bf Beamline}\\
{\bf } & {\bf } & {\bf [MeV/c]} & {\bf [s$^{-1}$]} & \\
\hline
MEG ($\mu\to e\gamma$) \cite{Adam:2013} & $\mu^+$ & 29.8 & $3 \cdot 10^7$ & $\pi$E5 at PSI\\
\hline
MuLan \cite{Tishchenko:2012ie} & $\mu^+$ & 29.8 & $8 \cdot 10^6$ & $\pi$E3 at PSI\\
\hline
TWIST \cite{Hillairet:2012} & $\mu^+$ & 29.8 & $<5 \cdot 10^3$ & TRIUMF\\
\hline
MEG upgrade$^*$ ($\mu\to e\gamma$) \cite{Baldini:2013ke} & $\mu^+$ & 29.8 & $7 \cdot 10^7$ & $\pi$E5 at PSI\\
\hline
Mu2e$^*$ ($\mu^- \to e^-$) \cite{Abrams:2012er} & $\mu^-$ & $\sim 40$ & $10^{10}$ & FNAL\\
\hline
$\mu^+ \to e^+e^-e^+$ (Phase 1)$^*$ \cite{Blondel:2013ia} & $\mu^+$ & $29.8$ & $<1 \cdot 10^8$ & $\pi$E5 at PSI\\
\hline
$\mu^+ \to e^+e^-e^+$ (Phase 2)$^*$ \cite{Blondel:2013ia} & $\mu^+$ & $29.8$ & $2 \cdot 10^9$ & HIMB at PSI\\
\hline
\end{tabular}}
\end{center}
\end{table}

In addition to the listed particle and nuclear physics experiments in Table \ref{table:cl:muonexperiments}, surface muon beams are highly polarized, making them suitable for material science applications via the Muon Spin Rotation (muSR) technique. Several facilities worldwide currently provide surface muon beams. Table \ref{table:cl:muonrates} shows the laboratories (PSI, J-PARC, and RCNP Osaka University) which currently have the highest rates (up to a few $10^8$\,s$^{-1}$) as well as future estimated rates including a possible facility at Fermilab with Project~X beams. Other facilities like TRIUMF, KEK, RAL-ISIS, and Dubna have rates $< 10^7$ \,s$^{-1}$ and were omitted in this table.

\begin{table}[ht]
\begin{center}
\caption{Summary of some current and planned muon beam facilities at various worldwide laboratories. Table reproduced in part from \cite{Blondel:2013ia} \label{table:cl:muonrates}}
{ \begin{tabular}{llll}\\
\hline
Laboratory / & Energy / & Present Surface  & Future estimated  \\
Beam Line & Power & $\mu^{+}$ rate (Hz) & $\mu^{+}/\mu^{-}$ rate (Hz) \\
\hline
\bf{PSI (CH )} & (590 MeV, 1.3 MW, DC) &  & \\
LEMS & `` & $4 \cdot 10^8$ &  \\
$\pi$E5 & `` & $1.6 \cdot 10^8$  & \\
HiMB & (590 MeV, 1 MW, DC) & & $4\cdot 10^{10} (\mu^{+})$ \\
\hline
\bf{J-PARC (JP)} & (3 GeV, 1MW, Pulsed) & & \\
 & currently 210 KW &  & \\
 MUSE D-line & `` & $3 \cdot 10^7$ & \\
 MUSE U-line & `` &  & $2\cdot 10^8 (\mu^{+})(2012)$ \\
 COMET & (8 GeV, 56 kW, Pulsed) & & $10^{11} (\mu^{-})(2019/20) $\\
 PRIME/PRISM & (8 GeV, 300 kW, Pulsed ) & & $10^{11-12}(\mu^{-}) ( > 2020)$ \\
 \hline
 \bf{FNAL (USA)} & &  & \\
 Mu2e & (8 GeV, 25 kW, Pulsed) & & $5\cdot 10^{10} (\mu^{-})(2019/20) $\\ 
 Project~X Mu2e & (3 GeV, 750 kW, DC to pulsed) & & $2\cdot 10^{12} (\mu^{-})(>2022)$ \\
\hline 
\end{tabular}}
\end{center}
\end{table}

Next generation CLFV experiments with a surface muon beam ($\mu^+ \to e^+\gamma$ \cite{Baldini:2013ke} and $\mu^+ \to e^+e^-e^+$ \cite{Blondel:2013ia}) are in the design and planning stages at PSI. Since these experiments have to suppress accidental backgrounds which scale with the square of the beam rate, they require continuous beams. The current $\pi$E5 beamline at PSI with conventional focusing quadrupole elements can deliver about $2\cdot10^8$\,muons/s at the proton beam power of 1.3\,MW. However, this beamline views target station E \cite{Heidenreich:2002zz} at PSI where only about 20\,\% of the proton beam interacts. Most of the beam is transmitted to the neutron spallation target at the end of the proton beamline. PSI is currently studying the possibility of a high intensity muon beamline (HIMB) at the neutron spallation target based on a large capture solenoid concept. The conceptual design is shown in the right panel of Figure \ref{fig:cl:muonrates}. About 70\,\% of the 1.3 MW proton beam is coming from the left and directed upwards into the SINQ target of lead-filled zircaloy tubes surrounded by a cooling D$_2$O layer. Surface muons from the aluminum window (see red framed inlet) could then be collected in the downwards direction with newly installed solenoids (blue) into a new dedicated beamline. Realistic Monte Carlo simulations give an estimate of about $1\cdot10^{11}$\,muons/s below 29.8\,MeV/c corresponding to a rate of $3\cdot 10$\,s$^{-1}$ for 10\,\% momentum bite around 28\,MeV/c \cite{Blondel:2013ia}.

Table \ref{table:cl:muonrates} also shows the different beam energies at the 3 facilities. While PSI employs a DC, 590\,MeV proton beam of about 1.3\,MW total power, J-PARC (pulsed mode) and Fermilab with its future Project~X\footnote{http://projectx.fnal.gov/} accelerator infrastructure (DC and pulsed mode) have 3\,GeV beams with about 1\,MW of power. Original studies at ISIS \cite{Bungau:2013hd} found that the muon yields per MW of beam power are 3-7 times lower at 3\,GeV compared to PSI's 590\,MeV. However, recently corrected pion cross sections in GEANT4 demonstrated that the yields are rather similar \cite{striganov:12}. Therefore, an optimized surface muon beamline in the Project~X era at Fermilab with 750\,kW beam power could be a competetive option to the HIMB at PSI. 

As described in the following sections, the feasibility of next generation experiments ($\mu^+ \to e^+\gamma$ and $\mu^+ \to e^+e^-e^+$) would require the availability of a high rate (sub-)surface muon beam. With the prospect of the high power 3\,GeV beams at Project~X surface muon beams could become available to a multitude of applications at Fermilab such as particle physics and material science with muSR. Besides the available beam power of about 750\,kW for a muon program, the flexibility of the accelerator beam structure is an additional benefit to serve a multitude of experiments with different requirements. The design of a future surface muon beamline will need to take into account a variety of parameters for the optimization:
\begin{itemize}
\item Proton target: The design of the proton target by itself has many design parameters. The target can either be at the end of the proton beamline in conjunction with a planned neutron spallation target similar to aforementioned HIMB strategy at PSI. Alternatively, the proton target could be in the proton beam and only using a fraction of the available beam. The material choice and target shape influence the surface muon yields, mechanical stability and determine the cooling requirements. Other aspects of the design include the minimization of secondaries (e.g. $e^\pm$, $\pi^\pm$, $\gamma$, and $n$) and ideally a low activation throughout operation. 
\item Muon beam channel: The development of a specific concept of the muon beam channel should be based on the experimental requirements such as required muon rates, polarization, momentum and momentum bite as well as the beam spot at the stopping target. As mentioned before, there are two major concepts based on either a more conventional design with focusing quadrupole elements or large solid angle capture solenoids. Specific elements such as a $E\times B$ separator will influence the beam contamination with electrons and the length of the beamline and pulsed mode are important for the remaining pion flux.
\item The muon stopping target: While this element is not necessarily part of the beam channel design, it is necessary to include its optimization together with the available tuning parameters of a beamline (such as momentum, momentum bite and beam spot). Design parameters include the target material and shape. 
\end{itemize}

In addition to the possibility of a dedicated new surface muon beamline that simultaneously serves the particle physics and muSR communities, one should also investigate the possibility of whether the planned Mu2e proton target and capture solenoids could be adapted to supply a surface muon beam.
Initial Monte Carlo simulations indicate that surface muons would be transported through the solenoids and could be efficiently stopped in a thin walled tubular target. For pion suppression, the Mu2e setup operates in pulsed mode which is not ideal for experiments like $\mu^+ \to e^+\gamma$ and $\mu^+ \to e^+e^-e^+$ as explained above due to the accidental backgrounds. While a simple estimate of a pulsed mode only shows a slight loss compared to an ideal DC beam, a detailed investigation with a full simulation is required to compare an adapted Mu2e beamline to a dedicated surface muon facility. In addition, the current Mu2e setup does not include a separator for electron suppression.

In summary, the prospect of a high power 3\,GeV proton beam available for a future muon program at Project~X offers the opportunity to provide a world leading surface muon beam competitive with the rates at other facilities (specifically the planned HIMB beamline at PSI). This could facilitate the next generation of CLFV experiments in addition to a material science-oriented muSR facility. However, given the variety of optimization parameters, the development of a concept would require significant Monte Carlo and other design studies.


\subsubsection{$\mu \to e \gamma$}
\vskip -12pt

The current limit on the $\mu^+\to e^+ \gamma$ branching fraction is 
$5.7\times 10^{-13}$ at 90\% confidence level from MEG at 
PSI~\cite{Adam:2013mnn}, using 
$3.6\times 10^{14}$ stopped muons, from data taken in 2009--2011. 
Their sensitivity is dominated by 
accidental background, which is related to the muon stop rate
$R_\mu$ and various experimental resolutions~\cite{Baldini:2013ke}:
\begin{equation}
N_{\rm acc} \propto R_{\mu}^2 \times \Delta E_\gamma^2 \times \Delta P_e 
\times \Delta \Theta^2_{e\gamma} \times \Delta t_{e\gamma} \times T ,
\end{equation}
where $\Delta t_{e\gamma}$, $\Delta P_e$, 
$\Delta E_\gamma$, and $\Delta\Theta_{e\gamma}$ are the resolutions of 
detector timing, positron momentum, photon energy, and positron-photon angle,
respectively, and $T$ is total data acquisition time. MEG will conitnue 
taking data through 2013. They expect to
approximately double their published dataset~\cite{GLimArgonne2013}.

The MEG upgrade plans to improve the experiment sensitivity by a factor of 10. 
They will increase the intensity of the surface beam, and use a thinner or
active stopping target. The detector upgrade includes a larger drift chamber with
thinner wires and smaller cells, an improved timing counter, and a larger LXe
calorimeter with SiPM readout. Data-taking is planned in the years 2016--2019.

The photon energy
resolution is a limiting factor in a $\mu^+\to e^+ \gamma$ search.  A pair spectrometer, based on reconstruction of $e^+e^-$ pair tracks produced in a
thin converter can provide improved photon energy resolution, at a sacrifice in efficiency. Even though only a small fraction of photons will convert,
the much higher power beam at Project X can
compensate for the loss of statistics~\cite{Fritz:2012aaa}. The thickness of the converter affects the energy 
resolution due to multiple scattering. Aa detailed study is thus required to prove that this approach does indeed provide an overall
improvement, as well as to optimize the converter thickness and to study the utility of making the converter active.

We have conducted an initial study of this concept using a fast simulation tool
(FastSim) originally developed for the Super$B$ experiment using the  \babar\ software
framework and analysis tools. FastSim allows us to model detector components
as two-dimensional shells of simple geometries. Particle scattering, energy
loss, secondary particle production (due to Compton scattering, Bremsstrahlung,
conversion, EM or hadron showers, {\it etc.}) are simulated at the intersection
of particles with detector shells. Tracks are reconstructed with a Kalman filter
into piece-wise trajectories. 

\begin{figure}[htbp]
   \centering
   \includegraphics[width=0.5\textwidth]{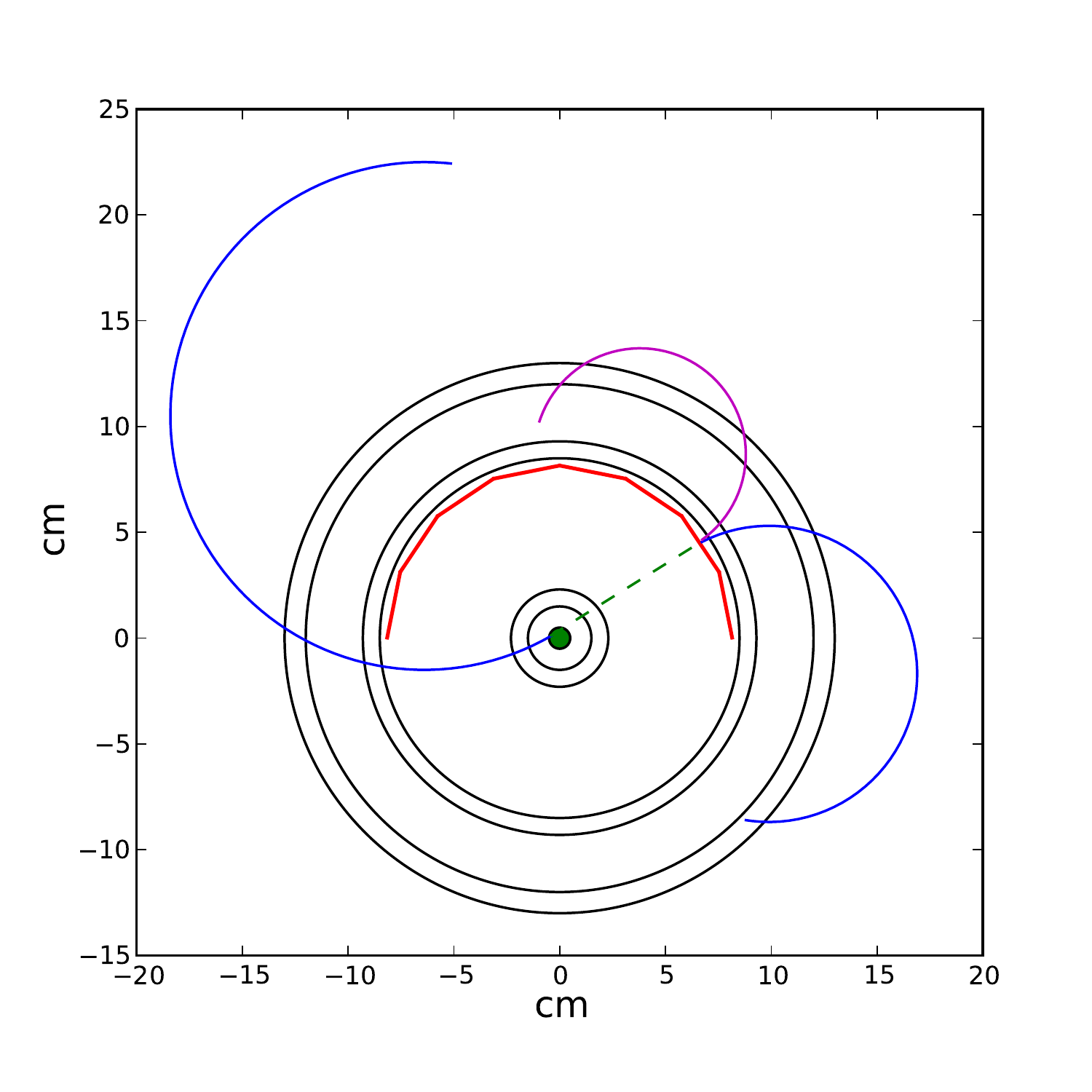} 
   \caption{Schematic drawing (view transeverse to the muon beam axis) of the $\mu\to e\gamma$ detector used in the 
   FastSim model. }
   \label{fig:muegamma-scheme}
\end{figure}

The FastSim model consists of a thin aluminum stopping target and a six-layer
cylindrical silicon detector. A 0.56 mm thick lead (10\%~$X_0$) half cylinder
covering 0--$\pi$ in azimuthal angle at $R=80$~mm serves as the photon converter. 
The target consists of two cones connected at their base; each cone is 30 mm 
high, 5 mm in radius, and 50~$\mu m$ thick.
Two silicon detector cylinders are placed close the target for better vertexing resolution;
two layers are placed just outside the Pb converter, and two layers a few cm away. The layout is shown in
Fig.~\ref{fig:muegamma-scheme}. The entire detector is placed in a 1~T
solenoidal magnetic field. 

\begin{figure}[htbp]
   \centering
   \includegraphics[width=0.48\textwidth]{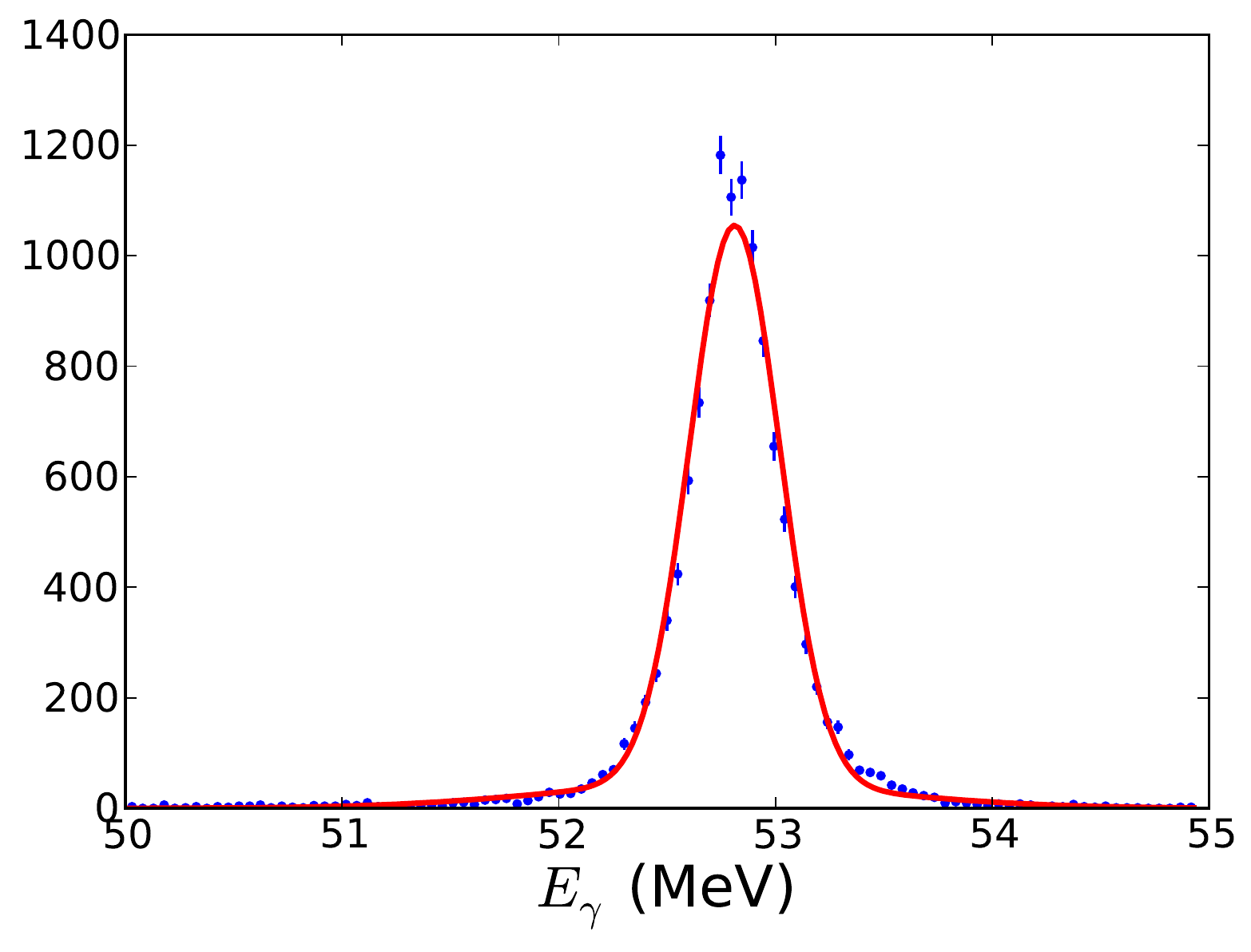} 
   \includegraphics[width=0.48\textwidth]{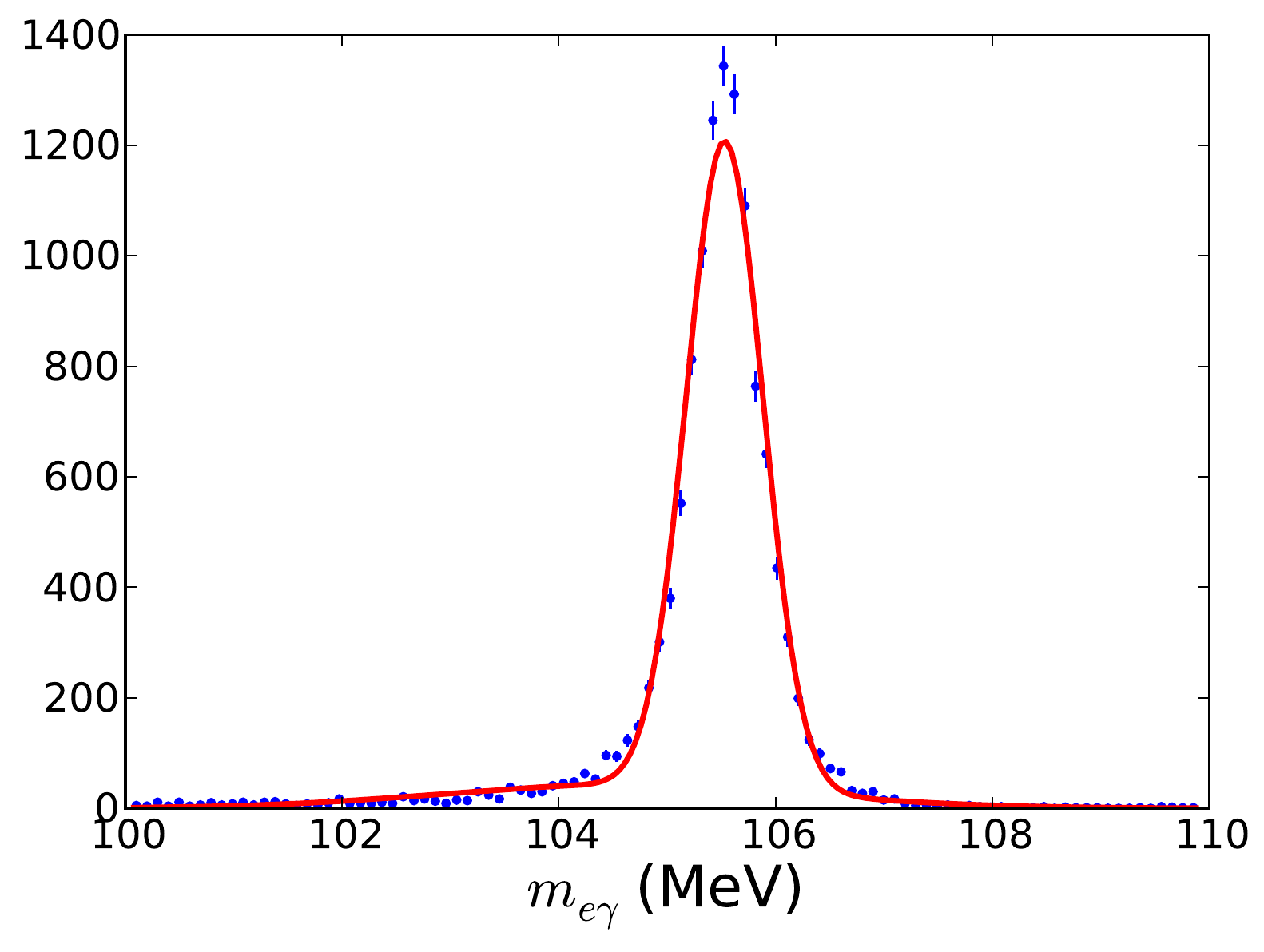} 
   \caption{Photon energy and muon candidate mass resolutions in 
   $\mu^+\to e^+\gamma$ FastSim study. Fitted curve is a double-Gaussian 
   distribution.}
   \label{fig:muegamma-resolutions}
\end{figure}

We generate muons at rest and let them decay via $\mu^+\to e^+\gamma$ to study
the reconstruction efficiency and resolution. Approximately 1.3\% of generated
signal events are well-reconstructed, passing quality and fiducial selections criteria.
The photon energy resolution is approximately 200~keV, similar to the positron momentum
resolution, which corresponds to $0.37\%$ for 52.8~MeV photons. This is a great
improvement compared to the 1.7\%--2.4\% resolution of the current MEG and the  1.0\%--1.1\% resolution goal of the MEG
upgrade. The muon candidate mass resolution is 340~keV (85\% Gaussian core).
Figure~\ref{fig:muegamma-resolutions} shows the photon energy and muon candidate 
mass resolutions. The positron energy resolution is better than that of MEG, but not
as good as what is expected in the MEG upgrade. Angular resolution is similar to  the
current MEG.

We then perform toy studies to estimate the levels of accidental background
and radiative muon decay background, and the sensitivities to signal branching
fraction, under different running conditions. In addition to the five variables
used in the likelihood fit in MEG ($E_e$, $E_\gamma$, $\theta_{e\gamma}$, 
$\phi_{e\gamma}$, and $t_{e\gamma}$), we also take advantage of direction 
information of the converted photon, which provides excellent discriminating
power against accidental background. We use a cut-and-count approach to
optimize the sensitivity. 

\begin{figure}[htbp]
   \centering
   \includegraphics[width=0.48\textwidth]{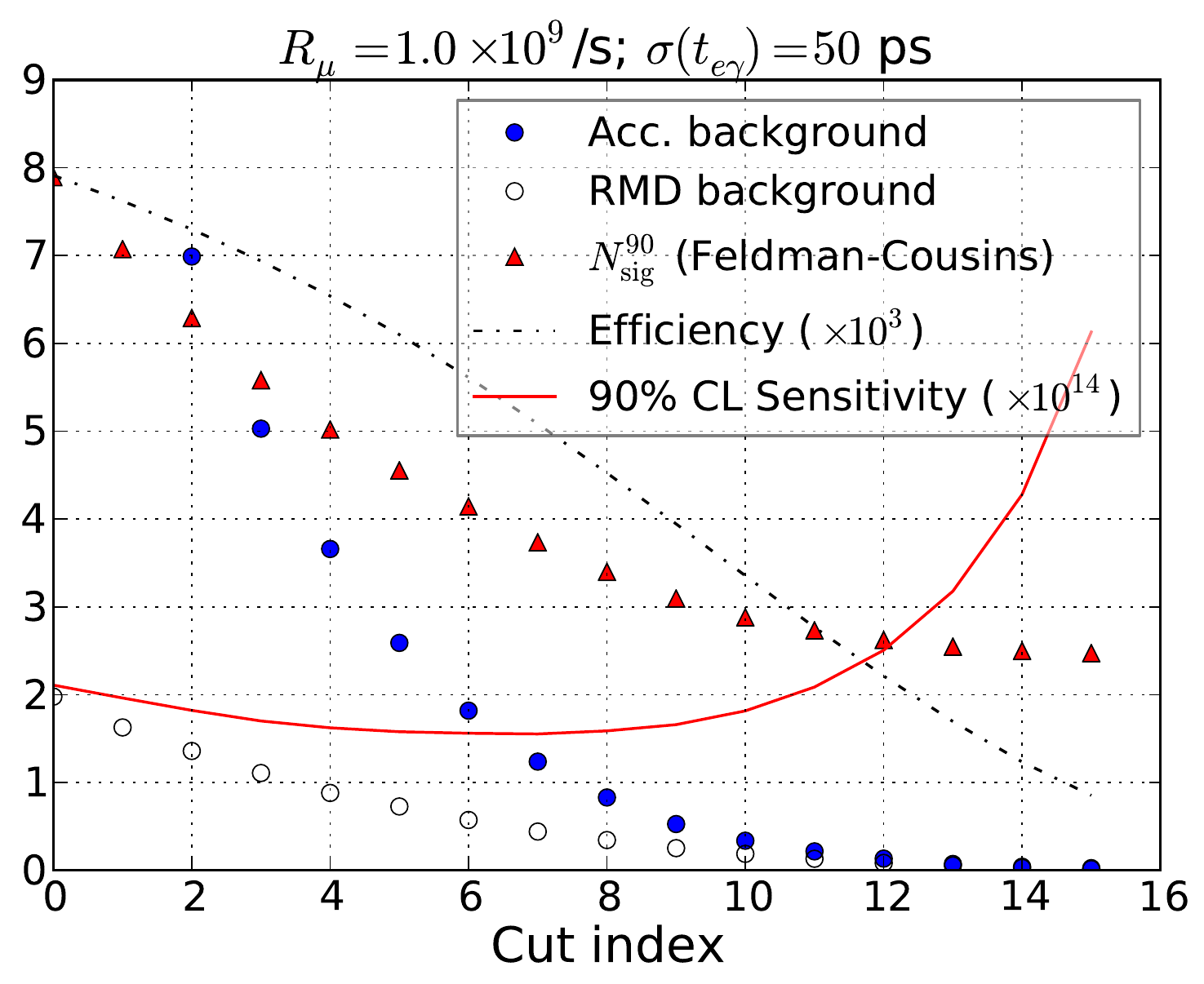} 
   \includegraphics[width=0.48\textwidth]{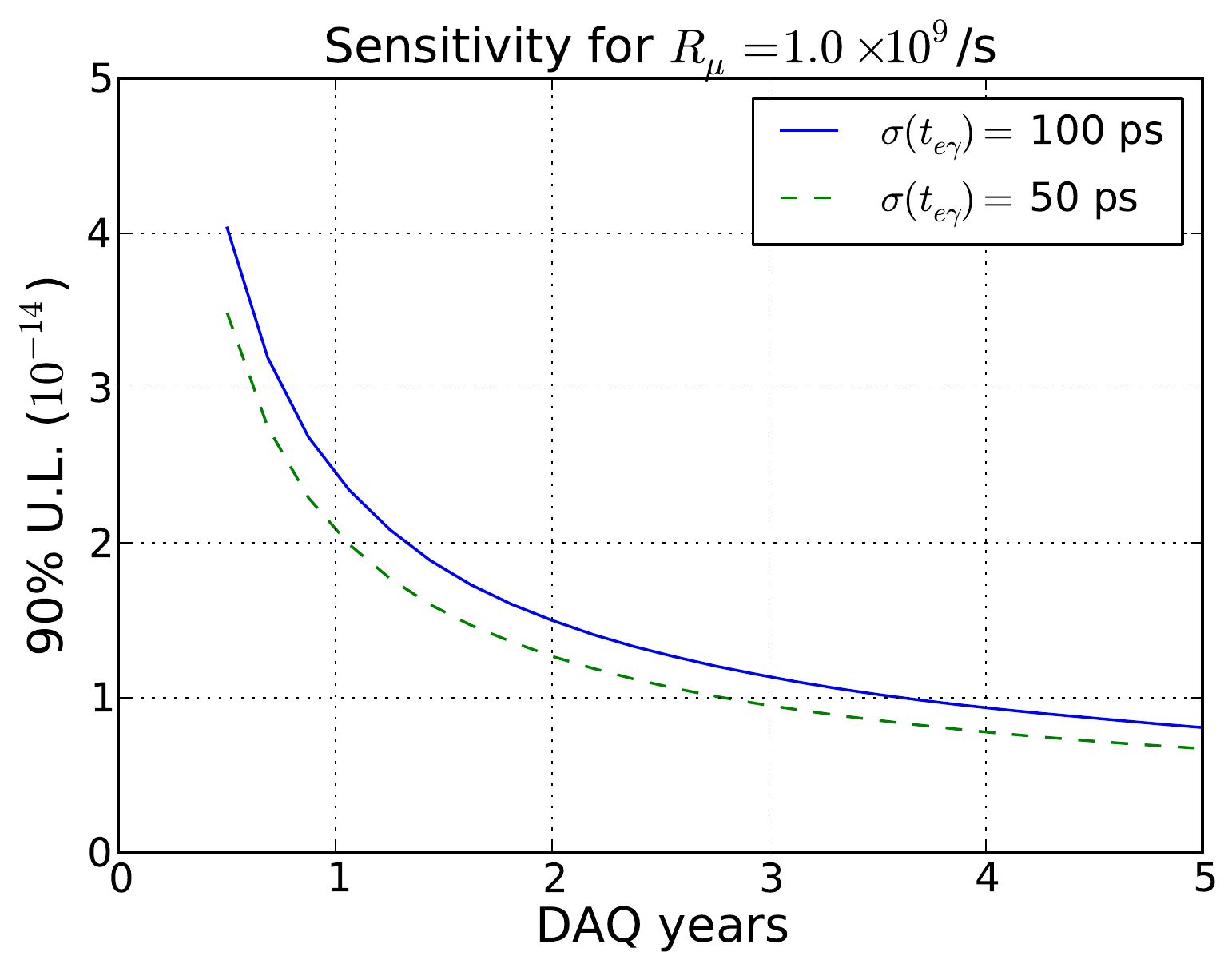} 
   \caption{Left: $B(\mu^+\to e^+\gamma)$ sensitivity optimzation. 
Best sensitivity is $1.6\times 10^{-14}$. Right: sensitivity as a function of integrated DAQ time for both 50 ps and 100 ps $t_{e\gamma}$ resolutions.}
   \label{fig:muegamma-sensitivity}
\end{figure}

Figure~\ref{fig:muegamma-sensitivity} shows the background levels, 
signal efficiency, and 90\%
C.L. sensitivity under various selection cuts for 
$R_\mu=1\times 10^{9}$~muons/s, 50 ps resolution on $t_{e\gamma}$, for
an integrated DAQ time of 1.5 years, as well as the sensitivity reach
as a function of integrated DAQ time for both 50 ps and 100 ps timing
resolutions. 

\begin{figure}[htbp]
\centering
\includegraphics[width=0.48\textwidth]{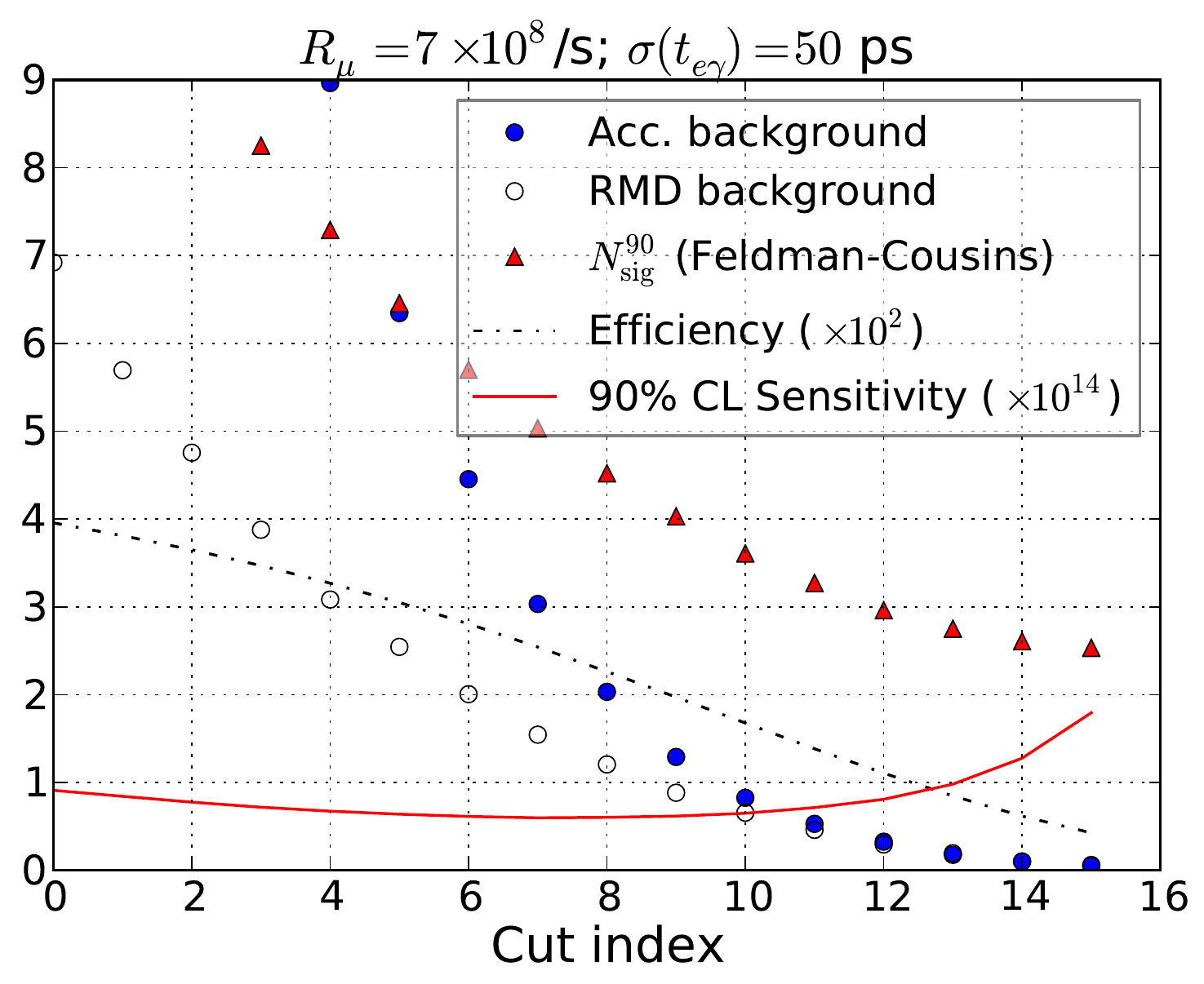}
\caption{Left: $B(\mu^+\to e^+\gamma)$ sensitivity optimzation with 5-times
higher signal sensitivity and lower $R_\mu$ than that in 
Fig.~\ref{fig:muegamma-sensitivity}. Best sensitivity is $6\times 10^{-15}$.}
\label{fig:muegamma-sens-5x}
\end{figure}

Increase the muon rate futher could improve the sensitivity. However,
it quick moves away from ${\cal O}(1)$ background regime because the accidental
background grows $\sim R_\mu^2$. A better approach is to increase the
efficiency and reduce the muon rate to keep the background level low. 
Figure~\ref{fig:muegamma-sens-5x} 
shows a scenario where the signal efficiency is 5-times higher
and muon stopping rate is slightly reduced to $R_\mu=7\times 10^{8}$. 
One can reach a sensitivity of $B(\mu^+\to e^+\gamma)<6\times 10^{-15}$.

Using a converted photon to increase the $\mu^+\to e^+\gamma$ detection
sensitivity thus appears to be a promising approach. Further studies are 
needed to quantify the requirements in detail to improve upon the MEG upgrade
sensitivity by an order of magnitude or more.

An alternative version of the photon conversion approach to a $\mu \to e \gamma$ experiment has also been discussed. In this version, consider a large volume solendoidal magnet, such as the KLOE coil, which has a radius of 2.9m, run at a field of perhaps 0.25T. A large volume, low mass cylindrical drift chamber provides many ($\ge$100) layers of tracking, utilizing small cells and having a total number of sense wires approaching $10^5$. Interspersed every ten layers is a 0.5 mm W converter shell. There are a sufficient number of points on the $e^+$ and $e^-$ tracks from converted photons behind each converter to reach a total conversion efficiency of perhaps 80\%, with excellent photon mass resolution.

\subsubsection{$\mu \to 3e$}
\vskip -12pt


The $\mu \rightarrow eee$ decay is a charged lepton flavor violating process strongly suppressed in the Standard Model. New physics mediated either via virtual loop or three diagrams can enhance these rates to values accessible by the next generation of experiments. An interesting feature of this process is 
the possibility to determine the chirality of New Physics, should it be observed with sufficient statistics~\cite{Okada:1999zk}. The current limit ${\cal B}(\mu^+\to e^+e^-e^+) < 1\times 10^{-12}$ has been set by the SINDRUM experiment at PSI~\cite{Bellgardt:1987du}. The Mu3e experiment~\cite{Blondel:2013ia} has been proposed to improve this bound by four orders of magnitude.

\begin{figure}[htb]
\begin{center}
\includegraphics[width=0.45\textwidth]{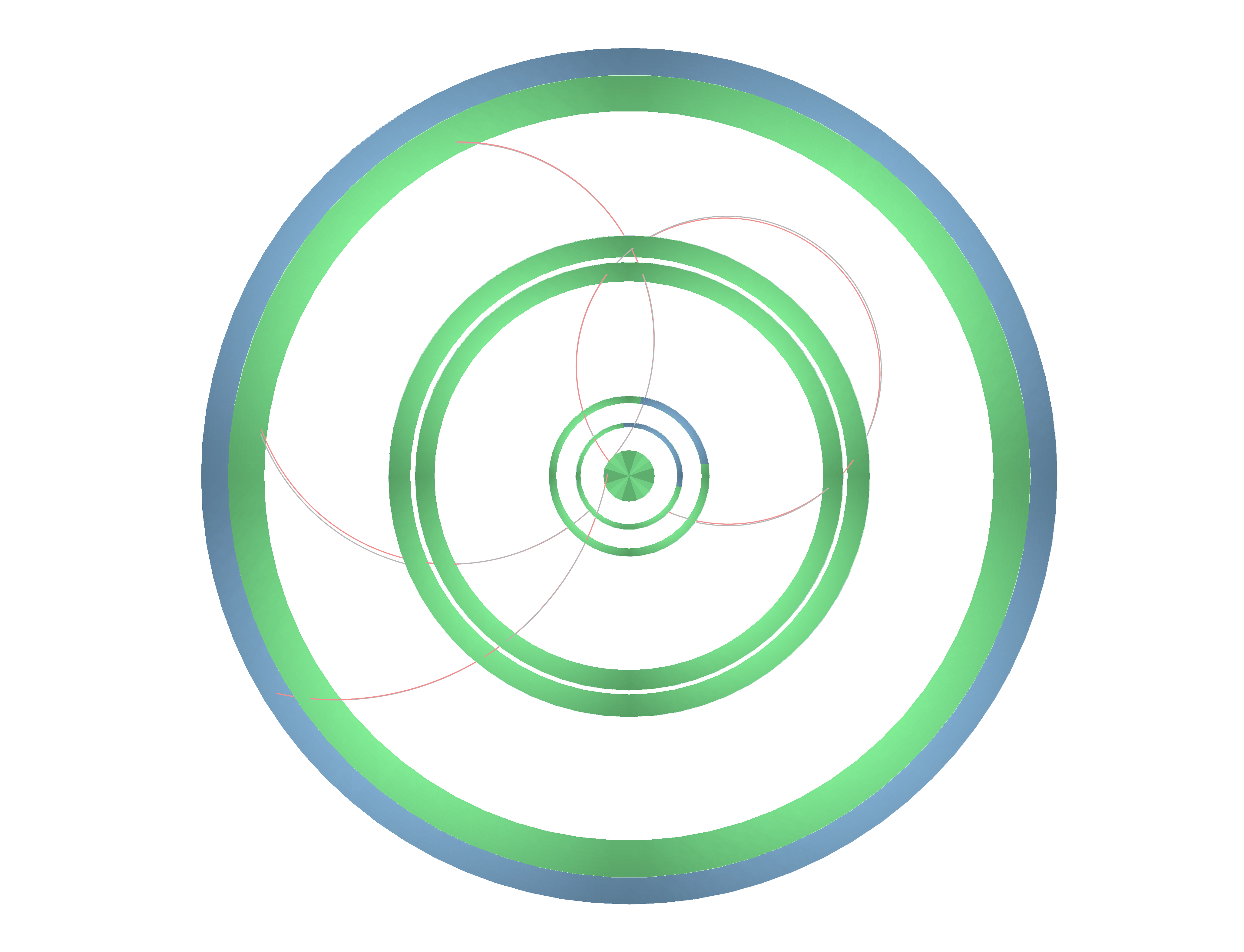}\includegraphics[width=0.45\textwidth]{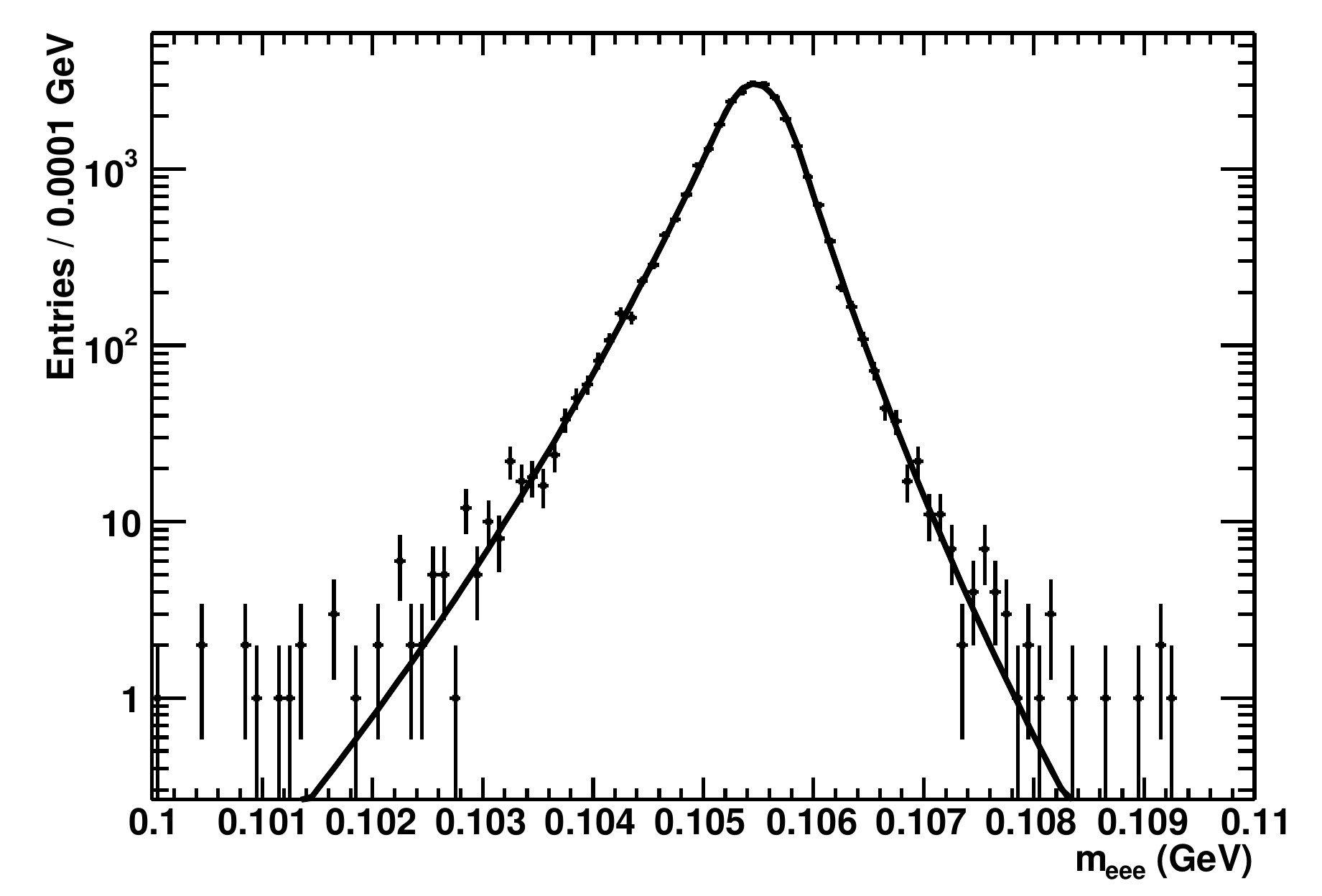}
\end{center}
\caption{Left: Display of the experimental setup, together with a simulated $\mu^+ \rightarrow e^+e^-e^+$ event. 
Right:  The $e^+e^-e^+$ invariant mass distribution after all selection criteria are applied fitted by a 
sum of two Gaussian functions.}
\label{Fig::mu3e}
\end{figure}

We present a detector concept to search for $\mu \rightarrow eee$ decay using the FastSim simulation package. The experimental setup  consists of a compact tracker made of 6 layers of cylindrical silicon detectors, each composed of 50 $\mu$m thick silicon sensors mounted on 50 $\mu$m of kapton. The target is formed of two hollow cones, having each a length and radius of 5 cm and 1 cm, respectively. Contrary to Mu3e, we consider an active target made of silicon pixel detectors, assuming a pixel size of 50 $\mu$m by 50 $\mu$m. Although not included, a time-of-flight system should be installed as well, providing a time measurement with a resolution of 250~ps or better. The apparatus is displayed in Fig.~\ref{Fig::mu3e}, together with a simulated $\mu^+ \rightarrow e^+e^-e^+$ event.

We generate $\mu^+ \rightarrow e^+e^-e^+$ events according to phase space, and constrain the tracks to originate from the same pixel in the active target. To further improve the resolution, we require the probability of the constrained fit to be greater than 1\%, and the reconstructed muon momentum less than $1 ~\Mev$. The absolute value of the cosine of the polar angle of each electron must also be less than 0.9. The resulting $e^+e^-e^+$ invariant mass distribution is shown in Fig.~\ref{Fig::mu3e}, and peaks sharply at the muon mass. We extract the resolution by fitting this spectrum with a double-sided Crystal Ball function (a Gaussian with power-law tails on both sides). The Gaussian resolution is found to be 0.3~\Mev\ for a signal efficiency of 27\%.  

To achieve a single event sensitivity at the level of $5\times\sim 10^{-18}$ after a 3-year run, a stop muon rate of the order of $8\times 10^{9}$ is needed. For comparison, the estimated stop muon rate at Mu3e with the HiMB beam is expected to be $2\times 10^{9}$~\cite{Blondel:2013ia}.

For the purpose of estimating background contributions, we define a signal window as $ 104.9 < m_{eee} < 106.5 ~\Mev$, containing approximately 90\% of the signal. The accidental background arise from $\mu \rightarrow e^+e^-e^+ \nu\bar\nu$ events where the two neutrinos carry almost no energy. We estimate its contribution to be about 7.5 events by convolving the branching fraction with the resolution function and integrating in the signal region. However, this background depends strongly on the tail resolution, and small improvements translate into large background reductions. For example, decreasing the thickness of the silicon sensors and the supporting kapton structure by 20\% (40\%) reduces the background down to $\sim 4$ ($\sim 1$) events. 

We consider accidental backgrounds produced by the combination of a Michel decay and a radiative Michel decay (2M$\gamma$ decays), or three simultaneous Michel decays (3M decays), where one positron is misreconstructed or produces an electron by interacting within the detector. In both cases, we assume that
the decays occur within the same pixel in the active target and within the same time window. This yields position and time suppression factors $\delta S = 7.8\times 10^{-7}$ and $\delta t = 2.5\times 10^{-10}$, respectively. The background rate is given by:
$$N_{2M\gamma} = {R_\mu}^2 \delta S \delta t {B(\mu^+ \rightarrow e^+ \nu_e \bar\nu_\mu)}^2 B(\mu^+ \rightarrow e^+ \nu_e \bar\nu_\mu \gamma) P(\gamma \rightarrow e^+ e^-)  P_\mu  \simeq 0.33 P_\mu$$
$$N_{3M} = {R_\mu}^3(\delta S)^2 {B(\mu^+ \rightarrow e^+ \nu_e \bar\nu_\mu)}^3 (\delta t)^2 P_\mu \simeq 0.02 P_\mu$$
where $P(\gamma \rightarrow e^+ e^-)\sim 0.18\%$ is the probability of photon conversion in the target and $P_\mu$ denotes the probability to reconstruct a muon candidate after all selection criteria are applied. We estimate the factors $P_\mu \sim {\cal O}(10^{-8})$ for 2M$\gamma$ decays and $P_\mu \sim {\cal O}(10^{-9})$ for 3M decays with our simulation. For a 3-year run and with a rate of $8\times 10^{9}$ stopped muons per second, both backgrounds are found to be less than an event.

In summary, we outline the requirements needed to improve by an order of magnitude the projected $\mu \rightarrow eee$ 
sensitivity of the mu3e experiment. We study a similar design, with the addition on an active target instead of a passive one. Assuming a 3-year run, a rate of $8\times 10^{9}$ stopped muons in the target per second would be required. Relatively modest improvements on the resolution are also needed to maintain the irreducible background at an appropriate level, while an active target proves to be essentially in the reduction of accidental 
backgrounds.


\section{Tau Experimental Overview}\label{sec:cl:tauexp}
\vskip -12pt
In contrast to muon CLFV searches, in which a single dedicated experiment is required for a given decay,  $\tau$~lepton CLFV searches are conducted using the large data sets collected in comprehensive $e^+e^-$ or hadron collider experiments. The relative theoretical parameter reach of $\mu$ and $\tau$ decay experiments is model-dependent, and thus comparisons of limits or observations in the two cases can serve to distinguish between models.
Tests with taus can be more powerful on an event-by-event basis than those using muons, since the large $\tau$ mass greatly decreases 
Glashow-Iliopoulos-Maiani (GIM) suppression, correspondingly increasing new physics partial widths (typically by a factor of $\geq 500$ in $\cal{B}(\tau \rightarrow \mu \gamma)$ or $e \gamma$ {\it vs.} $\cal{B}(\mu \rightarrow$$e \gamma$)).  The difficulty is that one can typically produce $\sim 10^{11}$ muons per second, while the samples from \babar\ and Belle collected over the past decade together total $\sim 10^{10}$ events.  

The new generation of super $B$ or $\tau$/$c$ factories, \cite{superkekb},\cite{Blinov:2009zzd},\cite{tc} promise to extend the
experimental reach in $\tau$ decays to levels that sensitively
probe new physics in the lepton sector. Since CLFV is severely suppressed in the Standard Model, CLFV $\tau$ decays
are especially clean probes for BSM Physics
effects.  The
super flavor factories can access $\tau$ CLFV decay rates two orders of magnitude smaller than current limits for the cleanest channels
({\it e.g.}, $\tau\to 3\ell$), and at least one order of magnitude smaller for other
modes that have irreducible backgrounds, such as $\tau\to \ell\gamma$. Super flavor factories thus have a sensitivity for CLFV decays that directly confronts many BSM models. 

Polarized beams at an $e^+e^-$ collider can provide further
experimental advantages.   Belle II at \hbox{SuperKEKB} will not have a polarized beam, but both the proposed BINP and Tor Vergata $\tau$/$c$ factories will have polarized electron beams.  Polarization of the taus thus produced provides several advantages. It allows reduction of backgrounds in certain
CLFV decay modes, as well as providing sensitive new
observables that increase precision in other important measurements, including
searches for $C\!P$ violation in $\tau$ production and decay, the measurement
of $g-2$ of the $\tau$, and the search for a $\tau$ EDM.  Preliminary studies indicate that polarization improves the sensitivity on these quantities by a factor of two to three. Should the CLFV decay $\tau \rightarrow 3\ell$ be found, a study of the Dalitz plot of the polarized $\tau$ decay can determine the Lorentz structure of the CLFV coupling.

The provision of polarization requires a polarized electron gun, 
a lattice that supports transverse
polarization at the desired CM energy, a means of interchanging transverse polarization in the
ring and longitudinal polarization at the interaction point and a means of monitoring
the polarization, typically a Compton polarimeter to monitor the
backscattering of circularly polarized laser light. Achieving useful longitudinal polarization at the interaction point requires sufficiently long depolarization time of the machine lattice, which is highly dependent on the details of the lattice and the beam energy.

Provision of a
polarized positron beam is difficult and expensive; it is generally
also regarded as unnecessary, as most of the
advantages of polarization for the measurements cited above can be
accomplished with a single polarized beam.

\begin{figure}[htb]

\begin{center}
\includegraphics[width=8cm]{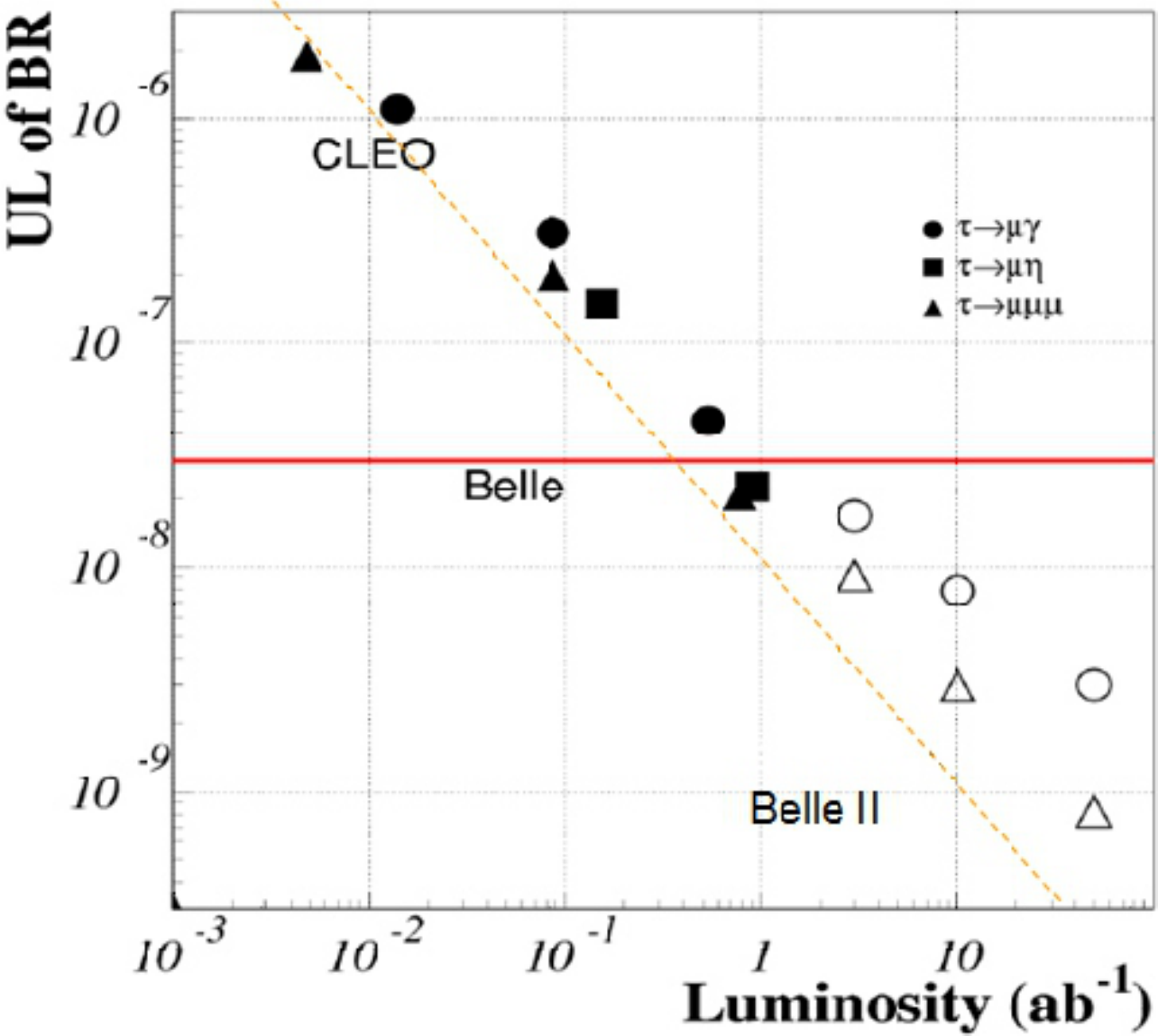}
\smallskip
\caption{\label{CL:Belle}Extrapolation of the 90\% upper limit sensitivity of Belle-II (open symbols) from existing limits (filled symbols). For $\tau\rightarrow \mu\gamma$, which has irreducible backgrounds, the limit scales as $1/{\sqrt{\!\int{\!\cal{L}}dt}}$. For $\tau\rightarrow \mu\mu\mu$, which is essentially background-free, the limit scales as $1/{\int{\!\cal{L}}dt}$.}
\end{center}
\end{figure}

The sensitivity of $\tau$ CLFV searches at SuperKEKB has been estimated by extrapolating from current CLEO, Belle and \babar\  
limits (see Figure~\ref{CL:Belle}). The optimization of search sensitivities depends on the size
of the sample as well as on the sources of background. For SuperKEKB, the
extrapolation for the (largely background-free) $\tau\to\ell\ell\ell$
modes assumes $1/{\mathcal L}$ scaling up to 5 ab$^{-1}$; that for $\tau\to\ell\gamma$
modes scales as $1/{\sqrt{\mathcal L}}$. 
The expected sensitivities for several modes are shown for the Belle II experiment in Table~\ref{tab:LFVExptSensitivities-BelleII}~\cite{Abe:2010sj}.  

\begin{table}[!b]
  \caption{
    \label{tab:LFVExptSensitivities-BelleII}
    Expected $90\%$ CL upper limits
    on $\tau\to\mu\gamma$, $\tau\to \mu\mu\mu$, and $\tau\to \mu\eta$
    with $5 \ {\rm ab}^{-1}$ and  $50 \ {\rm ab}^{-1}$ data sets from Belle II and  Super KEKB.
  }
  \begin{center}
    \begin{tabular}{lll}
      \hline \hline
Process & $5\ {\rm ab}^{-1}$ & $50\ {\rm ab}^{-1}$  \\
      \hline
      $\BR(\tau \to \mu\,\gamma) \rule{0pt}{2.6ex}$ &  $10 \times 10^{-9}$ &  $3 \times 10^{-9}$  \\
      $\BR(\tau \to \mu\, \mu\, \mu)$ &  $3 \times 10^{-9}$ & $1 \times 10^{-9}$  \\
      $\BR(\tau \to \mu \eta)$             &  $5 \times 10^{-9}$ & $2\times 10^{-9}$    \\
 \hline\hline
    \end{tabular}
  \end{center}
\end{table}

These CLFV sensitivities directly confront a large variety of new
physics models. Of particular interest is the correlation between
$\tau$ CLFV branching ratios such as $\tau\to \mu\gamma$ and $\tau\to e
\gamma$, as well as the correlation with $\mu\to e \gamma$ and the
$\mu\to e$ conversion rate, all of which are diagnostic of particular
models.  A polarized electron beam potentially allows the possibility of determining the helicity structure of CLFV couplings from Dalitz plot analyses of, for example, $\tau \to 3\ell$ decays.

The experimental situation at a $\tau/c$ factory is somewhat different. The luminosity of the proposed projects is $10^{35}$cm$^{-2}$s$^{-1}$, a factor of eight below the eventual SuperKEKB luminosity. The $\tau$ production cross section is, however, larger: $\sigma_{\tau\bar{\tau}}(3.77\  \gev)/\sigma_{\tau\bar{\tau}}(10.58\  \gev)=3$, and both have a polarized electron beam. In addition, while a Super $B$ factory is likely to spend the bulk of its running time at the $\Upsilon(4S)$, a $\tau/c$ factory will take data more evenly throught the accessible energy range.  A study for the BINP machine~\cite{bobrov}, with 1.5 ab$^{-1}$ at 3.686 GeV,
3.5 ab$^{-1}$ at 3.770 GeV, and 2.0 ab$^{-1}$ at 4.170 GeV, 
corresponding to  $2.5\times 10^{10}$ produced $\tau$ pairs, quotes a 90\% confidence level limit on $\cal{B}(\tau\rightarrow\mu\gamma)$= $3.3\times 10^{-10}$, provided the detector has $\mu/\pi$ rejection of a factor of 30. This is nearly an order of magnitude improvement over the SuperKEKB expectation at 50 ab$^{-1}$.


The LHC is a prolific source of $\tau$ leptons with an expected
production cross section of about 0.1mb, the majority coming from 
decays of $D_s$ mesons and $B$ hadrons.  
The LHCb experiment~\cite{Alves:2008zz} can profit from this large $\tau$ lepton
production rate thanks to its forward geometry and the flexible
trigger system. The LHCb collaboration has published a search for the
decay $\tau \to \mu\mu\mu$ with 1/fb of data at 7 TeV which  obtained a 90\% CL
limit of $8 \times 10^{-8}$~\cite{Aaij:2013fia}. The total dataset recorded in run 1 of the
LHC is 3/fb at an energy of 7 and 8 TeV.

The expected future sensitivity in this channel can be estimated
conservatively as scaling with the square root of the luminosity. By the end
of run 2 of the LHC (2018), an additional 5/fb are expected to be
collected at a beam energy of 13TeV. The sensitivity using this
dataset should be competitive with the current Belle sensitivity. 
The upgraded LHCb detector will start data taking in 2018 and is
expected to collect a dataset of 50/fb at 14TeV.
The sensitivity in the $\tau \to \mu\mu\mu$ search using this dataset is $8\times
10^{-9}$, assuming conservatively a scaling with the root of the
luminosity. In constrast, when assuming improvements on the analysis
strategy, the optimistic assumption of linear scaling can be made. The
sensitivity on $\tau \to \mu\mu\mu$ decays would then be $7 \times 10^{-10}$ with the
upgraded LHCb experiment. 

As the upgraded LHCb experiment will have a very efficient trigger
system also for softer hadrons, LFV $\tau$ decays with one or several
hadrons in the final state can reach sensitivities that are only slightly reduced
with respect to the purely muonic decay.



\chapter{Flavor-Conserving Processes}\label{sec:cl:fcp}
\medskip
\section{Magnetic and Electric Dipole Moment Theory Overview}\label{sec:cl:fct}
\vskip -12pt





The muon provides a unique opportunity to explore the properties of a
second-generation particle  with great precision. Several muon properties make
these measurements possible.  It has a long
lifetime of $\simeq 2.2~\mu$s,  it is produced in the weak decay
$\pi^- \rightarrow \mu^- \bar \nu_\mu$ providing copious numbers of
polarized muons, and the weak decay
$\mu^- \rightarrow e^- \nu_\mu \bar \nu_e $ is self-analyzing providing information on the muon spin direction at the time of decay.

In his famous paper on the relativistic theory of the electron,
Dirac\cite{Dirac28} obtained the correct magnetic moment for the
electron, and he also mentioned the possibility of an electric
 dipole moment, which like the magnetic dipole moment,
would be directed along the electron spin direction.
  The magnetic dipole (MDM) and electric dipole (EDM)
moments are given by
\begin{equation}
\vec \mu = g \left( \frac{Qe}{ 2m}\right) \vec s\, , \qquad
 \vec d = \eta  \left(\frac {Qe  }{ 2
     mc}\right)
\vec s \, ,
\label{eq:MDM-EDMdef}
\end{equation}
where  $Q =  \pm 1$ and $e>0$. Dirac theory predicts $g \equiv 2$,
but radiative corrections dominated by the
lowest-order (mass-independent) Schwinger contribution $a_{e,\mu,\tau} =
 \alpha/(2\pi)$~\cite{Schwinger48} make it necessary to
write the magnetic moment as
\begin{equation}
\mu = \left(1 + a\right)\frac{Qe \hbar }{ 2m}\quad {\rm with} \quad
a = \frac{{g - 2} }{ 2}.
\label{eq:muon-MDM}
\end{equation}

The muon played an important role in our discovery of the generation
structure of the Standard Model (SM) when
 experiments at the Nevis
cyclotron
showed  that $g_\mu$ was consistent with 2~\cite{Garwin57}.
Subsequent experiments at Nevis and CERN showed  that
$a_\mu \simeq \alpha/(2\pi)$~\cite{Garwin60,Charpak61},
implying that in a magnetic field, the muon
behaves like a heavy electron.
The SM value of the muon anomaly is now known
to better than half a part per million (ppm), and has
been measured to a similar precision~\cite{Bennett06}.

The quantity $\eta$ in Eq.~\ref{eq:MDM-EDMdef}
 is analogous to the $g$-value for the magnetic dipole
moment. An EDM violates both {\sl P} and {\sl T}
symmetries~\cite{Purcell50,Landau57,Ramsey58}, and since $C$ is conserved,  ${ C\!P}$ is violated as well.  Thus
searches for EDMs provide an important tool in our quest to
find non-Standard Model ${ C\!P}$ violation.

The measured value of the muon anomalous magnetic moment is in apparent
disagreement with the expected value based on the
SM.  The BNL E821 experiment finds~\cite{hep-ex/0602035}
\begin{equation} \hspace*{-23pt}
    a_\mu(\textrm{Expt}) = 116\,592\,089(54)(33)\times10^{-11},
    \label{eq:e821}
\end{equation}
where $a_\mu=(g-2)/2$ is the muon anomaly, and the uncertainties are
statistical and systematic, respectively.  This can be compared with
the SM prediction~\cite{arXiv:1010.4180,931465}
\begin{equation}
    a_\mu(\textrm{SM})   = 116\,591\,802(42)(26)(02)\times10^{-11},
    \label{eq:SM}
\end{equation}
where the uncertainties are from the $\mathrm{O}(\alpha^2)$ hadronic vacuum
polarization (HVP) contribution, $\mathrm{O}(\alpha^3)$ hadronic
contributions (including hadronic light-by-light (HLbL) scattering),
and all others (pure QED, including a 5-loop
estimate~\cite{arXiv:1110.2826}, and electroweak, including
2-loops~\cite{hep-ph/0212229}). The hadronic contributions dominate
the uncertainty in $a_\mu(\rm SM)$.  The discrepancy between the
measurement and the SM stands at
\begin{equation}
\Delta a_\mu=287(80)\times 10^{-11}
\end{equation}
(3.6 standard deviations ($\sigma$)), when based on the $e^+e^-\to\rm
hadrons$ analysis for the HVP
contribution~\cite{arXiv:1010.4180}. When the HVP analysis is
complemented by $\tau\to\rm hadrons$, the discrepancy is reduced to
2.4$\sigma$~\cite{arXiv:1010.4180}. However, a recent re-analysis,
employing effective field theory techniques, of the $\tau$
data~\cite{arXiv:1101.2872} shows virtual agreement with the
$e^+e^-$-based analysis, which would solidify the current discrepancy
at the 3.6$\sigma$ level. $\Delta a_\mu$ is large, roughly two times
the EW contribution~\cite{hep-ph/0212229}, indicating potentially
large new physics contributions.


The anomalous magnetic moment of the muon is sensitive to
contributions from a wide range of physics beyond the standard
model. It  will continue to place stringent restrictions on all of
the models, both present and yet to be written down. If  
physics beyond the standard model is discovered at the LHC 
or other experiments,
$a_\mu$  will constitute an indispensable tool to discriminate
between very different types of new physics, especially since it is
highly sensitive to parameters which are difficult to measure at the
LHC. If no new phenomena are found elsewhere, then it represents one of the few ways
to probe physics beyond the standard model. In either case, it will play an
essential and complementary role in the quest to understand physics
beyond the standard model at the TeV scale. 

The muon magnetic moment has a special role because it is
sensitive to a large class of models related and unrelated to electroweak symmetry breaking and
because it combines several properties in a unique way: it is a
flavor- and $C\!P$-conserving, chirality-flipping and loop-induced 
quantity. In contrast, many high-energy collider observables at the
LHC and a future linear collider are chirality-conserving, and many
other low-energy precision observables are $C\!P$- or
flavor-violating. These unique properties might be the reason why the
muon $(g-2)$ is the only one of the mentioned observables that shows a 
significant deviation between the experimental value and the SM
prediction.  Furthermore, while $g-2$ is sensitive
to leptonic couplings, 
$b$ or $K$ physics more naturally probe the hadronic couplings of new
physics. If charged lepton-flavor violation exists, observables such
as $\mu\to e$ conversion can only determine a combination of the
strength of lepton-flavor violation and the mass scale of new
physics. In that case, $g-2$ can help to disentangle the nature of the
new physics.

Unravelling the existence and the properties of such new physics
requires experimental information complementary to the LHC.
The muon $(g-2)$, together
with searches for charged lepton flavor violation, electric dipole
moments, and rare decays, belongs to a class of complementary
low-energy experiments.

In fact, 
The role of $g-2$ as a discriminator between very different standard
model extensions is well illustrated by a relation stressed by
Czarnecki and Marciano~\cite{czmar}. It holds in a wide range of
models as a result of the chirality-flipping nature of both  $g-2$ and
the muon mass: If a new
physics model with a mass scale $\Lambda$
contributes to the muon mass $\delta m_\mu(\mbox{N.P.})$, it also
contributes to $a_\mu$, and the two contributions are related as
\begin{equation}
\label{CzMbound} a_\mu(\mbox{N.P.})={\cal O}(1)\times
\left(\frac{m_\mu}{\Lambda}\right)^2 \times \left(\frac{\delta
m_\mu(\mbox{N.P.})}{m_\mu}\right). 
\end{equation}

The ratio $C(\mbox{N.P.})\equiv\delta m_\mu(\mbox{N.P.})/{m_\mu}$
cannot be larger than unity unless there is fine-tuning in the muon
mass. Hence a first consequence of this relation is that new physics
can explain the currently observed deviation 
only if
$\Lambda$ is at the few-TeV scale or smaller.

In many models, the ratio $C$ arises from one- or even two-loop
diagrams, and is then suppressed by factors like  $\alpha/4\pi$ or
$(\alpha/4\pi)^2$. Hence, even for a given $\Lambda$, the
contributions to $a_\mu$ are highly model dependent.

It is instructive to classify new physics models as follows:
\begin{itemize}
\item Models with $C(\mbox{N.P.})\simeq1$: Such models are of interest
  since the muon
  mass is essentially generated by radiative effects  at some
scale $\Lambda$.
A variety of such models  have been discussed in~\cite{czmar}, including
extended technicolor or generic models with naturally vanishing bare
muon mass. For examples of radiative muon mass generation within
supersymmetry, see e.g.\ 
\cite{Borzumati:1999sp,Crivellin:2010ty}.  In these models the
new physics contribution to $a_\mu$ can be very large, 
\begin{equation} 
a_{\mu}
(\Lambda) \simeq {m^2_{\mu} \over \Lambda^2}\simeq
1100\times10^{-11}\left(\frac{1\mbox{ TeV}}{\Lambda}\right)^2. 
\end{equation}
and the difference between experiment and prediction can  be used to place a lower
limit on the new physics mass scale, which is in the few TeV
range~\cite{elp,Crivellin:2010ty}.
\item Models with $C(\mbox{N.P.})={\cal O}(\alpha/4\pi)$:
Such a loop suppression happens in many models with new weakly
interacting particles like $Z'$ or $W'$, little Higgs or certain extra
dimension models.  As examples, the contributions to $a_\mu$ in a
model with $\delta=1$ (or
2) universal extra dimensions (UED)~\cite{AppelqDob} and the Littlest Higgs
model with T-parity (LHT)~\cite{Blanke:2007db} are given by
 with $|S_{\rm KK}| _{\sim}^{<}1$~\cite{AppelqDob}.
A difference as large as
the current deviation between experiment and prediction is very hard to accommodate unless the mass scale
is very small, of the order of $M_Z$, which however is often excluded
e.g.\ by LEP measurements.
So typically these models predict very small contributions to $a_\mu$
and will be disfavored if the current deviation will be confirmed by
the new $a_\mu$ measurement.

Exceptions are provided by models where new particles
interact with muons but are otherwise hidden from searches. An example
is the model with a new gauge boson associated to a gauged lepton
number $L_\mu-L_\tau$ \cite{LmuLtau}, where a gauge boson mass of
${\cal O}(100\mbox{ GeV})$ and large $a_\mu$ are viable.
\item Models with intermediate values for $C(\mbox{N.P.})$ and mass
  scales around the weak scale: In such
  models, contributions to $a_\mu$ could be as large as
  the current deviation or even larger, or smaller, depending on the
  details of the model. This implies that a more precise
  $a_\mu$-measurement will have significant impact on such models and
  can even be used to measure model parameters. Supersymmetric (SUSY) models
  are the best known examples, so muon $g-2$ would have substantial
  sensitivity to
 SUSY particles.
Compared to generic perturbative models,
supersymmetry provides an enhancement to $C(\mbox{SUSY})={\cal
  O}(\tan\beta\times\alpha/4\pi)$
and to $ a_\mu(\mbox{SUSY})$ by a factor $\tan\beta$ (the ratio of
the vacuum expectation values of the two Higgs fields). Typical SUSY
diagrams for the magnetic dipole moment, the electric dipole moment,
and the lepton-number violating conversion process $\mu \rightarrow
e$ in the field of a nucleus contain the SUSY partners of
the muon, electron and the SM U(1)$_Y$ gauge boson, $\tilde{\mu}$,
$\tilde{e}$, $\tilde{B}$. The full SUSY contributions involve also the
SUSY partners to the neutrinos and all SM gauge and Higgs bosons. In a
model with SUSY masses equal to $\Lambda$ 
the SUSY contribution to $a_{\mu}$ is given
by~\cite{czmar} \begin{equation} \label{amususy}
 a_{\mu}({\rm SUSY})\, \simeq \,{\rm sgn}\, (\mu) \ 130 \times 10^{-11}\ \tan \beta\
\left({100\ {\rm GeV}  \over \Lambda}\right)^2
\end{equation}
which indicates the dependence on $\tan \beta$,
and the SUSY mass scale,  as well as the sign of the
SUSY $\mu$-parameter. The formula still approximately applies even if
only the smuon and chargino masses are of the order $\Lambda$
but e.g.\ squarks and gluinos are much heavier. However the SUSY
contributions to $a_\mu$ depend strongly on the details of mass
splittings between the weakly interacting SUSY particles.
Thus muon $g-2$ is sensitive to  SUSY models with SUSY masses
in the few hundred GeV range, and it will help to measure SUSY
parameters. 

There are also non-supersymmetric models with similar
enhancements. For instance, lepton flavor mixing can help. An example
is provided in Ref.\ \cite{BarShalom:2011bb} by a model with two Higgs
doublets and four generations, which can accommodate large
$\Delta a_\mu$ without violating constraints on lepton flavor
violation. In variants of Randall-Sundrum models
\cite{Davoudiasl:2000my,Park:2001uc,Kim:2001rc} and large
extra dimension models \cite{Graesser:1999yg}, large
contributions to $a_\mu$ might be possible from exchange of
Kaluza-Klein gravitons, but the theoretical evaluation
is difficult because of cutoff dependences. A recent evaluation of the
non-graviton contributions in Randall-Sundrum models, however,
obtained a very small result \cite{Beneke:2012ie}.

Further examples
include scenarios of unparticle physics
\cite{Cheung:2007zza,Conley:2008jg} (here a more precise
$a_\mu$-measurement would constrain the unparticle scale dimension and
effective couplings), generic models with a hidden sector at the weak
scale \cite{McKeen:2009ny} or a model with the discrete flavor 
symmetry group $T'$ and Higgs triplets \cite{Ho:2010yp} (here
a more precise $a_\mu$-measurement would constrain hidden sector/Higgs
triplet masses and couplings), or the model proposed in
Ref.\ \cite{Hambye:2006zn}, which implements the idea that neutrino
masses, leptogenesis and the deviation in $a_\mu$ all originate from
dark matter particles. In the latter model, new leptons and scalar
particles are predicted, and $a_\mu$ provides significant constraints
on the masses and Yukawa couplings of the new particles.
\end{itemize}

The following types of new physics scenarios are quite different from
the ones above:
\begin{itemize}
\item Models with extended Higgs sector but without the
  $\tan\beta$-enhancement of SUSY models. Among these models are the
  usual two-Higgs-doublet models. The one-loop contribution
  of the extra Higgs states to $a_\mu$ is suppressed by two additional powers of
  the muon Yukawa coupling, corresponding to $a_\mu(\mbox{N.P.})\propto
  m_\mu^4/\Lambda^4$ at the one-loop level. Two-loop effects from
  Barr-Zee diagrams can be larger \cite{Krawczyk:2002df}, but typically the
  contributions to   $a_\mu$ are negligible in these models.
\item Models with additional light particles with masses below the
  GeV-scale, generically called dark sector models: Examples are
  provided by the  models of   Refs.\
  \cite{Pospelov:2008zw,Davoudiasl:2012qa}, where additional light
  neutral gauge bosons can affect electromagnetic interactions. Such
  models are intriguing since 
  they completely decouple $g-2$ from the physics of EWSB, and since
  they are hidden from collider searches at LEP or LHC (see however
  Refs.\ \cite{Essig:2009nc,Davoudiasl:2012ig} for studies of possible
  effects at dedicated low-energy colliders and in Higgs decays at the
  LHC). They can lead to
  contributions to $a_\mu$ which are of the same order as the deviation
  between experiment and prediction. Hence the new $g-2$ measurement will provide
  an important test of such models.
\end{itemize}
To summarize:
many well-motivated models can accommodate larger contributions to
$a_\mu$ --- if any of these are realized $g-2$ can be used to constrain
model parameters; many well-motivated new physics models
give tiny contributions to $a_\mu$ and would be disfavored if the
more precise $g-2$ measurement confirms the current deviation. There are also examples of models which lead to
similar LHC signatures but which can be distinguished using $g-2$.

\section{Muon $g-2$:  Experiment}
\vskip -12pt





Measurements of the magnetic and electric dipole moments make use of the
torque on a dipole in an external field, $\vec \tau = \vec\mu \times
\vec B + \vec d \times \vec E$. All muon MDM experiments except the original
Nevis ones used polarized muons in flight, and
 measured the rate at which the spin turns relative to the momentum,
$\vec \omega_a =\vec \omega_S - \vec \omega_C$, when a
 beam of polarized muons is injected into a magnetic field.
The resulting frequency, assuming that $\vec \beta \cdot \vec B = 0$,
 is given
by~\cite{Thomas26,Bargmann59}
\begin{equation}
\vec{\omega}_{a\eta}= \vec \omega_a + \vec \omega_\eta = -
 \frac {Qe}  {m}
\left[
a \vec{B}
- \left( a - \left(  \frac {m} {p} \right)^2  \right)
 \frac {\vec{\beta} \times \vec{E}} {c} \right] -  \eta \frac {Qe}{2m}
 \left[ \frac {\vec{E}} {c}  +  \vec{\beta} \times \vec{B} \right] .
 \label{eq:omegaa-edm1}
\end{equation}
Important features of this equation are the motional magnetic and
electric fields:
$\vec \beta \times \vec E$ and $\vec \beta \times \vec B$.

The E821 Collaboration working at the
Brookhaven AGS used an electric quadrupole field
to provide vertical focusing in the storage ring, and shimmed the magnetic
field to 1 ppm uniformity on average.  The storage ring was operated
 at the `` $g-2$'' momentum, $p_{ g-2} = 3.094$~GeV/c,
($\gamma_{ g-2}= 29.3$),
so that $a_\mu = (m/p)^2$ and the electric field did not
contribute to $\omega_a$.
They obtained\cite{Bennett06}
\begin{equation}
  a_\mu^{(\mathrm{E821})} = 116\,592\,089(63) \times
  10^{-11}~~\mbox{(0.54\,ppm)}
\end{equation}
 The final uncertainty of
0.54~ppm consists of a 0.46~ppm statistical component and a 0.28~ppm
systematic
component.  

The present limit on the muon EDM also comes from E821~\cite{Bennett08-edm}
\begin{equation}
d_\mu = (0.1 \pm 0.9) \times 10^{-19} e  \cdot {\rm cm}; \
 \vert d_\mu \vert < 1.9 \times 10^{-19}  e  \cdot {\rm cm}\ (95\%\ {\rm C.L.})\, ,
\label{muedm-result}
\end{equation}
so the EDM contribution to the precession is very small.  In the muon $g-2$
experiments, the motional electric field dominates the $\omega_\eta$ term,
which means that $\vec \omega_a$ and $\vec \omega_\eta$ are orthogonal.
The presence of an EDM in the  $g-2$ momentum experiments has two effects:
the measured frequency is the quadrature sum of the two frequencies,
 $\omega = \sqrt{\omega_a^2 + \omega_\eta^2}$, and the EDM causes a tipping of
 the plane of precession, by an angle
 $\delta = \tan ^{-1}[ \eta \beta/(2a_\mu)]$. This tipping results in
 in an up-down oscillation of the decay
 positrons relative to the midplane of the storage ring with frequency
$\omega_a$ {\it out of phase by} $\pi/2$ with the $a_\mu$ precession.

The E989 collaboration
at Fermilab will move the E821 muon storage ring to Fermilab, and
will use the  $g-2$ momentum technique to measure $a_{\mu^+}$.
 New detectors and electronics, and a
beam handling scheme that increases the stored muon rate per hour
 by a factor of 6
over E821 will be implemented.  The goal is at least 21 times the
statistics of E821, and a factor of four overall uncertainty reduction, with
equal systematic and statistical uncertainties of $\pm 0.1$ ppm.

The scope of Project X includes 50-200kW of beam power at 8 GeV,
about three to fifteen times the beam power of E989.  This large step in beam
power could be used to measure $g-2$ for negative muons,
and provide muon beams with lower emittance thereby reducing
experimental systematics. 

Given the high impact of the E821 result and the 
crucial role the value of $g-2$ plays in interpreting energy frontier results, 
it is imperative to have a second measurement with at least equal 
precision but with a complementary approach to the measurement. 
An alternate approach planned for J-PARC~\cite{JPARC-Lg2} uses a much lower muon
energy, and does not use the  $g-2$ momentum technique. A surface muon beam
produced by  the low energy Booster is brought to rest in an aerogel
target, where muonium (the $\mu^+ e^-$ atom) is formed.  The muonium
is ionized by a powerful laser which produces a very slow muon beam with
extremely small emittance. This low emittance beam is then accelerated by a
linac to 300 MeV, and injected into a  $\sim 1$~m diameter
solenoidal magnet with point to point
uniformity of $\pm 1$ ppm, approximately 100 times better than at the Brookhaven experiment.  
The average uniformity is expected to be known to better then 0.1 ppm.  The decays are detected 
by a full volume tracker consisting of an array of silicon
detectors.  This provides time, energy, and decay angle information for every positron, maximizing
 the sensitivity to separate the  $g-2$ and EDM precession frequencies.  The expected   $g-2$ sensitivity 
 is comparable to the
 Fermilab experiment but will have very different systematic uncertainties and the combined results 
 from the two experiments should bring the precision to below the 100 ppb level.

\section{Muon $g-2$:  Expected Improvements in the Predicted Value}
\vskip -12pt



\begin{figure}[bp]
    \centering
    \includegraphics[width=0.3\columnwidth]{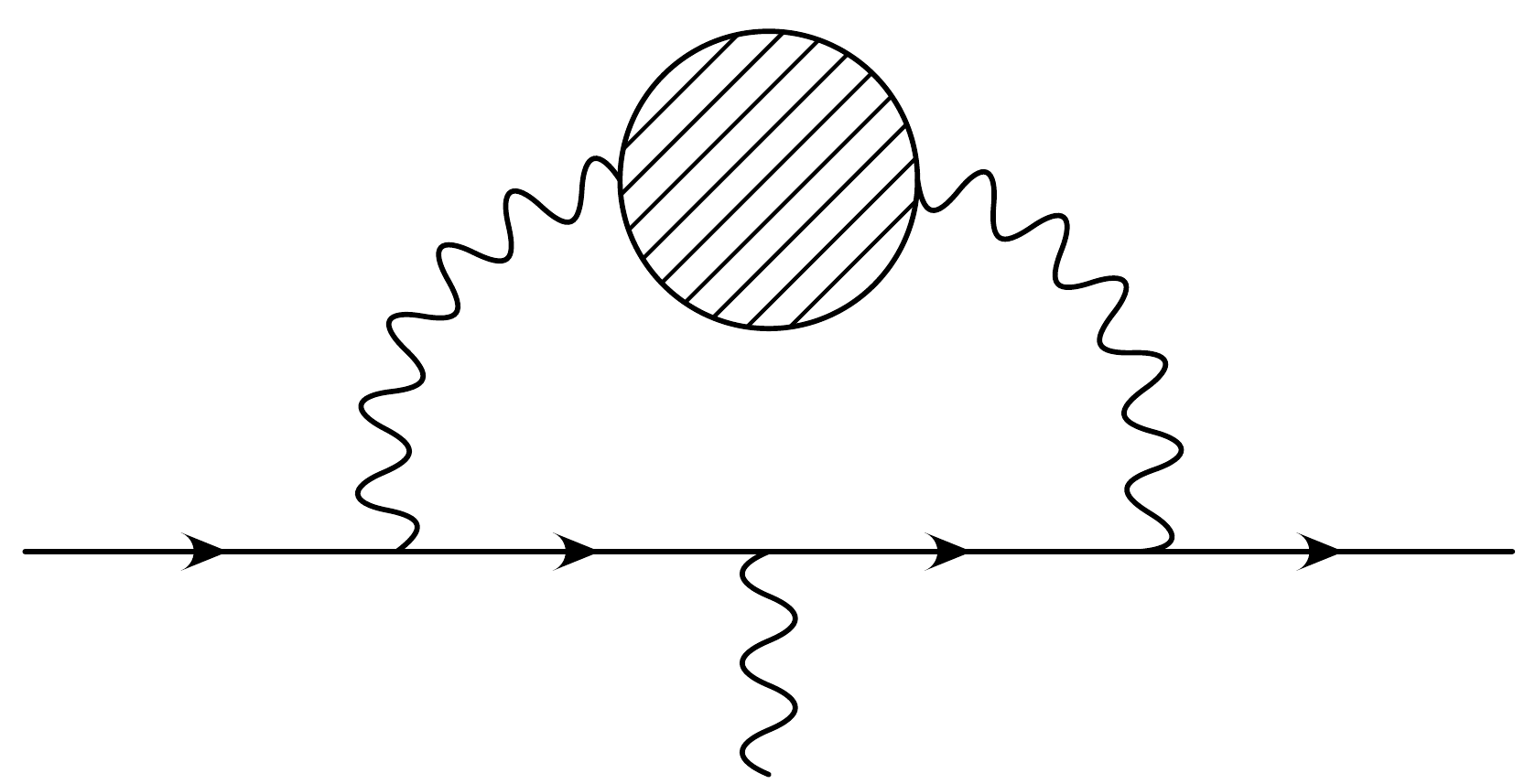}\hskip 1cm
    \includegraphics[width=0.3\columnwidth]{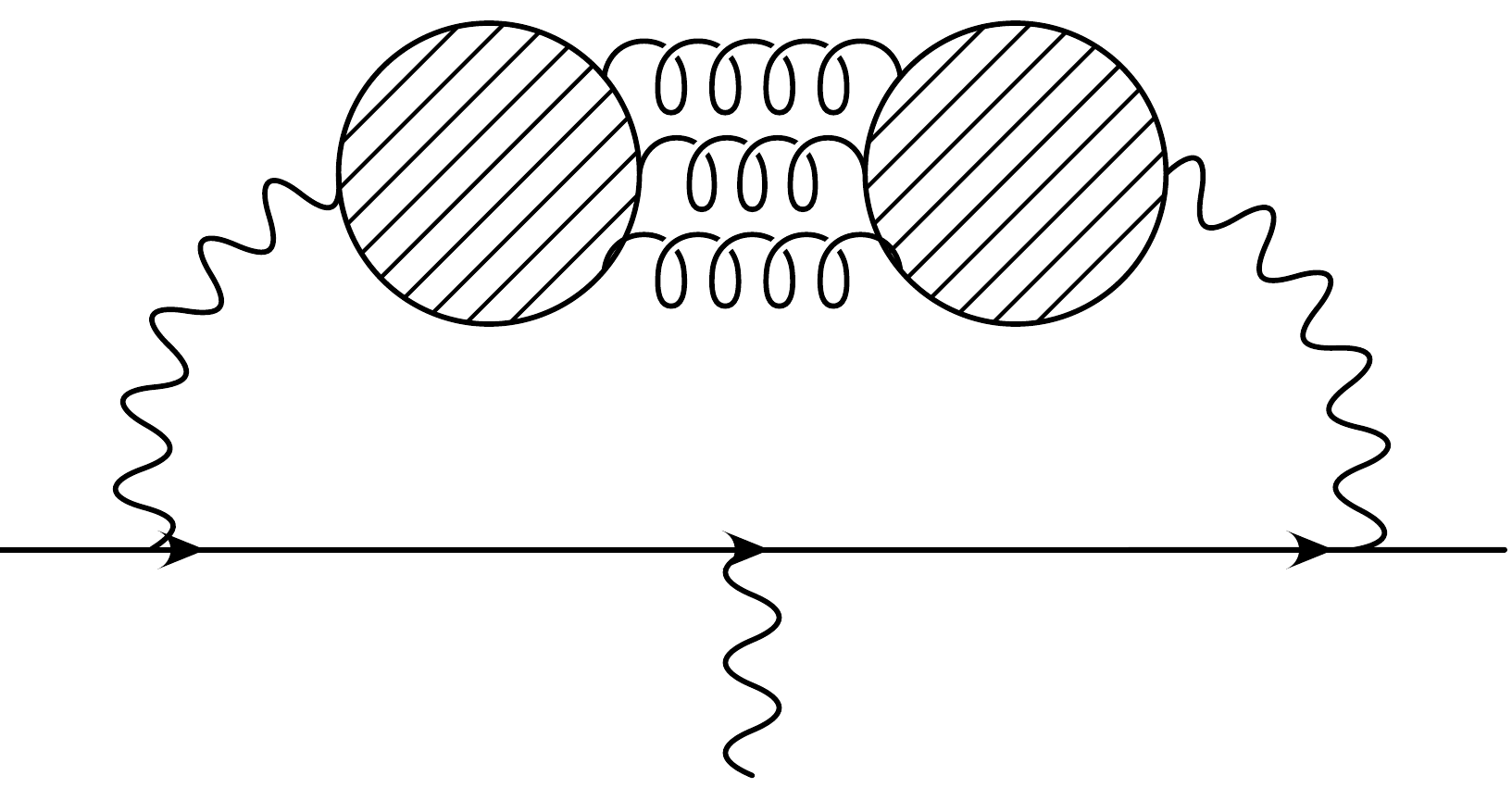}
\caption{Hadronic vacuum polarization diagrams contributing to the SM muon anomaly. The horizontal lines represent the muon. (Left panel) The blob formed by the quark-antiquark loop represents all possible hadronic intermediate states.  (Right panel) Disconnected quark line contribution. The quark loops are connected by gluons.}

    \label{fig:hvp}
\end{figure}

The QED and electroweak contributions to $g-2$ can be calculated from first principles
and are regarded as robust.  The two dominant QCD contributions are hadronic vacuum polarization (HVP) and hadronic light-by-light (HLBL).
The HVP contribution to $a_\mu$ can be determined from the
cross-section for $e^+e^-\to\rm hadrons$ (and over a certain energy
range, by $\tau\to\rm hadrons$) and a dispersion relation. It can also
be computed from purely first principles using lattice QCD to
calculate the HVP directly~\cite{hep-lat/0212018}. The two methods are
complementary and can be used to check each other. The current best
uncertainty comes from the first method,
\begin{equation}
a_\mu(\rm HVP)=(692.3\pm4.2)\times 10^{-10},
\end{equation}
or about 0.61\%~\cite{arXiv:1010.4180} when only $e^+e^-$ data are used. 
If $\tau$ data are included, $a_\mu(\rm HPV)=(701.5\pm4.7)\times 10^{-10}$, or 0.67\% 
(but see~\cite{arXiv:1101.2872} for the analysis that brings the $\tau$ 
into good agreement with $e^+e^-$). In the next 3-5 years the uncertainty on 
$a_\mu(\rm HVP)$ is expected to drop by roughly a factor of 2, relying 
on new results from {\babar}, Belle, BES, and VEPP2000.
The lattice calculations presently have an uncertainty of 
about 5\%~\cite{hep-lat/0608011, arXiv:1103.4818, Boyle:2011hu,  DellaMorte:2011aa}, which is 
expected to decrease to 1-2\% in the next 3-5 years~\cite{USQCD}. At the 
one-percent level contributions from dynamical charm quarks and quark-disconnected 
diagrams (right panel, Fig.~\ref{fig:hvp}) enter. Both are currently under investigation.

\begin{figure}[bp]
    \centering
\vspace*{-2pt}
    \includegraphics[width=0.3\columnwidth]{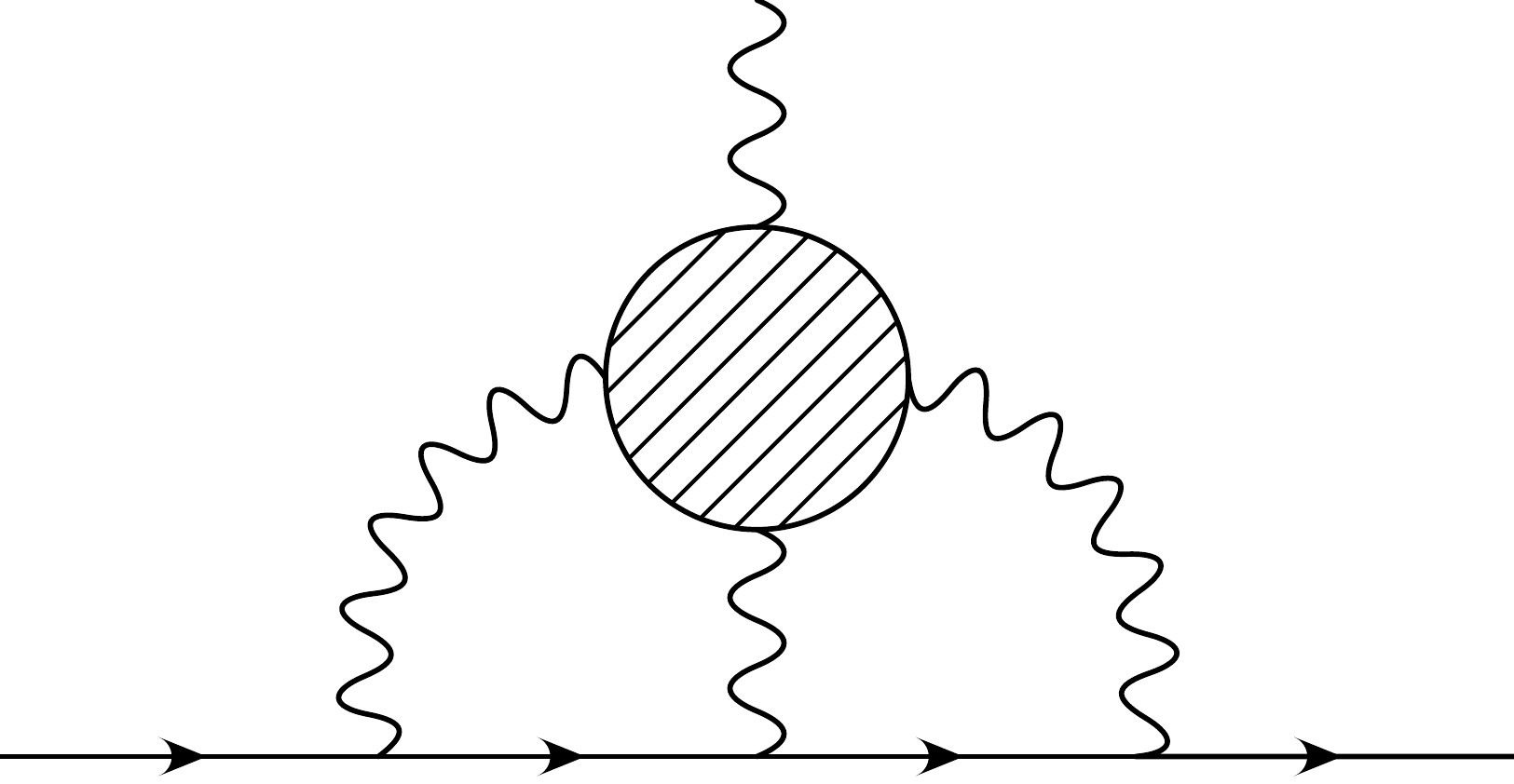}\hskip 1cm
    \includegraphics[width=0.3\columnwidth]{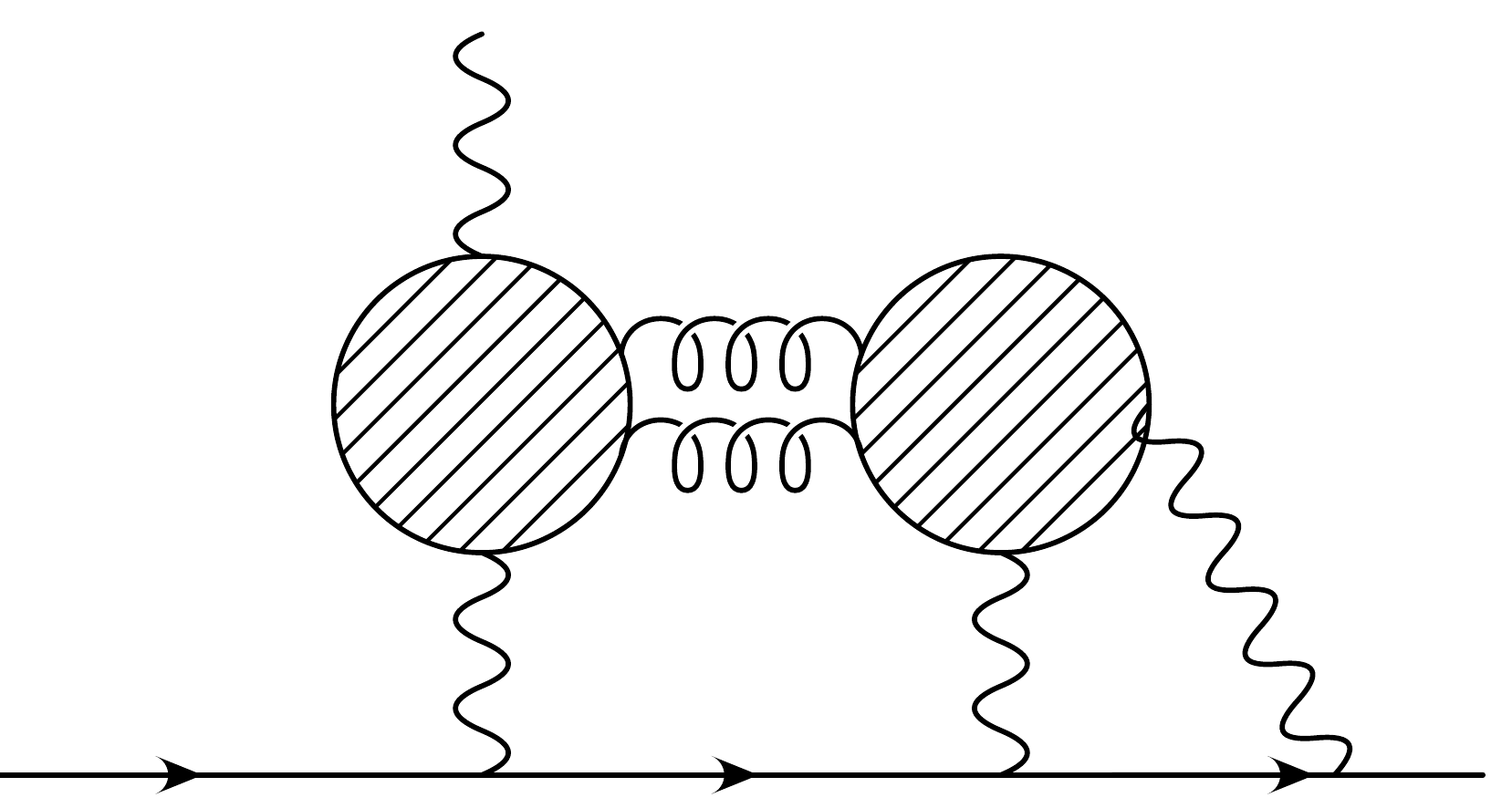}
    \caption{Hadronic light-by-light scattering diagrams contributing to the SM muon anomaly. The horizontal lines represent the muon. (Left panel) The blob formed by the quark loop represents all possible hadronic intermediate states.  (Right panel) One of the disconnected quark line contributions. The quark loops are connected by gluons.}
    \label{fig:hlbl}
\end{figure}

The HLBL scattering amplitude, shown in
Fig.~\ref{fig:hlbl}, is much more challenging.  The contribution
\begin{equation}
    a_\mu(\textrm{HLBL}) = (10.5\pm 2.6) \times 10^{-10},
    \label{eq:PRV}
\end{equation}
is not well known. It is based on the size of various hadronic contributions estimated in
several different models~\cite{arXiv:0901.0306}.
Its uncertainty, though estimated to be less than that in a(HVP) by about 50\%, is less reliable and will be difficult to reduce with current methods.
Finding a new approach, such as lattice QCD, in which uncertainties are systematically improvable,
is crucial for making greatest use of the next round of experiments.
With this in mind, a workshop was recently convened at the Institute for Nuclear Theory~\cite{INTws}.
Workshop participants discussed how models, lattice QCD, and data-driven methods could be exploited to reduce the
uncertainty on $a_\mu(\textrm{HLBL})$.
The outcome of this workshop is that 
a SM calculation of the HLBL contribution with a total uncertainty of around 10\% may be possible on the time scale of the planned experiment.
A detailed discussion of the computation of $a_\mu(\rm HLBL)$ in lattice QCD is given in the USQCD Collaboration white paper on $g-2$~\cite{USQCD}.
%


The currently most promising approach using lattice methods is to compute the entire amplitude on the lattice, including the muon, in a  combined QED+QCD gauge field~\cite{hep-lat/9602005,hep-lat/0509016,827504}. The method has passed several non-trivial tests. First, it has been successfully checked against perturbation theory in pure QED. Large finite volume effects (the photons are long range) appear manageable. Preliminary calculations in full QED+QCD, at unphysical quark and muon mass and momentum transfer $q^2$, show a statistically significant result. The method requires a non-perturbative subtraction of leading order in $\alpha$ contributions which has been checked by varying the strength of the electric charge in the calculations and observing the expected scaling, before and after the subtraction. Disconnected contributions like the one shown in the right panel of Fig.~\ref{fig:hlbl} have not been included yet, but will be once the simpler first diagram (left panel, same figure) is fully under control. Calculations on a larger volume with smaller masses are in progress.

In addition to these direct approaches, there is other ongoing work on lattice-QCD calculations that check or
supplement the model calculations.
For example, it is well-known that the pion pole (namely, $\gamma\gamma^*\to\pi^0\to\gamma^*\gamma^*$)
provides the largest contribution to the QCD blob in Fig.~\ref{fig:hlbl}.
Just as experiments are being mounted to examine this physics (\emph{e.g.}, PrimEx at JLab and KLOE at LNF),
several groups~\cite{arXiv:0810.5550,arXiv:0912.0253,XFeng} are using lattice QCD to compute the amplitudes for
$\pi^0\to\gamma\gamma^*$ and $\pi^0\to\gamma^*\gamma^*$ (with one or two virtual photons).

If the SM and experiment central values do not change while both experiment and theory uncertainties are reduced, the discrepancy between the two would grow to well over 5$\sigma$ in significance. The improvement expected from E989 (0.14 PPM) by itself improves $\Delta a_\mu$ to 5$\sigma$. A simultaneous decrease in the HLBL uncertainty to 10\% from the current 25\% pushes it to 6$\sigma$, and finally, reducing the uncertainty on the HVP contribution by a factor of two increases it to 9$\sigma$. Such a large and clear difference between experiment and the Standard Model for the muon $g-2$ will be extremely  discriminating between new physics scenarios responsible for this discrepancy.

\section{$\tau$ $g-2$ and EDM}
\vskip -12pt

The experimental discrepancy with the Standard Model prediction for
the muon anomalous magnetic moment heightens interest in the
possibility of measuring $g-2$ of the $\tau$ lepton using angular
distributions in $\tau$-pair production. This can be done at a super flavor factory, with or without electron polarization.  

The best current bound on the $\tau$ anomalous moment  $a_{\tau}=(g-2)/2$ is indirect, derived from the LEP2 measurement of the total cross section for $e^+e^- \rightarrow e^+e^- \tau^+\tau^-$: $-0.052 < a_\tau < 0.013$ @ 95\% CL. This is well above the Standard Model prediction: $a_\tau^{SM} = 1177.21(5) \times 10^{-6}$, but even so provides a model-independent bound on New Physics contributions:  $-0.007 < a_\tau^{NP} < 0.005$ @ 95\% CL.~\cite{ref:bern2}

This measurement can be done in $e^+ e^- \rightarrow \tau^+\tau^-$ production with unpolarized beams. Determination of the real part of the form factor requires the measurement of correlations between the decay products of both polarized taus. 

With polarized taus one can access new observables, $A_T^\pm$ and $A_L^\pm$, the transverse and longitudinal polarizations of the outgoing $\tau$s,
\[\begin{array}{l}
A_T^ \pm \, = \,\frac{{\sigma _R^ \pm |{P_e}\,\, - \,\sigma _L^ \pm |{P_e}}}{\sigma }\, = \, \mp \,{\alpha _ \pm }\frac{{3\pi }}{{8(3 - {\beta ^2})\gamma }}\left[ {{{\left| {{F_1}} \right|}^2}\, + \,(2 - {\beta ^2}){\gamma ^2}{\mathop{\rm Re}\nolimits} \left\{ {{F_2}} \right\}} \right]\\
\,\,A_L^ \pm \, = \,\frac{{\sigma _{{\rm{FB}}}^ \pm ( + )|{P_e}\, - \,\sigma _{{\rm{FB}}}^ \pm ( - )|{P_e}}}{\sigma }\, = \, \mp {\alpha _ \pm }\frac{3}{{4(3 - {\beta ^2})}}\left[ {{{\left| {{F_1}} \right|}^2}\, + \,2{\mathop{\rm Re}\nolimits} \{ {F_2}\} } \right]\\
{\rm{                  Re}}\left\{ {{F_2}(s)} \right\}\, = \, \mp \frac{{8(3 - {\beta ^2})}}{{3\pi \gamma {\beta ^2}}}\frac{1}{{{\alpha _ \pm }}}\left( {A_T^ \pm \, - \,\frac{\pi }{{2\gamma }}A_L^ \pm } \right).
\end{array}\]
that are estimated by
Bernab\'eu {\it et al.}\cite{ref:bern1} to increase the sensitivity to
$\Re(F_2)$ by a factor of three, to $\sim2\times 10^{-6}$ with 80\%
electron polarization, which could allow a measurement of the Standard
Model moment to a precision of several percent with a data sample of
75 ab$^{-1}$. 

Observation of a $\tau$ EDM would be evidence of $T$ violation.  $T$-odd observables can be isolated by the study of $\tau$ angular
distributions using unpolarized beams. The best current limit is from Belle~\cite{bellemom}:
\vskip -12pt
\[ - 0.22\,e{\rm{cm}}\,\, < \,{\mathop{\rm Re}\nolimits} \{ d_\tau ^\gamma \} \, \times \,{10^{16}}\, < \,0.45\,e{\rm{cm}}\,@\,95\% \,\,C\!L.\]
\vskip -8pt
Having a polarized electron
beam allows these investigations to be done using the decay products
of individual polarized taus. We can define a new, more sensitive $C\!P$-odd $T$-odd observable:
\vskip -12pt
\[A_N^{C\!P}\, = \,\frac{1}{2}\left( {A_N^ + \, + \,A_N^ - } \right)\, = \,{\alpha _h}\frac{{3\pi \gamma \beta }}{{8(3 - {\beta ^2})}}\frac{{2{m_\tau }}}{e}{\mathop{\rm Re}\nolimits} \left\{ {d_\tau ^\gamma } \right\}\]
where the azimuthal asymmetry for the two polarizations is 
\[A_N^ \mp \, = \,\frac{{\sigma _L^ \mp \, - \,\sigma _R^ \mp }}{\sigma }\, = {\alpha _ \mp }\frac{{3\pi \gamma \beta }}{{8(3 - {\beta ^2})}}\frac{{2{m_\tau }}}{e}{\mathop{\rm Re}\nolimits} \left\{ {d_\tau ^\gamma } \right\}\]
\vskip -12pt
The upper-limit sensitivity for the real part of the
$\tau$ EDM has been estimated to be to be $|\Re{d_\gamma}|\simeq 3\times 10^{-19} \ e \cdot {\rm cm}$
with 50 ab$^{-1 }$ at Belle II and $|\Re{d_\gamma}|\simeq 7\times 10^{-20} \ e \cdot {\rm cm}$
with 75 ab$^{-1 }$ with a polarized electron beam at Super$B$\cite{GonzalezSprinberg:2000mk}.

A $C\!P$-violating asymmetry in $\tau$ decay would be manifest evidence
for physics beyond the Standard Model. \babar\ has recently published
a 3$\sigma$ asymmetry in $\tau\to\pi K_S^0(\ge 0\pi^0)$
decay\cite{BABAR:2011aa}. The super flavor factories have the sensitivity to
definitely confirm or refute this measurement, and, further, provide
access to new $C\!P$-odd observables that increase the sensitivity in the
search for a $C\!P$ asymmetry to the level of $\sim 10^{-3}$.

\section{Storage Ring EDMs}
\vskip -12pt



As detailed above, the equation for the spin precession frequency of a charged particle in a storage ring is
\begin{equation}
\vec{\omega}_{a\eta}= \vec \omega_a + \vec \omega_\eta = -
 \frac {Qe}  {m}
\left[
a \vec{B}
- \left( a - \left(  \frac {m} {p} \right)^2  \right)
 \frac {\vec{\beta} \times \vec{E}} {c} \right] -  \eta \frac {Qe}{2m}
 \left[ \frac {\vec{E}} {c}  +  \vec{\beta} \times \vec{B} \right] .
 \label{eq:omegaa-edm2}
\end{equation}
The discussion above focused on experiments operating at the magic momentum, $p = \sqrt{a }/m$ that cancels the effect of the focusing $E$ field.  The precession frequency is then by far dominated by the $aB$ term.  
The key to extracting sensitivity to the EDM term $\eta$ is to find ways of reducing or eliminating the motion due to the magnetic term $a$.  

The first method is to use a magnetic storage ring, such as the E821/E989 storage ring, to extract a 
limit on the muon EDM.  In the muon rest frame, the muon sees a strong motional electric field
 pointing towards the center of the ring adding a small horizontal component to the precession 
 frequency vector that tilts the rotation plane.  For a positive EDM, when the spin is pointing into the 
 ring it will have a negative vertical component and when the spin is pointing to the outside of the ring 
 it will have a positive vertical component.  Since the positrons are emitted along the spin direction, 
 this asymmetry maps into the positron decay angle. Since the asymmetry is maximized when the spin 
 and momentum are perpendicular, the angular asymmetry is 90 degrees out of phase with the $g-2$ precession frequency.  
Searches for this asymmetry have been used to set limits on the muon EDM both at the CERN and Brookhaven $g-2$ experiments. 

 A number of the Fermilab Muon $g-2$ detector stations will
be instrumented with straw chambers to measure the decay positron
tracks. With this instrumentation, a simultaneous EDM
measurement can be made during the $a_\mu$ data collection,
improving on the  Brookhaven muon EDM~\cite{Bennett08-edm}
 limit by up to two orders of magnitude down to
$\sim 10^{-21}\,  e \cdot {\rm cm}$.   The primary detector element of the J-PARC muon $g-2$ proposal  is a silicon tracker that will provide
 decay angle information for all tracks and expects similar improvements.

Going beyond this level for the muon will require a dedicated EDM experiment that
uses
the ``frozen spin'' method~\cite{Farley04,Roberts2010}.
 The idea is to operate a
muon storage ring off of the  magic momentum and to use a radial electric field
to cancel the $\omega_a$ term in Eq.~\ref{eq:omegaa-edm1},
 the $g-2$ precession.  The  electric field needed to freeze the spin is
$E \simeq aBc\beta\gamma^2$.
Once the spin is frozen, the EDM will cause a steadily increasing
out-of-plane motion of the spin vector. One stores polarized muons in a ring
with detectors above and below the storage region and forms the asymmetry
(up - down)/(up + down).  To reach a sensitivity of $10^{-24}e \cdot {\rm cm}$ would
require $\sim 4 \times 10^{16}$ recorded events~\cite{Farley04}.
 Preliminary discussions have begun on a frozen spin experiment
using the ~1000 kW beam power available at the Project X rare process campus.
  
It is possible to make a direct measurement of the electron EDM using a storage ring analogous to way $g-2$ is extracted for the muon.  
Here, the key to removing the magnetic precession is to use an electrostatic storage ring so that any spin precession can be attributed to an EDM.  Stray radial magnetic fields 
would lead to a false signal.  The effects of these can be measured and controlled by using counter rotating beams. Several years of 
R$\&$D have already been invested into this technique either for searching for a proton, deuteron, or muon EDM.  Fortuitously, the ratio
 of the $g-2$ value to the mass of the electron is very similar to this ratio for the proton.  Many systematic effects scale with this ratio so 
 that many studies already performed for the proton can be used for the electron.  Conversely, performing a storage ring EDM 
 experiment on the electron would be an excellent test bed for a (more expensive) proton EDM experiment.
 
 Also fortuitous for the electron is that its magic momentum is 15 MeV, requiring a relatively small and inexpensive storage ring.  The technology for the polarized source,
  electrostatic magnets, and beam position monitors all seem to be available.  Concepts for the polarimeter are still being 
  developed and are expected to be the limiting factor in the ultimate sensitivity of the experiment.  A polarimeter with high 
  analyzing power would most likely lead to a sensitivity comparable with model independent extractions of the electron 
  EDM from atoms and molecules.

\section{Precision Measurements of the Muon Decay Spectrum}
\vskip -12pt

Improved measurements of muon decay parameters have potential to probe new physics.  While some terms of the effective Lagrangian of muon decay are tightly bound by the neutrino mass scale if naturalness is assumed for both Dirac~\cite{erwin:2006}
and Majorana~\cite{erwin:2007}
neutrinos, 6 out of 10 effective couplings are not constrained by these considerations, and their best limits come from experiments.

Moreover, $g^{S}_{LR}$ and $g^{S}_{RL}$ are only constrained for Majorana neutrinos.  If a deviation from the Standard Model is found, and if it is possible to establish that it is due to one of those couplings, it would suggest that neutrinos are Dirac particles.  This is a unique probe, as other known experiments positively test for the Majorana nature of neutrinos.

\section{Parity-Violating Experiments}\label{sec:cl:pve}
\vskip -12pt

Ultra-precise measurements of weak neutral current amplitudes in fixed target experiments provide a complementary indirect probe of new TeV-scale dynamics in flavor-conserving processes.
In order to match the sensitivity of future collider
searches and other indirect probes, it is necessary to measure amplitudes with an uncertainty
approaching $10^{-3}\times G_F$. 

A comprehensive discussion of leptonic and semi-leptonic weak neutral current processes, their respective 
sensitivitiies, and complementarity are discussed in the report of the Working Group on Nuclei and Atoms. 

The MOLLER experiment proposed at Jefferson Laboratory can improve on the E158~\cite{ref:cl:Anthony:2005pm}
measurement by more than a factor of 5, taking advantage of the energy upgrade of the high intensity
polarized electron beam to 11 GeV~\cite{ref:cl:Dudek:2012vr}. The goal is a measurement of $A_{PV}$ to a fractional accuracy
of 2.3\%. 
This opportunity can be summarized in three inter-related bullets:
\begin{enumerate}
\item The proposed $A_{PV}$ measurement is sensitive to new neutral current interaction amplitudes as 
small as $1.5\times 10^{-3}\cdot G_F$, which corresponds to a sensitivity of $\Lambda/g =  7.5$~TeV, where $g$ characterizes the strength and $\Lambda$ is the
scale of the new dynamics. This would be {\it the} most sensitive probe of new flavor and $C\!P$-conserving 
neutral current interactions in the leptonic sector until the advent of a new lepton collider or a neutrino factory. 
\item  The two most precise determinations of 
$\sin^2\theta_W$, carried out at the $Z^0$ pole, differ from each other by more than 3 standard deviations. 
The proposed $A_{PV}$ measurement, which aims to achieve
$\delta(\sin^2\theta_W) = \pm 0.00029$, may be able to resolve this discrepancy.
\item The proposed measurement would be carried out at $Q^2\ll M_Z^2$, far from the $Z^0$. A 
convenient way to parametrize a class of new physics effects, to which $Z^0$ resonance observables are 
insensitive, is via the parameter 
$X(Q^2)\equiv\alpha^{-1}(\sin^2\theta_W(Q^2)-\sin^2\theta_W(M_Z^2)$. The projected MOLLER sensitivity is 
$\delta(X)\approx 0.035$. This is by far the most sensitive reach among similar potential
measurements under discussion and probes a region of discovery space of new low
energy flavor-conserving effective amplitudes that might be induced, for example, by dark photons with 
a tiny admixture of the Standard Model $Z^0$ boson~\cite{ref:cl:darkz}. 
\end{enumerate}

The electroweak theory prediction at tree level in terms of the weak mixing angle is $Q^e_W = 1 - 4\sin^2\theta_W$; 
this is modified at the 1-loop level~\cite{ref:cl:Czarnecki:1995fw, ref:cl:Czarnecki:2000ic, ref:cl:Erler:2004in} and
becomes dependent on the energy scale at which the measurement is carried out, {\em i.e.} $\sin^2\theta_W$
``runs". At low energy, $Q^e_W$  is predicted to be 
$0.0469\pm 0.0006$, a $\sim 40$\%\ change of its tree level value of $\sim 0.075$ (when evaluated at $M_Z$).

The value of $A_{PV}$ at the MOLLER energy is 
$\approx 35$~parts per billion (ppb); the expected statistical precision is 0.73 ppb,
providing a 2.3\%\ measurement of $Q^e_W$. The reduction in the numerical value of $Q^e_W$ 
due to radiative corrections leads to increased fractional accuracy in the determination of the weak mixing
 angle, $\sim 0.1$\%, comparable to the two best such determinations from measurements of asymmetries in
$Z^0$ decays at LEP and SLC. 

Figure.~\ref{fig:cl:s2tw} shows
the four best measurements from studies of $Z^{0}$ decays~\cite{ref:cl:lepewwg}
as well as the projected uncertainty of the MOLLER proposal. Also shown
is the Standard Model prediction for a Higgs mass ($m_H$) of 126 GeV. The grand average of the
four measurements is consistent with the theoretical expectation, but the scatter in the measurements
somewhat large;  MOLLER would provide an additional measurement with comparable precision. 


At the level of sensitivity probed, the proposed measurement could be influenced by radiative loop effects of new 
particles predicted by the Minimal Supersymmetric Standard Model (MSSM). The impact on the weak charges of the
electron and the proton $Q_W^{e,p}$ have been analyzed in detail~\cite{ref:cl:Kurylov:2003zh}. A combined analysis of 
precision low energy measurements of both charged and neutral current processes can be found in a comprehensive review~\cite{ref:cl:RamseyMusolf:2006vr}, which has been recently updated~\cite{ref:cl:Erler:2013xha}. 
Inspecting a random scan over a set of MSSM parameters whose values are consistent with current precision 
measurements as well as the most recent LHC search limits from 7 and 8 TeV running, $A_{PV}$ would see 
in the effects in the range of 2 and 3 $\sigma$ at larger values of the MSSM parameter $\tan\beta$ (the ratio of vacuum expectation values of the model's two Higgs scalars) or if one of the superpartner 
masses is relatively light. 
If the assumption of R-parity conservation is relaxed (RPV), tree-level interactions could generate deviations 
in $A_{PV}$ of opposite sign and similar magnitude. Thus, if nature is supersymmetric, the proposed measurement
would shed light on an important followup question regarding the validity of R-parity symmetry.

{\vskip -12pt
\begin{figure}[ht]
\begin{minipage}[b]{0.48\linewidth}
\centering
\includegraphics[width=0.95\linewidth]{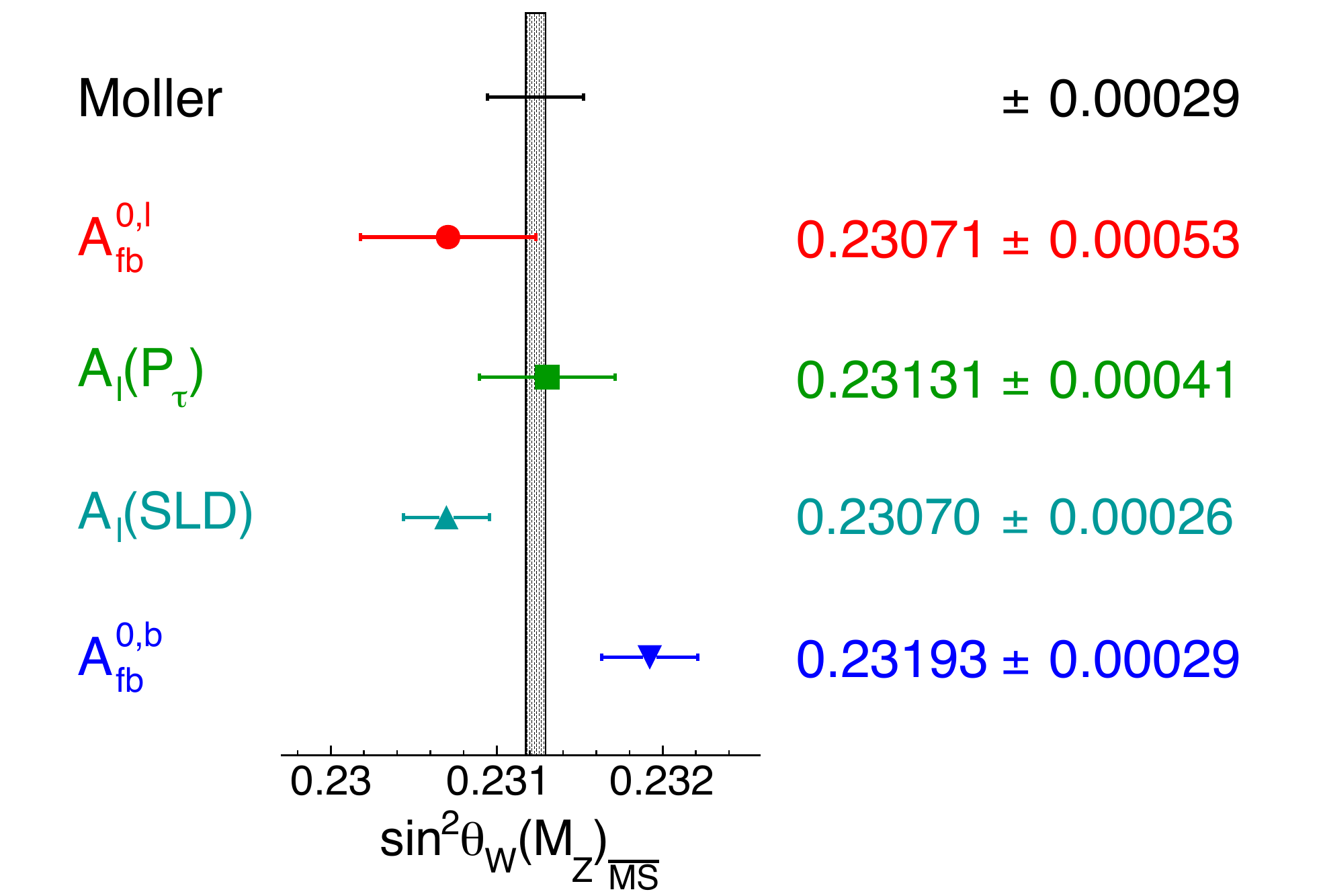}
  \caption{{\it The four best $\sin^2\theta_W$ measurements and the projected error of the MOLLER proposal.
  The black band represents the theoretical prediction for $m_H = 126$ GeV.}}
\label{fig:cl:s2tw}
\end{minipage}
\hspace{0.3cm}
\begin{minipage}[b]{0.48\linewidth}
\centering
    \includegraphics[width=3in]{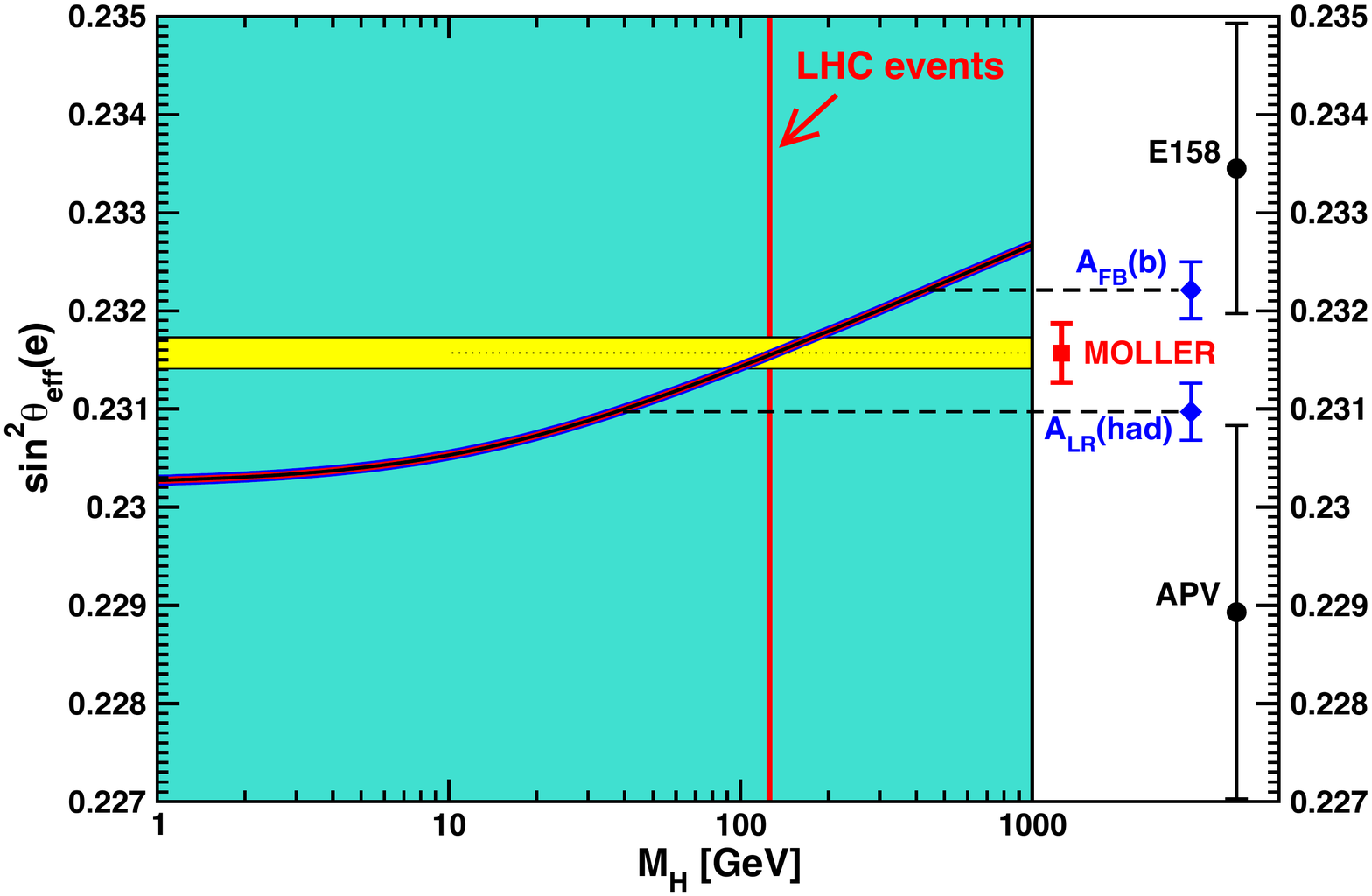}
  \caption{{\it $\sin^2\theta_W$ vs $m_H$. The yellow band shows the world average. The blue data points
  represent the two best high energy determinations while the black points are the most precise low energy
  determinations. The projected MOLLER error is shown in red. }}
  \label{fig:cl:s2twvsmh}
\end{minipage}
\end{figure}
}

A comprehensive analysis of the MOLLER sensitivity to 
TeV-scale $Z^\prime$s has been carried out~\cite{ref:cl:Erler:2011iw} 
for a class of family-universal models
contained in the $E_6$ gauge group.  While models with full $E_6$ unification are already
excluded by existing precision electroweak data, the $Z^\prime$ bosons in these models with the same electroweak charges to SM particles are still motivated, since 
they also arise in many superstring models as well as from a bottom-up approach~\cite{ref:cl:Erler:2000wu}. 
$A_{PV}$ probes
 $M_{Z^\prime} $ of order 2.5 TeV, comparable to the anticipated reach of 
early LHC running after the energy ramp-up to 13 TeV. The reach of $A_{PV}$ would be further 
enhanced in comparison 
to direct searches if one relaxes the model-dependent assumption of GUT coupling strength; indirect 
deviations scale linearly with other values of the coupling strength whereas dilepton production at colliders
has a much milder dependence on this parameter.

The measurement would be carried out in Hall A at Jefferson Laboratory, where a 11 GeV longitudinally polarized electron beam would be incident on a 1.5 m liquid hydrogen target.
The experimental techniques for producing an ultra-stable polarized electron beam, systematic
control at the part per billion level, calibration techniques to control normalization errors including the degree of electron 
beam polarization at the 1\%\ level have been continuously improved over fifteen years of development at JLab. The goal is to obtain construction funding for MOLLER by 2015, with 
the hope of installing the apparatus and commissioning the experiment in 2017/18.



\chapter{Conclusions}\label{sec:cl:gm2edmdisc}
\vskip 6pt
The enormous physics potential of the charged
lepton experimental program was very much in evidence at this Workshop. There are discovery opportunities both in experiments that will be conducted over the coming decade using existing facilities and in more sensitive experiments possible with future facilities such as Project X.
Sensitive searches for rare decays of muons and tau leptons, together with precision measurements of their properties will either elucidate the scale and dynamics of flavor generation, or
limit the scale of flavor generation to well above $10^4$ TeV.  This information
will be vital to understanding the underlying physics responsible for new particles discovered at the LHC.

The crown jewel of the program is the discovery potential of muon and tau decay experiments searching for charged lepton flavor violation with several orders-of-magnitude improvement in sensitivity in
multiple processes.  This is an
international program, with experiments recently completed, currently running, and
soon to be constructed in the United States, Japan, and Europe.  The potential program is interesting over the near term, with the completion of the MEG experiment measurement at PSI and the new Mu2e experiment at Fermilab, but is substantially improved by new facilities such as Super $B$ factories with polarized beams, and over the longer term, by experiments exploiting megawatt proton sources such as Project X.

Over the next decade
gains of up to four orders-of-magnitude are feasible in
muon-to-electron conversion and in the $\mu \to 3 e$
searches, while gains of at least two orders-of-magnitude
are possible in $\mu \to e\gamma$ and $\tau \to 3\ell$ decay and more than one order of magnitude in $\tau \to \ell\gamma$ CLFV
searches.  Existing searches already place strong constraints on
many models of physics beyond the standard model; the contemplated improvements increase these constraints significantly, covering substantial regions of the parameter space of many new physics models.
These improvements are important regardless of the outcome of new particle searches of the
LHC; the next generation of CLFV searches are an essential
component of the particle physics road map going forward.  If the LHC finds new
physics, then CLFV searches will confront the lepton sector in ways
that are not possible at the LHC, while if the LHC uncovers no sign of
new physics, CLFV may provide the path to discovery.

In general, muon measurements have the best
sensitivity over the largest range of the parameter space of many new
physics models. There are, however, models
in which  rare tau decays could provide the discovery
channel. It was clear from the discussion that as many different
CLFV searches as feasible should be conducted, since the best discovery
channel is model-dependent and the model is not yet known.  Should a
signal be observed in any channel, searches and measurements in as
many CLFV channels as possible will be crucial to determining the nature
of the underlying physics, since correlations between the rates
expected in different channels provide a powerful discriminator among
physics model.

The new muon $g\!\!-\!\!2$ experiment will measure the anomaly to close to 100 parts per billion precision. This will be an important measurement whether or not the LHC sees new physics. If the LHC sees SUSY-like new physics, $g\!\!-\!\!2$ will be used as a constraint in determining which model we see. The LHC will be particularly sensitive to color super-partners, while $g\!\!-\!\!2$ can pin down the flavor sector. The sensitivity of $g\!\!-\!\!2$ to $\tan\beta$ will provide a test of the universality of that parameter. If the LHC does not see new physics, then $g\!\!-\!\!2$ can be used to constrain other models, such as those involving dark photons and extra dimensions. Any new physics model will have to explain the discrepancy between the theoretical and experimental values of $g\!\!-\!\!2$.
The reduction of theory errors in the calculation of $g\!\!-\!\!2$ is thus also of great importance, particularly the contribution of light-by-light scattering, to maximize the new physics discovery potential.
New lattice calculations to be undertaken by, among others, members of the USQCD collaboration. as well as results from the KLOE, \babar\ and BES-III experiments, will help put the candidate models on firmer ground. A Super $B$ Factory with a polarized electron beam can measure, for the first time, the anomalous moment of the $\tau$, using new variables encompassing the $\tau$ polarization.

The search for EDMs will also play an important role in new physics
searches. The achievable limit on the electron EDM is the most stringent, but searches for muon and tau EDMs are nonetheless of interest, since new physics contributions scale as the lepton mass. These can be
important: if an electron EDM were to be found, the value of second and third generation EDMs would be of great interest.  Parasitic measurements with the new Fermilab $g\!\!-\!\!2$ experiment will improve the $\mu$ EDM limit by two
orders of magnitude. Improvement of this limit would also help to rule out
the possibility that the muon EDM is the cause of the current discrepancy in the
$g\!\!-\!\!2$ measurement. New dedicated experiments now being discussed
could bring the limit down to the $10^{-24}$ $e$cm level, making it
competitive with the electron EDM constraints. In the same vein, a Super $B$ Factory with a polarized electron beam can reach a sensitivity below $10^{-21}$ $e$cm.
Additional symmetry tests will also be possible, including sensitive searches for $C\!P$ violation in $\tau$ decay and 
tests of electroweak parity violation using electron scattering and $e^+e^-$ collisions. 

An exciting program of sensitive searches for new physics using the large samples of $\mu$ and $\tau$ decays in experiments at the intensity frontier awaits us. These experiments will likely be central to our understanding of physics beyond the Standard Model.




\begin{thebibliography}{99}
\bibitem{MoveLiveGate} We investigated the possibility of also
beginning the live gate at an earlier time, but this resulted in an
unacceptable increase in background events from radiative pion
captures.
\end{thebibliography}
\end{document}